\documentclass[linenumbers]{aastex631}

\nolinenumbers

\newcommand\red{\color{red}}
\newcommand\blue{\color{blue}}

\shorttitle{Extended outflows in different AGN spectral classes}
\shortauthors{Torres-Papaqui et al.}
\graphicspath{{./}{figures/}}

\begin{document}
\title{The role of AGN winds in galaxy formation: connecting AGN outflows at low redshifts to the formation/evolution of their host galaxies}

\author[0000-0002-8009-0637]{Juan Pablo Torres-Papaqui}
\affiliation{Departamento de Astronom\'{\i}a, Universidad de Guanajuato\\
Callej\'on de Jalisco S/N, Col. Valenciana CP: 36023 Guanajuato, Gto, M\'exico}

\author[0000-0001-6927-522X]{Roger Coziol}
\affiliation{Departamento de Astronom\'{\i}a, Universidad de Guanajuato\\
Callej\'on de Jalisco S/N, Col. Valenciana CP: 36023 Guanajuato, Gto, M\'exico}

\author[0000-0002-4216-7138]{Aitor C. Robleto-Or\'us}
\affiliation{Centro de Investigaci\'on de Astrof\'{\i}sica y Ciencias Espaciales (CIACE), Universidad Nacional Auton\'onoma de Nicaragua (UNAN-Managua), Rotonda Universitaria Rigoberto L\'opez P\'erez 150 metros al Este, C.P.: 14172 Managua, Nicaragua}

\author[0000-0002-1489-6229]{Karla A. Cutiva-Alvarez}
\affiliation{Departamento de Astronom\'{\i}a, Universidad de Guanajuato\\
Callej\'on de Jalisco S/N, Col. Valenciana CP: 36023 Guanajuato, Gto, M\'exico}

\author{Paulina Roco-Avilez}
\affiliation{Departamento de Astronom\'{\i}a, Universidad de Guanajuato\\
Callej\'on de Jalisco S/N, Col. Valenciana CP: 36023 Guanajuato, Gto, M\'exico}






\begin{abstract}
\nolinenumbers
Using SDSS spectra, we applied an automatic method to search for outflows (OFs) in three large samples of narrow-line AGN at low redshifts ($z < 0.4$), separated in three spectral activity classes: radio-loud RG, 15,793, radio-quiet, Sy2, 18,585, and LINER, 25,656. In general, the probability of detecting an OF decreases along the sequence Sy1$\to$Sy2$\to$LINER/RG and, independently of the AGN class, the wind velocity, traced by W80, increases with the AGN luminosity. Moreover W80 is systematically higher in RG or any of the other AGN class when detected in radio. These results support the idea that there are two main modes of production of OF, the radiative mode dominant in radio-quiet AGN and the jet mode dominant in radio-loud galaxies, although both modes could also happen simultaneously at different levels. From the spectra and SDSS photometry, the characteristics of the AGN host galaxies and their super-massive black holes (SMBHs) were also retrieved using the stellar population synthesis code STARLIGHT. This revealed that, independently of their spectral class, 1) galaxy hosts with OFs have systematically later morphological types and higher star formation rates than their counterpart without OF, 2) they occupy different positions in the specific diagnostic diagram, sSMBH vs. sSFR, which suggests they follow different evolutionary paths congruent with the morphology of their galaxy hosts, and 3) they show no evidence of AGN quenching or triggering of star formation. These results are consistent with a scenario explaining the different AGN classes as consequences of different formation processes of galaxies: early-type galaxies (LINER and RG) formed bigger bulges and more massive SMBHs, exhausting their reservoir of gas more rapidly than late-type galaxies (Sy2 and Sy1), quenching their star formation and starving their SMBHs. 


\end{abstract}



\section{Introduction} \label{sec:intro}

It seems now clear that outflows (OFs) are ubiquitous in AGN \citep{Woo2016,Fiore2017,Cresci2018,2020Torres-Papaqui}. However, what role these OFs play in the evolution of the AGN host galaxies is still an open question, as multi-frequencies observations are difficult to interpret following one unifying model \citep{Harrison2018,2022Smirnova-Pinchukova,2023Speranza,2023Harrison} and evidence of feedback are either negative \citep{2019Shin,2019Chen,2019Shimizu,2020Torres-Papaqui,2021Ruschel-Dutra,2022Molina,2022Tripodi,2023Kakkad} or ambiguous \citep{2019Fluetsch,2020Wylezalek,2020Scholtz,2021Luo,2022Tamhane,2023Veilleux,2023Kim,2023Salome,2023Veilleux,2023Kharb}.


The actual paradigm upholds that OFs triggered by supermassive black holes (SMBHs) at the center of galaxies regulate the mass of the stellar bulge of their host galaxies by quenching star formation, explaining the M$_{BH}$-$\sigma$ relation \citep{Magorrian1998,Ferrarese2000,Haring2004,Gultekin2009,Graham2011}.
The main argument in terms of energy is apparently compelling \citep{2012Fabian}. Comparing the binding energy of the stars in the bulge, ${\rm E}_{bulge} \sim {\rm M}_{bulge} \sigma^2$, where $\sigma$ is their velocity dispersion, with the energy radiated by the BH accretion disk, ${\rm E}_{BH} \sim \eta {\rm M}_{BH} {\rm c}^2$, where ${\rm M}_{BH}$ is the mass of the BH and $\eta$\ the radiative efficiency (c being the velocity of light), the latter easily dominates over the former (for example,  ${\rm E}_{BH}/{\rm E}_{bulge} = 225$, adopting standard values, $\eta = 0.1$, ${\rm M}_{BH}\sim 10^{-3}\ {\rm M}_{bulge}$, $\sigma \sim 200$ km s$^{-1}$). This implies that even a small percentage of the energy radiated by an AGN, assuming it founds a channel out of the central region, could perturb the dynamics of the gas, quenching star formation in its host galaxy \citep{2019HerreraCamus,2020Davies,2021Bischetti,2021Circosta,2021Vayner,2022Luo,2023Ayubinia}. This is the AGN feedback hypothesis \citep{Silk1998}. 

However, finding a channel out of the center of a galaxy could be more difficult than it seems, because of the differences in spatial scales \citep{1997Peterson}. A typical SMBH has a Schwarzschild radius of the order of $10^{-6}$~pc while the two most central structures, accretion disk and broad-line region, followed by the sublimation zone (a region depleted of dust due to the intensity of radiation; the central zone of maximum influence of AGN radiation) have generic scales of the order of only $10^{-4}$, $10^{-3}$ and $10^{-2}$~pc, respectively. Further away from the center, the next important structure is the obscuring torus, assumed to form in all AGN, which has a size that could vary between 0.1 to 10~pc. According to the unification scheme, the torus being rich in dust shields the BH at its center, making the radiation field of an AGN anisotropic, forcing OFs to follow the path of least resistance \citep{2020AHerreraCamus,2021Rupke,2021GarciaBurillo,2021AlonsoHerrero,2022Yu,2022Molina}. 
Finally, considering the common size of bulges in spiral galaxies \citep[tyically 1 kpc;][]{2008Kormendy}, 
the region where we are certain interstellar gas (ISM) is still abundant (remembering that OFs only affect the gas) is the narrow line region (NLR), an extended region of ionized gas that could reach 
a few 10~kpc or more \citep{2021Meena}. In all, this represent an increase in scale by at least $10^{5}-10^{8}$ compared to the accretion disk. This implies that the mechanisms necessary to make OFs efficient over a kpc scale, disturbing the gas dynamics and star formation, must use a significant fraction of the energy produced by the SMBH \citep{2021Luo,2021Wang,2021Shi,2021VidalGarcia,2022Brusa}. Alternatively, this might suggest that AGN feedback is mostly efficient at small scales, massive OFs hampering more gas to fall onto the accretion disk, starving the BH \citep[][]{2019Molyneux,2020Nomura,2021Lusso,2021Nomura,2022Tripodi}. 

There are two AGN feedback modes proposed in the literature: 1- the radiative mode (or quasar mode), which produces OFs driven directly by radiation pressure (the Eddington luminosity being the limit) or conservation of momentum at different scales and with different levels of energy \citep{2020Somalwar,2020Igo,2020Marasco,2020Smith,2021Sibasish,2021Mizumoto,2021Ishibashi}, and 2- the jet mode (or kinetic mode) which is more typical of radio-loud AGN \citep{2021Schulz}, although in radio-quiet AGN both modes could act at the same time, low-intensity radio jets acting at smaller scales  \citep{2020Santoro,2021Biny,2021Jarvis,2021Rosario}. However, which OF mode dominates in AGN with different spectral classes and how they affects the galaxy hosts with different morphological types are questions that need to be more thoroughly investigated \citep[e.g.][]{2020Rojas,2020Brownson,2020Sebastian}. 

In our previous study \citep{2020Torres-Papaqui}, we developed a method to automatically detect and measure ``resolved'' OFs in the emission line [OIII]$\lambda5007$ \AA\ (that is, Doppler structures with blue shifts higher than the spectral resolution used) and applied it to a sample of 3,896 Seyfert~1 AGN (Sy1) with $z \leq 0.4$. After confirming the high frequency of such structures---37\% of the galaxies in our sample---and
established through a multi-correlation analysis that OFs are consistent with AGN winds---radiatively launched and triggered by higher accretion rates---we then searched for evidence of direct feedback effect on the star formation of their host galaxies. Having found no such evidence, we then try to verify the delayed hypothesis, as suggested by \citet{Cresci2018}, which postulates that the time-scale of the effects are longer than the time-scale of the activity of the SMBHs \citep[e.g.,][]{2020Scholtz,2020Wylezalek,2021Luo}. However, considering the physical differences between Sy1B and Sy1N and their different ranges in redshifts \citep[see explanations in][]{2020Torres-Papaqui}, we determined that in Sy1 a typical delay would need to be longer than 3 or 4 Gyr, which seems unreasonably large compared to the rapid depletion time-scale of the gas (of the order of a few $10^8$ yr but lower than $10^9$ yr) due to its ejection and consumption by stars \citep[e.g.,][]{2019Loiacono,2019LeFevre}. Confronted with these negative results about the purported effect of AGN winds, we therefore postulated another explanation, which is that the feedback effects we are looking for might have happened after the main AGN activity phase, that is, once the galaxies are not recognizable anymore as Sy1 \citep[e.g.,][]{2019Fluetsch}. 
More specifically, assuming OFs could also starve their SMBHs, it should be possible to distinguish feedback effects in galaxies with lower levels of AGN activity, for example, Sy1 transforming into Sy2 (Seyfert~2) or LINER (Low-Ionization Nuclear Emission-Line Regions) or some other forms of low-luminosity AGN (LLAGN), where emission lines become so weak that they cannot be measured or detected. This led us to contemplate a possible evolution connection between all the AGN types, along the sequence QSO/Sy1$\to$Sy2$\to$LINER$\to$LLAGN/RG$\to$NoAGN. 


To better assess the probability of such evolutionary pattern and complete our view of AGN winds in nearby galaxies, we henceforth decided to extend our automatic search for OFs, started in \citet{2020Torres-Papaqui}, to a large samples of narrow-lines (type~2) AGN at low redshifts, with spectra available in SDSS DR8. Our sample is composed of 59,731 galaxies separated into three spectral activity classes: Sy2, LINER and RG. 
In Section~\ref{sec2:SmpDataOf}, we introduce our three samples and describe comprehensively how the galaxies were classified as AGN and how the characteristics of their SMBHs and host galaxies were retrieved. 
We then close Section~\ref{sec2:SmpDataOf} by a brief revision of how the OFs were detected and measured. In Section~\ref{sec3:RES}, we present and analyse our results, comparing in each AGN class the characteristics of the galaxy hosts and their SMBHs with and without OF. These comparisons include a discussion about the results of statistical tests (with figures in appendix), which were applied to confirm the statistical significance of the differences observed, and a discussion about the results of multi-correlations analyses. In Section~\ref{sec4:Disc}, we tentatively establish a connection between the different AGN spectral classes and the different formation/evolution processes of their respective host galaxies, consistent with our analysis and better delineate the role OFs could play during these processes. Our final conclusions can be find in Section~\ref{sec5:Con}. Note that all the physical parameters that depend on the distance were calculated assuming a $\Lambda$CDM cosmology, adopting the generic parameters $H_0 = 70$ km s$^{-1}$ Mpc$^{-1}$, $\Omega_{DM} = 0.30$, and $\Omega_{\Lambda} = 0.70$.

\section{Samples and data}
\label{sec2:SmpDataOf}

From the SDSS DR8 spectral catalog, we selected a large sample of galaxies with redshifts $z < 0.4$, which is the same limit we used for our previous study of resolved OFs in Sy1 \citep{2020Torres-Papaqui}. Using different scripts in Interactive Data Language (IDL), all these spectra were automatically treated and analysed. First, a correction for Galactic extinction is applied, using the extinction map of \citet{Schlegel1998} and the reddening law of \citet{Cardelli1989}, then each spectrum is corrected to its rest frame wavelength using its SDSS redshift. After that, STARLIGHT \citep{2005CidFernandes,2013Torres-Papaqui} is applied to fit templates of simple stellar populations (SSP), which are subtracted from the spectra, allowing the emission lines to be accurately measured. Keeping only those galaxies that have S/N~$\ge 3$ for the emission lines and S/N~$\ge 10$ for their adjacent continuum \citep{2012aJPTP,2013Torres-Papaqui}, we finally proceed in classifying the galaxies according to their spectral characteristics.  

\begin{figure*}[ht!]
\gridline{\fig{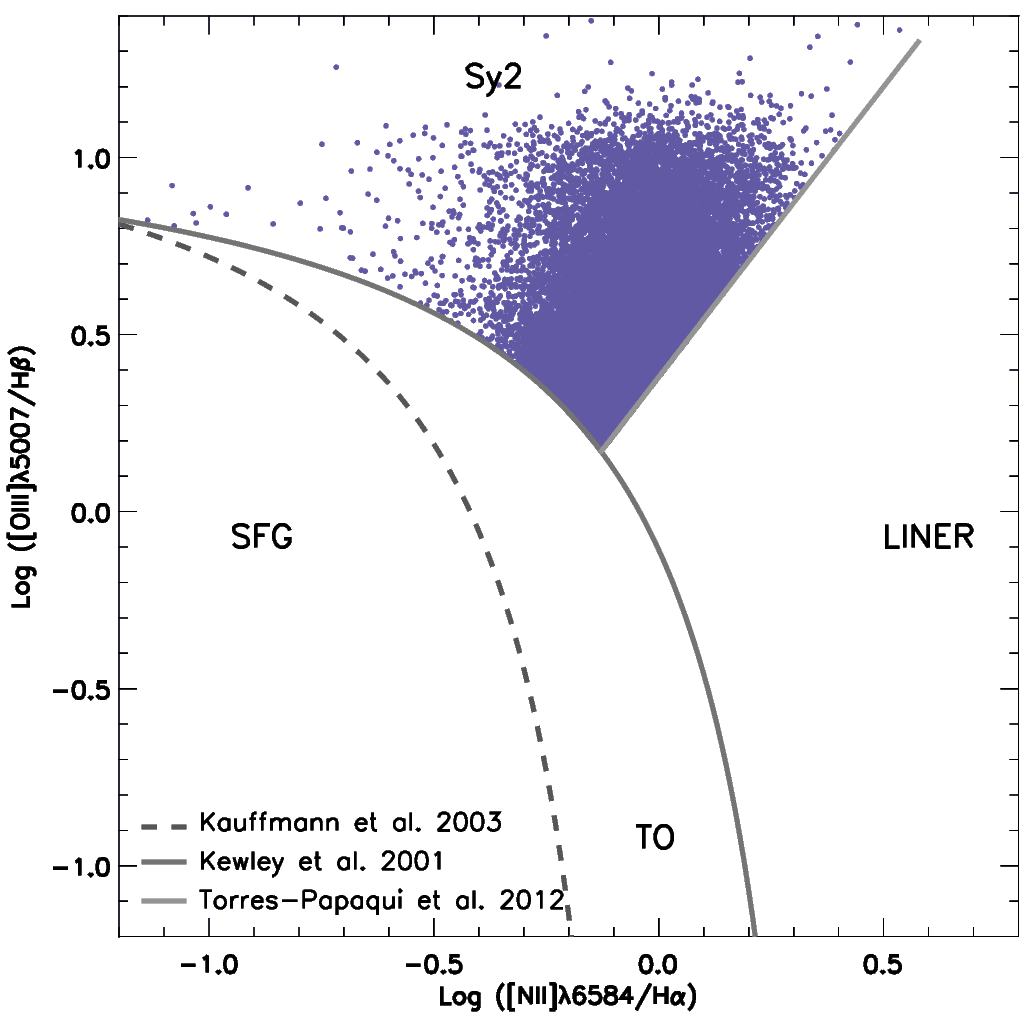}{0.4\textwidth}{(a)} 
\fig{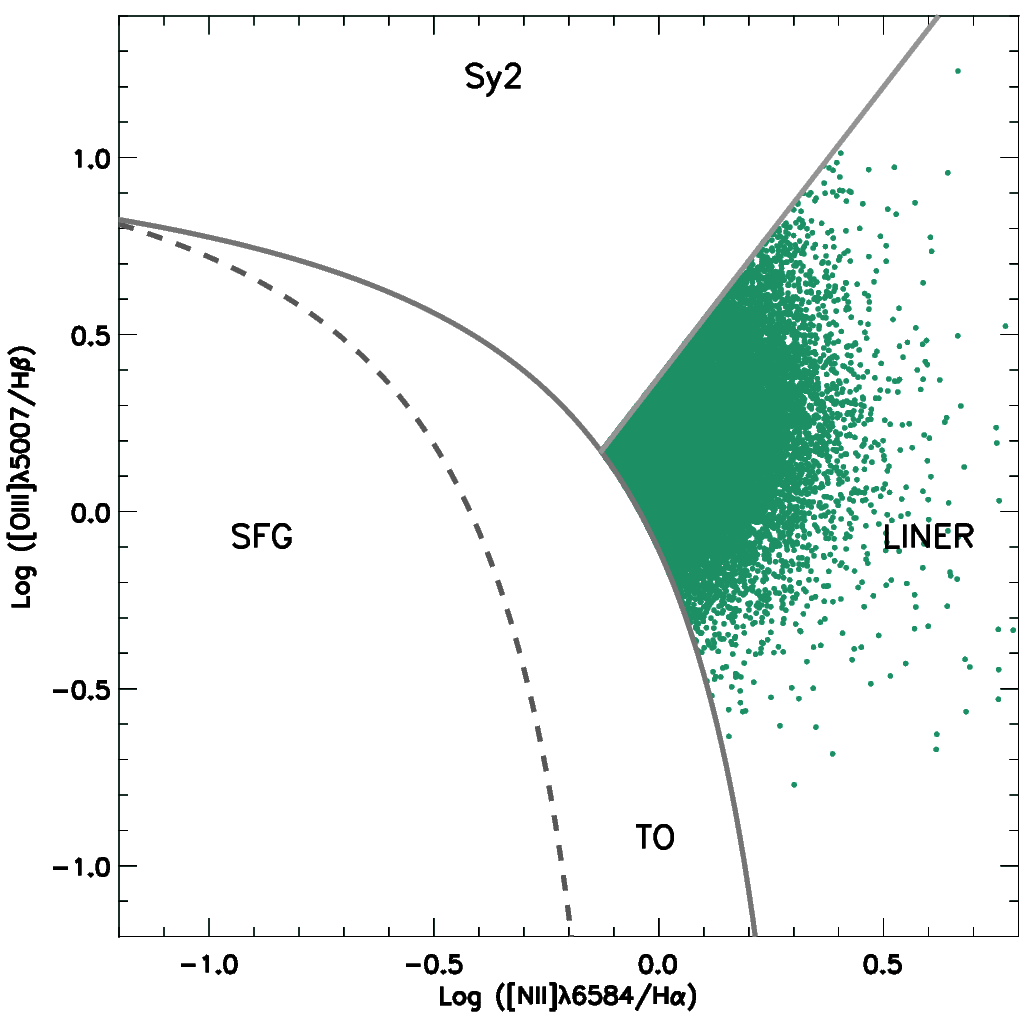}{0.4\textwidth}{(b)} }
\caption{Classification of AGN using standard (BPT) diagnostic diagram: (a) for Sy2 and (b) LINER. The different zones are as defined in \citet{2003Kauffmann}, \citet{2006Kewley} and \citet{2012aJPTP}.   
\label{fig1:EM_classification}
}
\end{figure*}

To define the different spectral classes of activity we used three standard diagnostic diagrams \citep{Baldwin1981,Veilleux1987}, applying the semi-empirical emission-line ratio relations as determined by \citet{2003Kauffmann},  \citet{2006Kewley} and \citet{2012aJPTP}. The results for Sy2 and LINER using one of these diagrams, [NII]/H$\alpha$ vs. [OIII]/H$\beta$ (hereafter BPT diagram), are shown in Figure~\ref{fig1:EM_classification}a and 1b, respectively. After applying the line ratio criteria, we identified 18,385 Sy2 and 25,656 LINER. Note that diagnostic diagrams based on different line ratios do not always concord in their classification, since whatever the line ratio criteria one uses, the separations between the types are somewhat arbitrary. This is particularly true in the case of Sy2 and LINER in Figure~\ref{fig1:EM_classification}, which actually trace a continuous sequence, suggesting these galaxies show more similarities than differences. 

\begin{figure*}[ht!]
\gridline{\fig{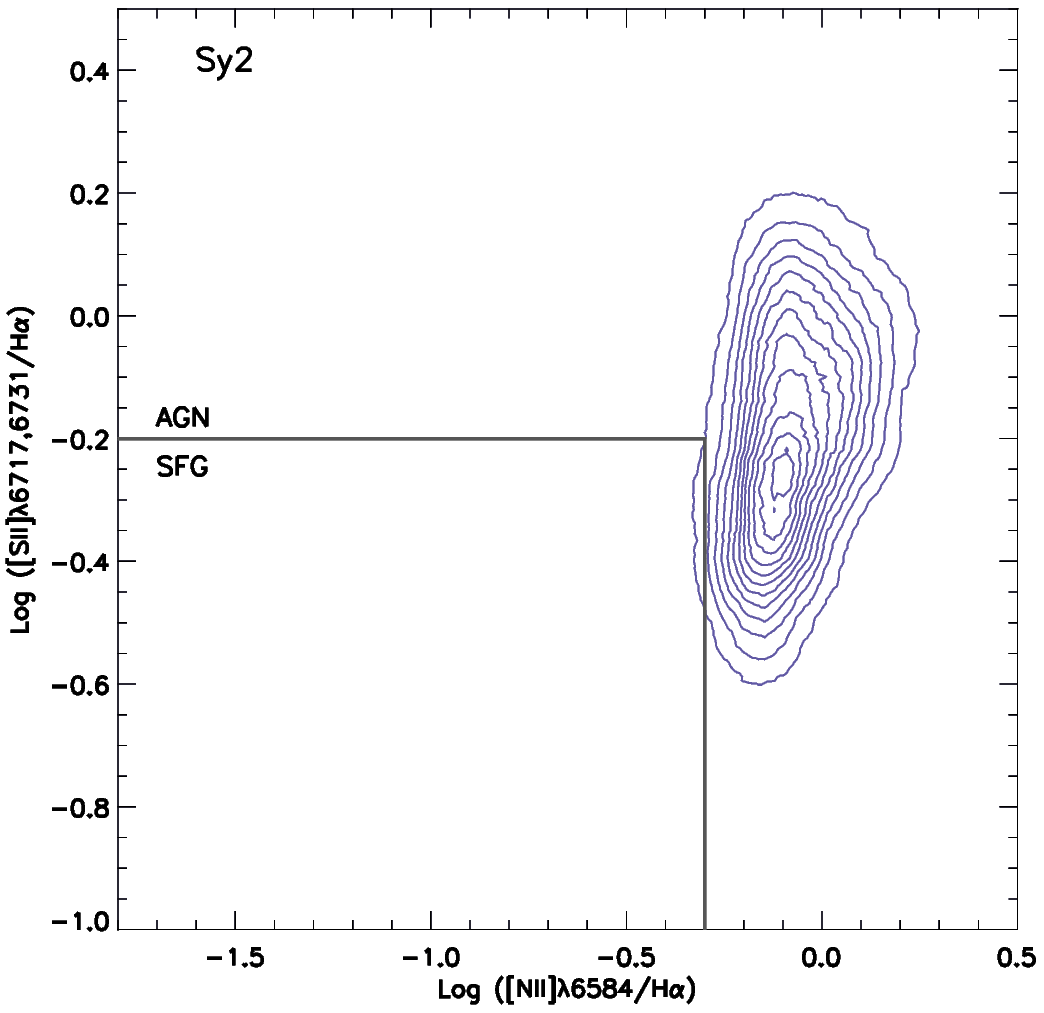}{0.4\textwidth}{(a)}
\fig{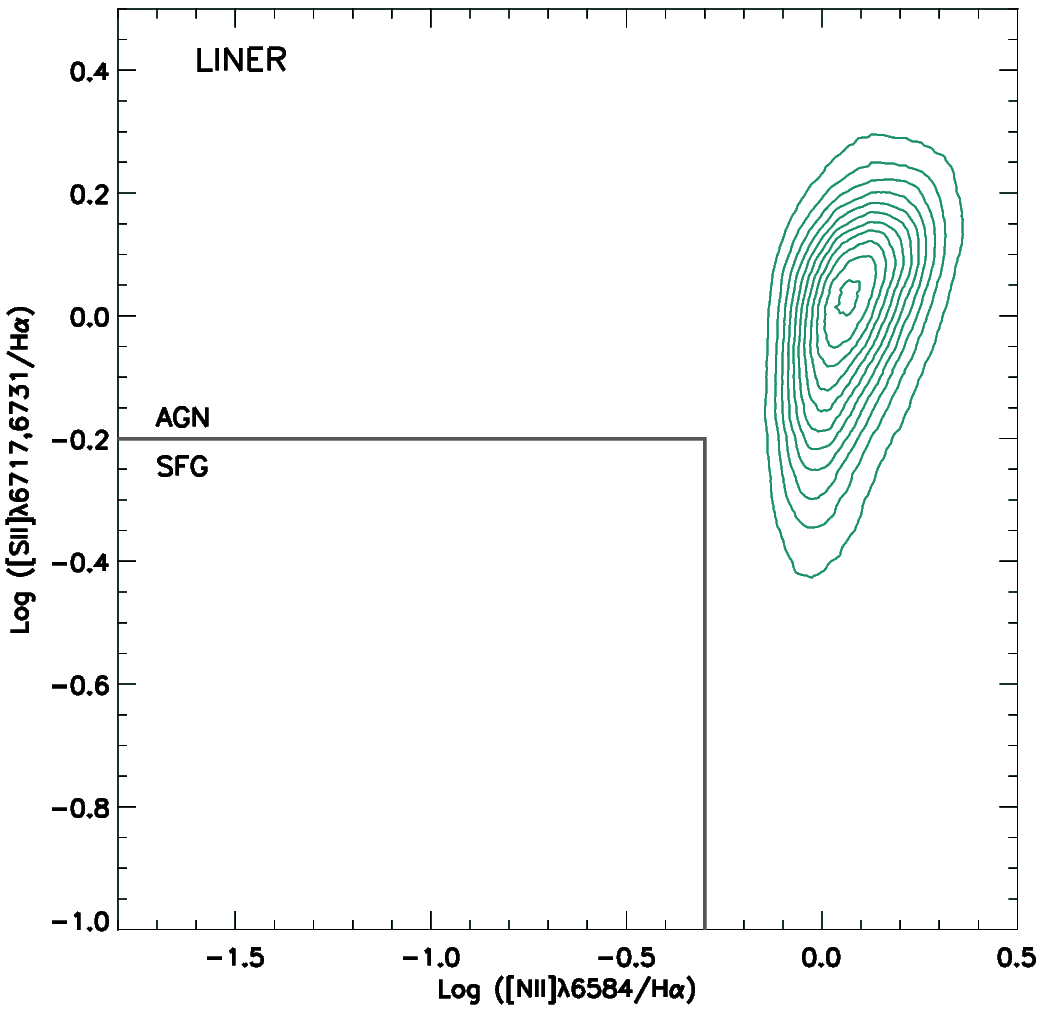}{0.4\textwidth}{(b)}  
}
\caption{Comparing the line ratios [SII]$\lambda 6717,6731/{\rm H}\alpha$ with [NII]$\lambda 6584/{\rm H}\alpha$ \citep{2012bJPTP}, and identifying  the zones with typical values for SFG and AGN, in (a) Sy2 and (b) LINER. Note the huge overlap between the distributions.\label{fig2:NIIvsII}
}
\end{figure*}

In Figure~\ref{fig2:NIIvsII}, we present the results for one different diagnostic diagram, which clearly favors AGN over star formation (in star-forming galaxies; SFG) as the main source of ionization of the gas in Sy2 and LINER. In this diagnostic diagram, AGN typically have higher ratios [NII]$\lambda 6584/{\rm H}\alpha$ and [SII]$\lambda 6717,6731/{\rm H}\alpha$ than SFG \citep{Coziol1998,2012bJPTP}; in general, LINER differs from Sy2 by the higher intensity of their low-ionization lines and their lower star formation rates \citep[SFRs;][]{2012aJPTP}.

For radio-loud galaxies (RG), we started with the sample of 18,282 radio-loud AGN as defined by \citep{2012Best}, which is available through the VizieR Online Data Catalog. After applying our redshift limit, the number of galaxies was reduced to 15,793, for which we retrieved their spectra in SDSS DR8 by cross-correlating their positions using the Vizier tool X-match.\footnote{http://cdsxmatch.u-strasbg.fr} At the moment of classifying our sample of RG using the same method as the other narrow-line galaxies, we were faced with an important difficulty, which is that even after subtracting the STARLIGHT templates, in 51.1\% of these radio sources (8,075 galaxies) the emission lines turned out to be either too weak to be measured or simply missing. Frequently, residual emission appear under the form of weak [NII] lines, while H$\alpha$ appear in absorption. This is a common trait of low-luminosity AGN
\citep[LLAGN;][]{Coziol1998,2014Coziol}, where Balmer absorption due to intermediate-age stellar populations (dominated by A type stars) is maximum. In our sample of RG with emission lines, 34\% (5,381 galaxies) have S/N $\geq3$ and 14.8\% (2,237 galaxies) have S/N $<3$. In order to classify all the RGs, therefore, we needed to use two different diagnostic diagrams: the BPT diagram for RGs with emission lines and an alternative diagnostic diagram proposed by \citet{2005Best} for those with no measurable or detected emission.  

\begin{figure*}[ht!]
 \gridline{\fig{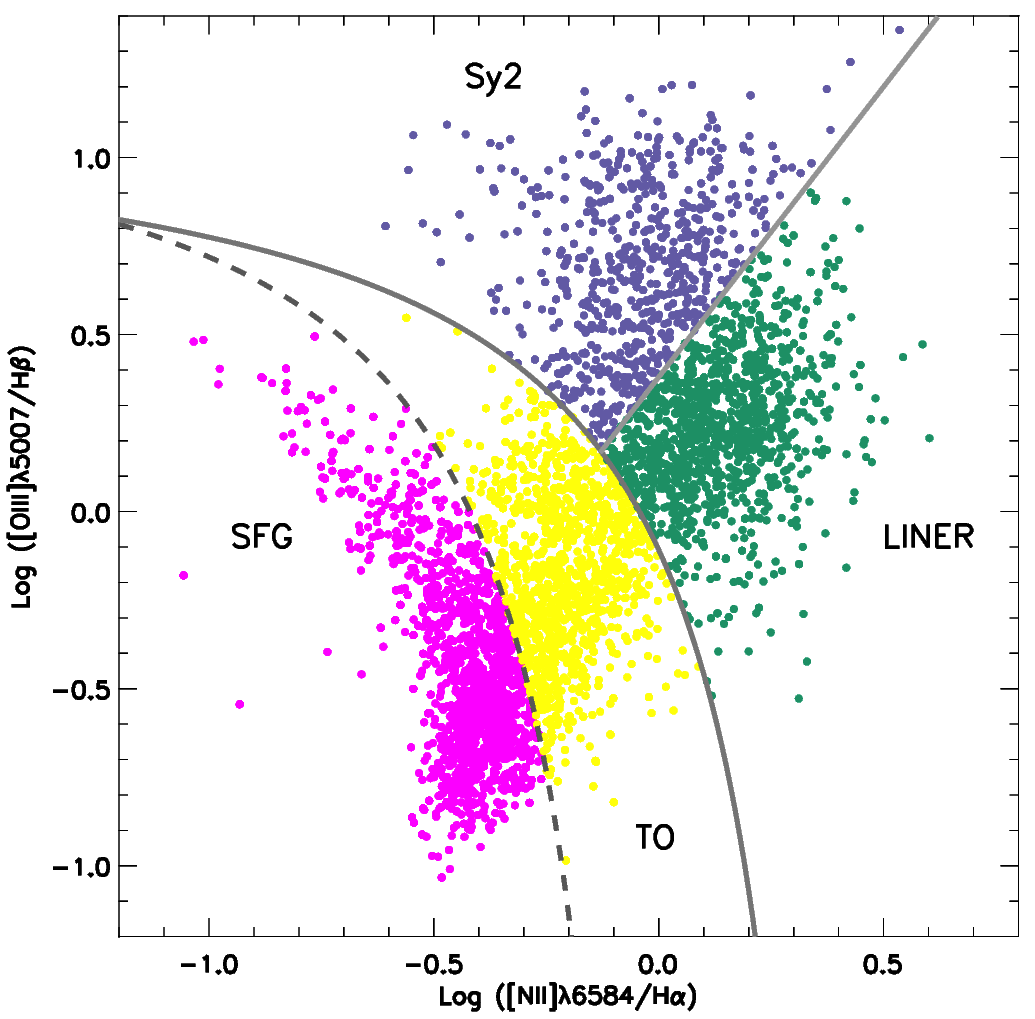}{0.38\textwidth}{(a)}
 \fig{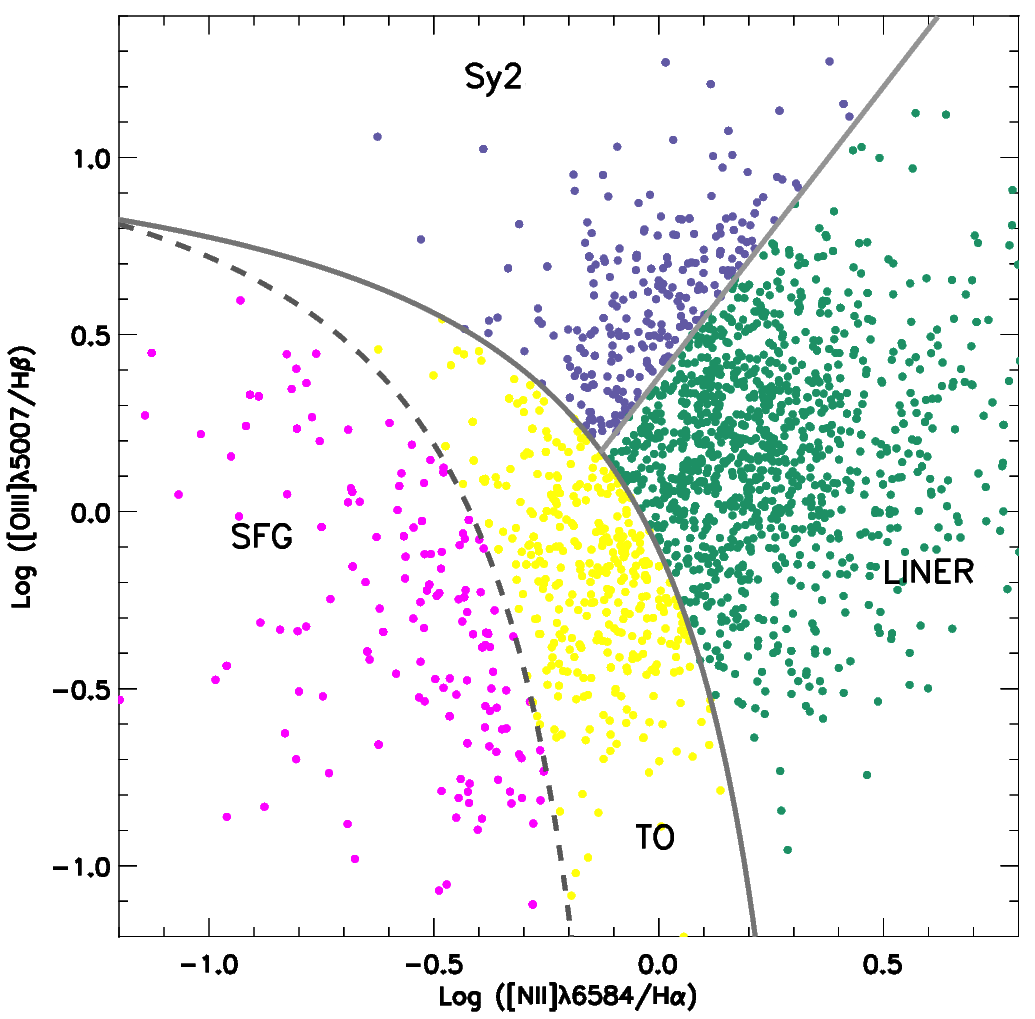}{0.38\textwidth}{(b)} }
\caption{BPT diagrams for RGs with emission lines, (a) with S/N $\ge 3$ and (b) S/N $<3$. The AGN type separations are the same as in Figure~\ref{fig1:EM_classification}.\label{fig3:BPT_RG}
}
\end{figure*}

\begin{figure*}[ht!]
\epsscale{0.5}
\plotone{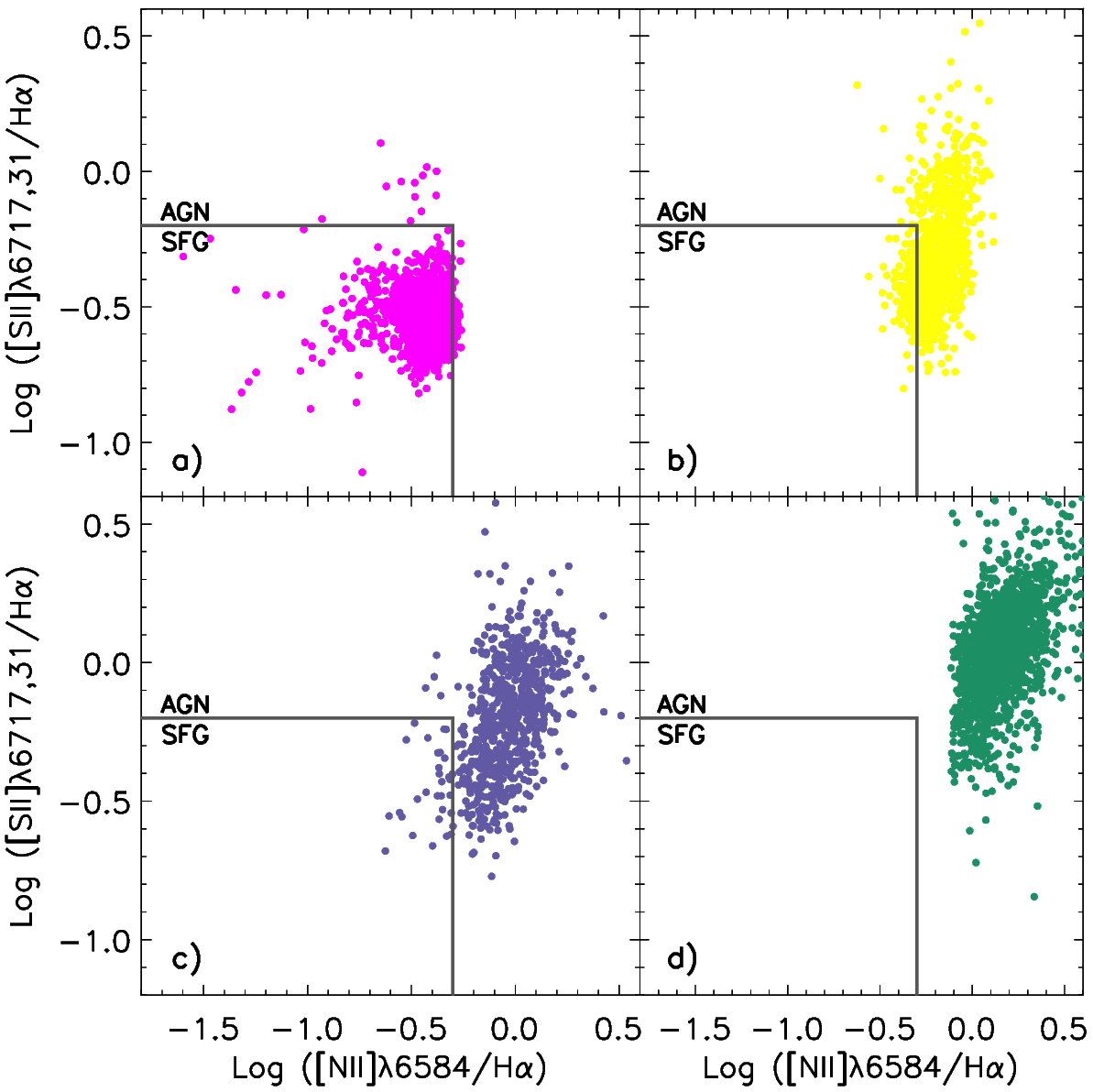}
\caption{Comparing the line ratios [SII]$\lambda 6717,6731/{\rm H}\alpha$ with [NII]$\lambda 6584/{\rm H}\alpha$ in all the RG with emission lines (both sub-samples with different S/N are used); a) SFG, b) TO, c) Sy2 and d) LINER. Note that there is almost no difference between Sy2 and TO in this diagram.
\label{Fig4:BPT_RG}
}
\end{figure*}

The BPT diagram for RGs with emission lines are presented in Figure~\ref{fig3:BPT_RG}. For RGs with high S/N in Figure~\ref{fig3:BPT_RG}a, 829 (15.1\%) are classified as Sy2, 1,460 (27.1\%) as LINER, 1,467 (27.3\%) as transition type objects \citep[TO;][]{Kewley2001} and 1,625 (30.2\%) as SFG. For those with low S/N in Figure~\ref{fig3:BPT_RG}b, 304 (13.0\%) are Sy2, 1,396 (59.7\%) LINER, 448 (19.2\%) TO and 189 (8.1\%) SFG. Consequently, there is an obvious trend in RGs with emission lines to be LINER-like as S/N decreases. This time, the second diagnostic diagram, [SII] vs. [NII], presented in Figure~\ref{Fig4:BPT_RG}, not only favors AGN as the main source of ionization of the gas in LINER and Sy2 but also in TO. However, to avoid ambiguities about the source of radio emission in TO and SFG, we will only considered as genuine radio-loud AGN those RG with emission identified as Sy2 and LINER. 

However, there is still the problem of determining the source of radio emission in the RG without optical emission lines. This is where the alternative diagnostic diagram for RG becomes important. This diagram consists in comparing the 4000 \AA\ break, D$_n(4000)$, with the radio luminosity at 1.4 GHz, L$_{NVSS}$, weighted by the stellar masses of the galaxies, M$_{*}$ \citep{2005Best}. In principle, this allows to distinguish AGN from SFG: because D$_n(4000)$ increases as SFR decreases, a high ratio L$_{NVSS}$/M$_{*}$ in a galaxy with high D$_n(4000)$, that is, in a galaxy where the star formation is either weak or nil (consistent with early-type galaxies), can only be explained by an active SMBH. Consequently, these could be securely classified as genuine radio-loud AGN.

\begin{figure}[ht!]
\epsscale{0.8}
\plotone{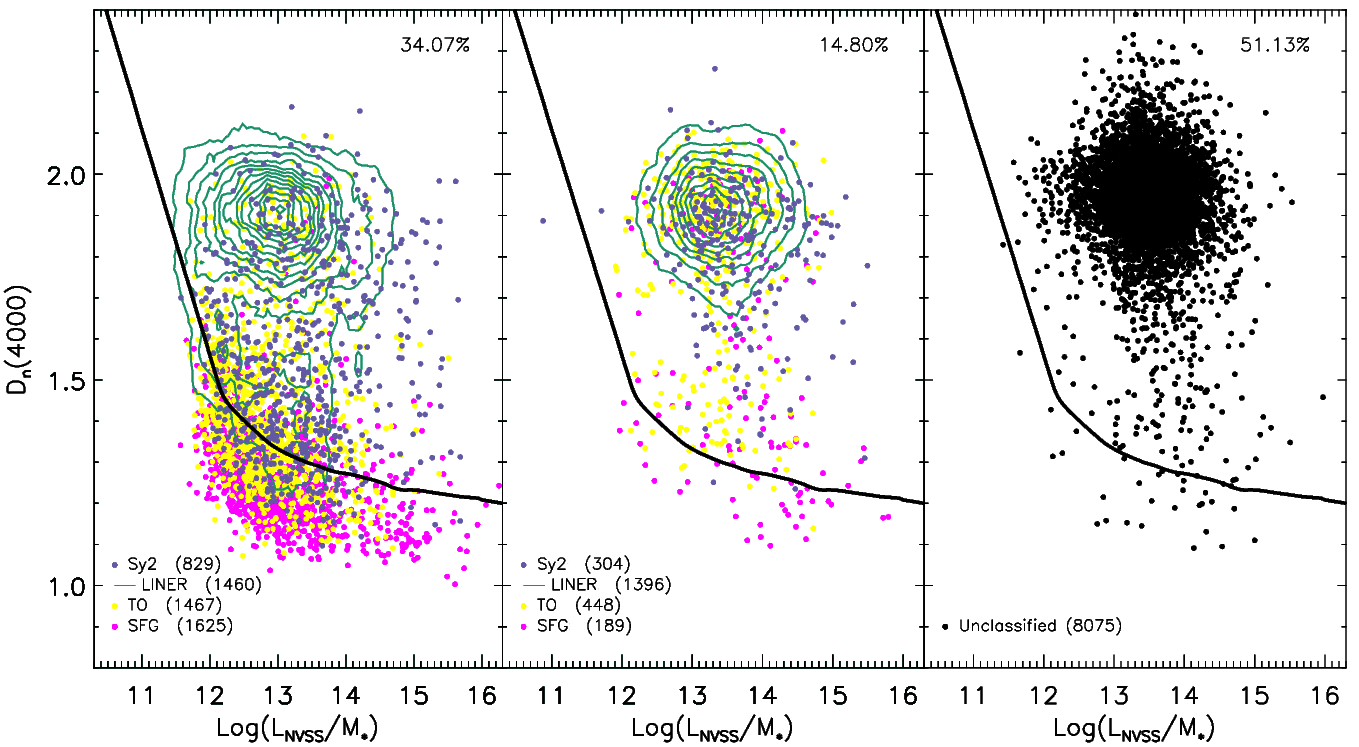}
\caption{Alternative classification for RG. In each panel, the solid curve separates SFG from AGN. Left panel, RG with intense emission lines (S/N $\ge 3$), classified using the BPT diagrams (Figure~\ref{fig3:BPT_RG}). Middle panel, RG with weak emission lines (S/N $ < 3$), but also classifiable using the BPT diagram. Right panel, RG with non-measurable or undetected emission lines. The percentages refer to the number of galaxies relative to the total sample. \label{Fig5:AltDiag_RG}}
\end{figure}

The value of D$_n(4000)$ in all our galaxies was measured automatically in the spectra of each RG using a routine written in IDL, the algorithm calculating the ratio of the average fluxes within the bands 4050-4250~\AA\ and 3750-3950~\AA\ \citep{2005Best}. The other parameters required for the alternative diagram are the radio luminosity at 1.4 GHz, as reported in the Vizier catalog, and the stellar mass, M$_*$, which was retrieved from the SSP templates obtained from STARLIGHT \citep{2005CidFernandes,2013Torres-Papaqui}. The program do it by multiplying the mass-luminosity ratio, M$_*$/L$_B$ by L$_B$ \citep{2003Kauffmann}: the latter is estimated using the synthetic B band luminosity, L$_B$, from the best fitted SSP, and the former is obtained by applying the relation between M$_*$/L$_B$ and $(B-R)$, as established by \citep{2003Bell}, using the transformation for the SDSS color $(B-R) = 1.506 (g-r) + 0.370$ as determined by \citet{1996Fukugita}. 

In Figure~\ref{Fig5:AltDiag_RG}, the alternative diagram for our three sub-samples of RG are shown. Note that the ratio L$_{\rm NVSS}/{\rm M}_*$ increases in those RG with non-measurable or undetected emission lines. According to this diagram, therefore, the source of radio emission in RG with weak or no emission in the optical could only be an active SMBH. This is also supported by the fact that the locus of these RG in this diagnostic diagram  is similar to the locus for LINER and Sy2 classified as AGN based on their emission lines in standard diagnostic diagrams (note however the few in the SFG region). This suggests that most of the RG in our sample (at least 77\%, but possibly as much as 88\% counting TO) have an active SMBH.  

\subsection{PAGB as an alternative to AGN}
\label{subsec2.1}

In principle, the presence of a SMBH in RG should not be viewed as ambiguous. Indeed, no other source could produce such powerful emission in radio (considering that only a few, 12\% in our sample, are SFG). However, since most of these galaxies only show weak or no emission lines, that is, apparently no evidence in the optical of a massive SMBH actively accreting matter, some researchers in the field find difficult to accept these galaxies as genuine AGN. In the past, this situation has introduced some ambiguities about the role AGN could play in the evolution of galaxies. 

For example, in 1980, \citet[][]{1980Heckman} discovered a huge numbers (30\%) of apparently normal spiral galaxies with low-ionization, narrow emission lines in their nucleus (calling them LINER), almost similar to what is observed in the more luminous Sy2 AGN. This important discovery led many researchers to conclude that SMBHs are ubiquitous in galaxies and that different levels of activity could be the key connection between galaxies and quasars, the latter being seen as an earlier phase in the evolution of galaxies. Therefore, when in 1986 \citet{1986Phillips} found numerous massive early-type galaxies in clusters with no star formation but weak emission lines in their nuclei, the simplest explanation was that these were evidence for a remnant of SMBH activity (akin to LLAGN). However, at the time many people were not ready to accept this idea, and more specifically, that SMBH could still be active in massive early-type galaxies in clusters. Instead, recognizing these galaxies are dominated by old stellar populations, \citet[][]{1994Binette} proposed that post-asymptotic giant branch stars (PAGB) could be the source of ionized gas in these galaxies. Still, in 1998 when \citet{Coziol1998} discovered in compact groups of galaxies that a huge number of early-type galaxies, like in clusters, show the same weak emission lines, more specifically, [NII]$\lambda$6548,6584 but no H$\alpha$ \citep[][]{1996Coziol,2011Coziol}, and that together with LINER they form the bulk of galaxies in these groups, again, the simplest explanation seemed that these were evidence for low-levels of SMBH activity (recognizing LLAGN and LINER as AGN). However, in \citet{2011CidFernandes} the PAGB idea was extended to also explain LINER in the field, changing the original interpretation of \citet[][]{1980Heckman} and thus unsettling the role SMBHs could play in the evolution process of galaxies. 

How does this affect our interpretation of RG as genuine AGN is the following. Although it seems clear PAGB cannot be at the source of radio emission in RG, a good number are LINER-like in the BPT diagram and in the alternative diagnostic diagram, and thus should might also be considered to be PAGB dominant. However, since they are deficient in gas, there are no evidence of PAGB in these galaxies. The connection of LINER-like RG with LINER that are are mostly radio quiet is consequently not evident. How can we distinguish between LINER that are PAGB dominant and those that are AGN is not at all obvious. 

\begin{figure}[ht!]
\epsscale{0.6}
\plotone{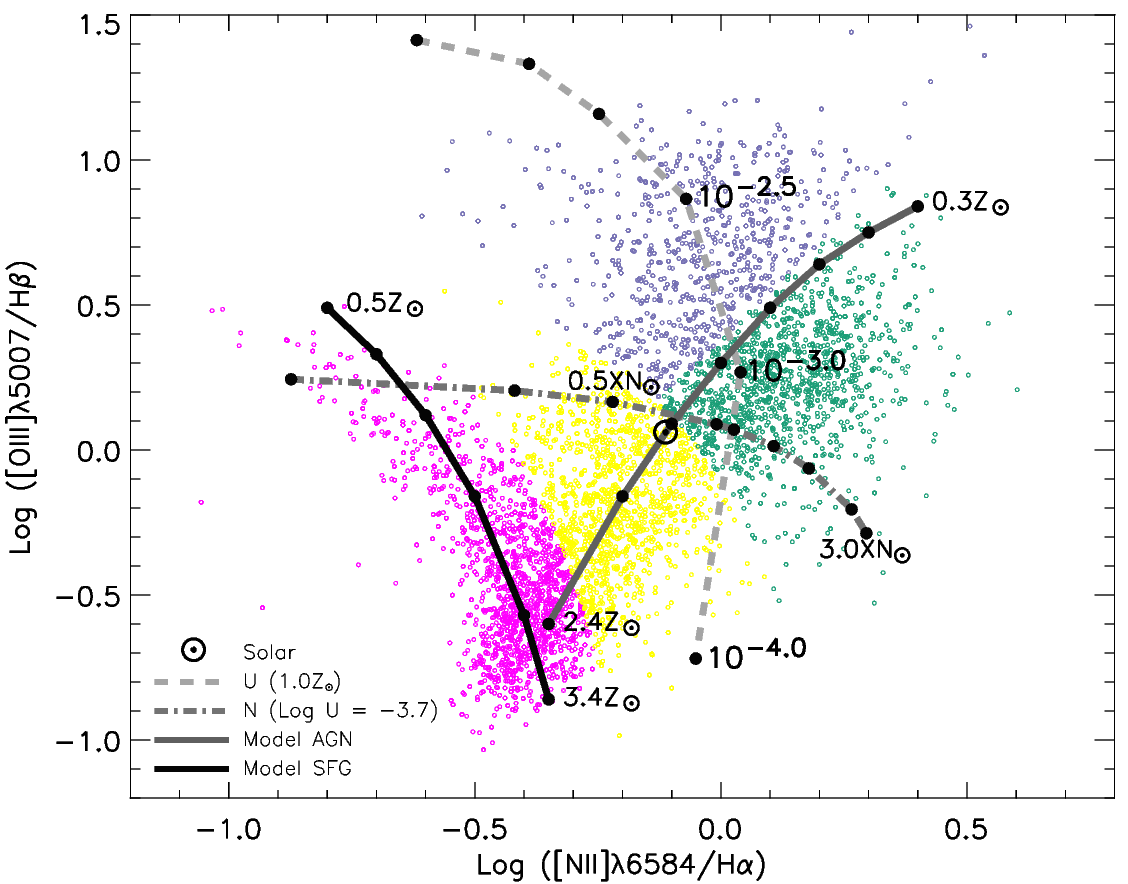}
\caption{BPT diagnostic diagram for the RG galaxies with emission lines compared with photo-ionization models calculated by \citet{2006aBennert}.  \label{Fig6:BPT_RG_model}}
\end{figure}

Actually, the debate about the AGN nature of LINER is not a new one, as the question was already considered and discussed at length in the seminal study of \citet[][]{Baldwin1981} and the followup by \citet{Veilleux1987}. To better illustrate the situation, we once again trace in Figure~\ref{Fig6:BPT_RG_model} the BPT diagram for RG with emission lines but without the empirical separations (we only keep the colors). In this diagram, emission-line galaxies trace a continuous distribution, having the form of the Greek letter $\nu$, with SFG on the left branch, with ratios Log([NII]$\lambda6584$/H$\alpha$)~$<-0.3$, and AGN on the right branch, with Log([NII]$\lambda6584$/H$\alpha$)~$\ge-0.3$, the branches separating at high metallicity (in SFG), corresponding to Log([OIII]$\lambda5007$/H$\beta$)~$\sim -0.6$ in this diagram. Using the photo-ionization models made by \citet{2006aBennert}, we show that the left branch is consistent with photo-ionization by massive stars, the ratio [OIII]$\lambda5007$/H$\beta$ decreasing as the metallicity increases \citep[see][and references therein]{2012bJPTP}, while the right branch is consistent with photo-ionization by accretion of matter onto a SMBH, where both ratios, [OIII]$\lambda5007$/H$\beta$ and [NII]$\lambda6584$/H$\alpha$ increase to the right as the metallicity decreases. 

Consequently, the first evidence favoring AGN in LINER is that they form a continuous sequence with Sy2, where the ratios [NII]$\lambda6584$/H$\alpha$ are higher than in SFG, and where the ionization parameter decreases, from $10^{-2.5}$ in Sy2 to $10^{-3.0}$ or even lower in LINER \citep{2006aBennert}, implying the ionizing sources of the gas in these two galaxies are similar, differing by the hardness of their radiation field (the SED of Sy2 producing more UV photons of high energy than LINER and/or the gas metallicity increases in the latter). In fact, this was the exact conclusion of \citet[][]{Veilleux1987}, who, considering the large overlap between Sy2 and LINER in any of the line ratios diagnostic diagrams (not only the one shown above), concluded that the dichotomy between high- and low-ionization AGN is insignificant, and that: “LINER probably constitute the lower part of a sequence of ionization of Seyfert 2”. Moreover, also according to \citet[][]{2006bBennert}, photo-ionization models predict that narrow line regions ionized by AGN could extend around the nucleus over kpc regions---with a concentration that decreases with the AGN luminosity, that is, narrow line regions becoming more extended in low- than in high-ionization AGN \citep[see also][]{2014Richardson}. Therefore, the fact LINER apparently show no bright nucleus but ``only'' extended narrow line regions cannot be taken as an argument to reject the presence of an active SMBH in these galaxies; even more so, considering that since LINER are type~2 AGN, the torus of gas and dust assumed around the nuclei in these galaxies are expected to hide it from our view. On the other hand, it seems clear that determining whether the gas in LINER is ionized by PAGB or by a SMBH accretion disk depends first and foremost on the emission line ratios.

\begin{figure}[ht!]
\epsscale{0.6}
\plotone{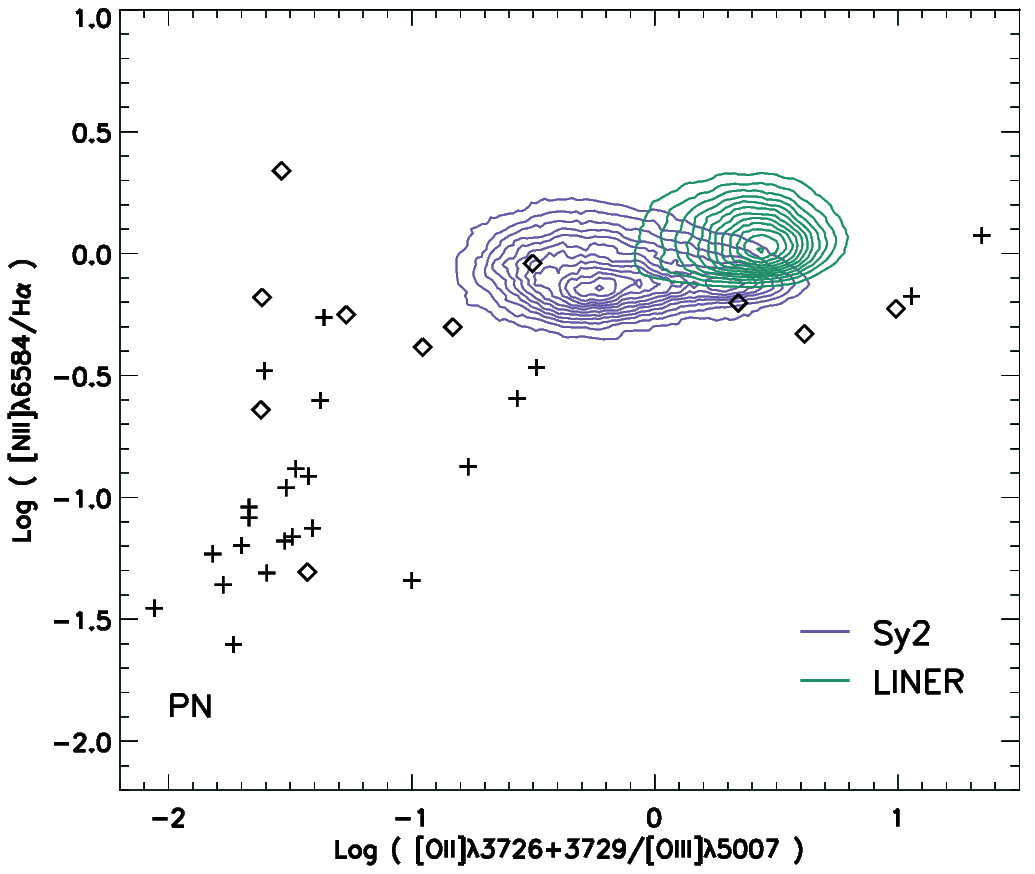}
\caption{BPT diagnostic diagram comparing Sy2 and LINER in our sample with two samples of PN: high-excitation (black crosses) and low-excitation (black diamonds).  \label{Fig7:BPT_PAGB}}
\end{figure}

Comparing the line ratios of different sources was already done by \citet{Baldwin1981}. Looking in various diagnostic diagrams for the presence of planetary nebulae (PN)---the most conspicuous form of PAGB capable of ionising the gas---these authors verified that such stars do not produce the typical high [NII]$\lambda6584$/H$\alpha$ ratio of AGN. On the contrary, in the diagnostic diagram comparing [NII]$\lambda6584$/H$\alpha$ with  [OII]$\lambda3726+3729$/[OIII]$\lambda5007$, they found the latter ratio to be typically low, consistent with the high temperatures of the central stars in PN (CSPN). In Figure~\ref{Fig7:BPT_PAGB}, we trace this diagram for our sample of Sy2 and LINER, comparing their line ratios with those of two samples of PN: one classified as high-excitation  \citep{1971Peimbert} the other as low-excitation \citep{1985AKondrateva}. Based on this comparison, we conclude that to be consistent with the normal ratios shown by Sy2 and LINER, the line ratios produced by PN would need to fall within a narrow range of low-excitation values, with atypically high [NII]$\lambda6584$/H$\alpha$ ratios. This is a severe constraint, considering that the transformation of a CSPN into a white dwarf is a complex process, closely linked to the formation and evolution of the ejected AGB envelope, which, depending on the winds, form low and high ionization shells, explaining some of the low-excitation line ratios \citep[e.g.,][]{1990Breitschwerdt}. Moreover, considering that the Sy2 and LINER distributions in Figure~\ref{Fig7:BPT_PAGB} show an important overlap, the few PN with the right line ratios in this diagram would not be specific to LINER. Therefore, and in good agreement with \citet[][]{Baldwin1981} seminal study, none of the diagnostic diagrams we used is consistent with the presence of a large number of PN ionizing the gas in our sample of LINER. 

\begin{figure*}[ht!]
 \gridline{\fig{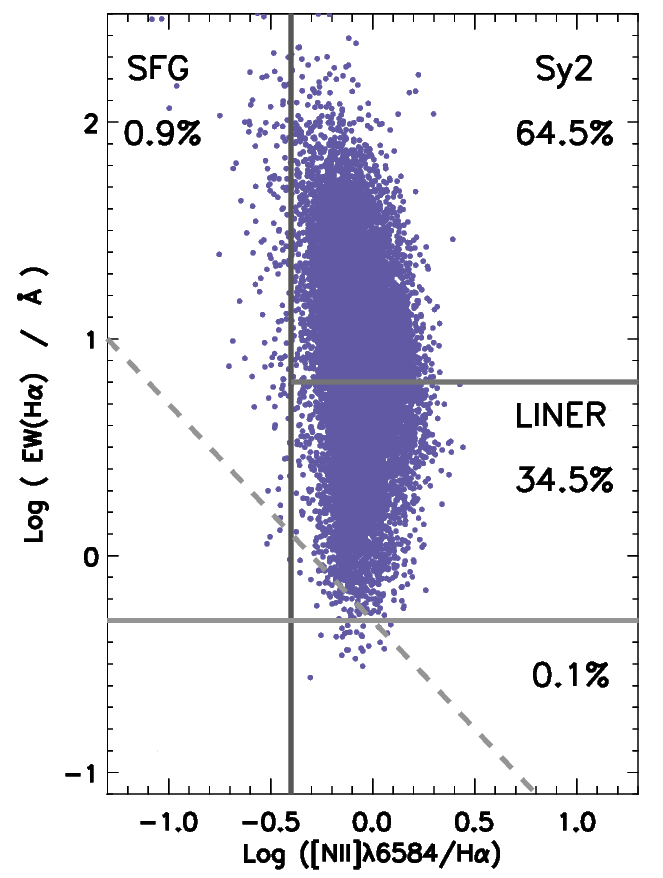}{0.4\textwidth}{(a)}
 \fig{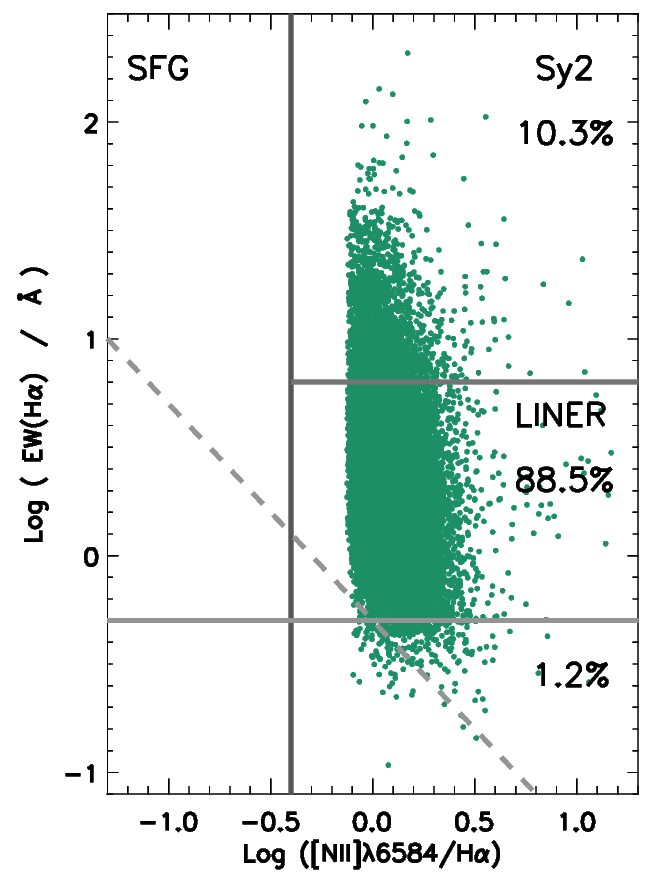}{0.4\textwidth}{(b)}
 }
\caption{(a) Alternative classification for Sy2 and (b) LINER using the WHAN diagram. \label{fig8:WHAN}
}
\end{figure*} 

However, not satisfied by the conclusions of \citet[][]{Baldwin1981} and \citet{Veilleux1987}, \citet{2011CidFernandes} produced a new ``diagnostic diagram'' that apparently would allow to unambiguously mark the presence of PAGB in LINER. This is the WHAN diagram, which consists in comparing the ratio of [NII]$\lambda6584$/H$\alpha$ with the equivalent width of the H$\alpha$ line, EW(H$\alpha$) in \AA. In Figure~\ref{fig8:WHAN}, we trace the WHAN diagram for the Sy2 and LINER in our sample. The problem with this diagram is twofold. First, it does not unequivocally separate Sy2 from LINER. This is obvious in Figure~\ref{fig8:WHAN}, where 34.5\% of the Sy2 and 10.3\% of the LINER in our sample are in the ``wrong'' regions. Second, contrary to the various BPT diagnostic diagrams, the WHAN diagram is a one-parameter diagnostic, since the ratio [NII]$\lambda6584$/H$\alpha$ is not predicted by any PAGB photo-ionization model. Therefore, the only discriminating criterion is EW(H$\alpha$), which is a parameter that in galaxies naturally decreases with the intensity of the H$\alpha$ emission, irrespective of what produces it \citep{1996Coziol}. In other words, this criterion is not specific to any source of ionization \citep[as was also shown in][]{2014Coziol}. 

In order to establish a quantitative criteria based on EW(H$\alpha$) alone, \citet{2011CidFernandes} used a SSP synthesis model for what they called ``retired'' galaxies, that is, galaxies having ceased to form stars a few $10^8$ years or Gyr before we observed them, which, according to these authors, would be the specific evolutionary phase of galaxy defining the LINER class. According to this model, a value of Log(EW(H$\alpha$) $< 0.5$ (3 \AA) was proposed as the signature of PAGB in these galaxies. However, it is obvious that applying blindly this criterion in Figure~\ref{fig8:WHAN} would not only eliminate most of the LINER in our sample, but also a good fraction of the Sy2 (in contradiction for the latter with the BPT diagram). Actually, many authors in the literature have already disputed this criterion, suggesting that PAGB cannot produce EW(H$\alpha$) higher than 1 \AA\ \citep[as reported by][]{2016Belfiore}. Again, it is obvious that lowering the value to 1 \AA\ instead of 3 \AA\ in Figure~\ref{fig8:WHAN} would make a huge difference for our sample. Consequently, as applied to our sample, the WHAN diagram does not allow us to unambiguously identify PAGB in our galaxies. Note that different groups of authors have subsequently tried to include PAGB in SSP synthesis models, testing different evolution scenarios and, in particular, post-starburst galaxies, which would have been consistent with a relative excess of nitrogen due to the effect of starburst winds. However, their results were generally negative, concluding that the PAGB contribution to the integrated light of galaxies is generally low \citep[see][and references therein]{2013Conroy}. 

\begin{figure*}[ht!]
 \gridline{\fig{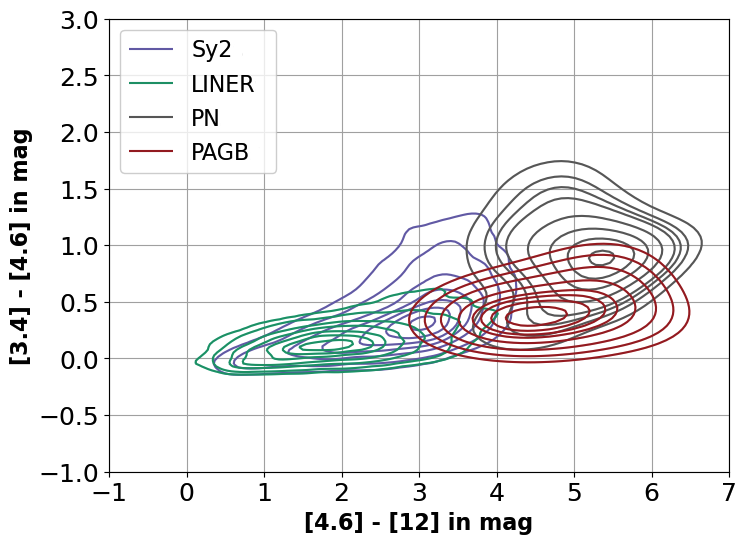}{0.45\textwidth}{(a)}
 \fig{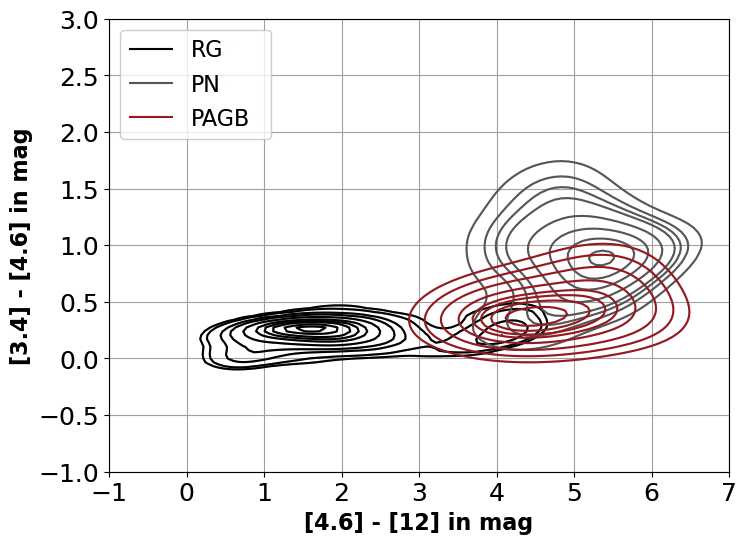}{0.45\textwidth}{(b)} }
\caption{Comparing the WISE colors of PN and PAGB with (a) Sy2 and LINER, (b) all the RG in our sample. Note that the few RG consistent with PAGB are in fact Sy2 in late-type spirals with high SFR.\label{fig9:WISE}
}
\end{figure*}

In other words, as it is, there is no clear or indisputable argument favoring the idea that PAGB are the main source of ionization of the gas in LINER. What is missing is independent observational evidence, unambiguously identifying PAGB in galaxies. But to go further we must first better understand in what consist the PAGB phase. According to \citet{2016Bertolami}: ``The transition between the asymptotic giant branch (AGB) and white dwarf phase is arguably one of the least understood phases of the evolution of low- and intermediate-mass single stars (M$_i \sim 0.8-8 M_\odot$). During this phase, stars are expected to evolve as OH/IR stars, proto-planetary nebula central stars and, under the right conditions, the central stars of planetary nebulae [CSPN].'' We already concluded that the BPT diagrams show no evidence of PN specific to LINER. However, we could also check for evidence of the first two phases in MIR \citep[most specifically at $12 \mu$m;][]{2010Kelson}. In \citet{2014Coziol} we have already assembled two samples of PAGB and PN, one observed in the MIR by \citet{2006Suarez}, 82 PAGB, and the other by \citet{2013Weidmann}, 210 PN, and retrieved their magnitudes in WISE \citep{Wright2010}. In figure~\ref{fig9:WISE} we compare their MIR colors with those of the galaxies in our sample. As one can see, neither the PN nor PAGB have colors consistent with LINER. Actually, the greater overlap in colors is between PAGB and Sy2. As we will show later, this similarity is consistent with a high amount of hot dust, heated by young stars in Sy2, which are missing in LINER. In figure~\ref{fig9:WISE}b we have also added the RG in our sample (including those with weak or no optical emission). Except for a small numbers of galaxies with colors [4.6]-[12] $\sim 4$, in general, RG have MIR colors similar to LINER, which, coincidentally, is also congruent with their LINER-like spectral characteristics. And this is despite the majority of RG having weak emission or none at all (cf. Figure~\ref{Fig5:AltDiag_RG}), consistent with old galaxies with early morphological types and very low or nil star formation activity \citep{2002Nagar}. 

Compared to Figure~12 in \citet{Wright2010}, the WISE colors of PAGB and PN are similar to those of luminous infrared galaxies (LIRG) or even ultra-luminous infrared galaxies (ULIRG), which, incidentally, are classified as either starburst or LINER. These galaxies are luminous in infrared because they are rich in hot dust, heated by a higher than normal star formation activity \citep{1998Kim} and frequently mixed with a dust-bury AGN \citep{Sanders1988,1999Lutz}. The reason why the colors of PAGB and PN are similar to these galaxies is consequently straightforward: the transition from PAGB to CSPN happens on a track (horizontal in the HR diagram) where the envelope ejected during the AGB phase is heated by the very hot central star that is rapidly shrinking in size on its way to become a white dwarf, the SED in MIR rapidly dominating the SED in the optical \citep{2015Vickers}.

On the other hand, the MIR colors of Sy2 and LINER in our sample are typical of normal spiral galaxies, differing by their levels of star formation, high in Sy2 and low in LINER, and, as we will show later in our analysis, the few RG with colors consistent with PAGB are late-type spirals with high SFR. This association of Sy2 and LINER with normal spirals is in full agreement with the original definitions of Seyfert galaxies by \citet{1977Adams} and LINER by \citet{1980Heckman} seminal studies. Moreover, compared to LIRG, the cooler MIR colors of LINER suggest continuous star formation over their all lifetime (10-13 Gyr), which is the standard mode of star formation in spiral galaxies. Consequently, the LINER in our sample cannot be considered to be ``retired galaxies''. 

The above results leave us with a serious problem when considering the PAGB hypothesis for LINER: they seem to leave no trace either as pre-PN (WISE diagram) or PN (BPT diagram). However, in \citet{2016Belfiore} the authors tried to bypass the problem, explaining that, in fact, and contrary to what \citet{2011CidFernandes} proposed, not all PAGB are involved, but only a special type they called ``lazy'' PAGB. According to these authors, those would be typical of AGB with initial mass lower than 1~M$_\odot$, which would reach their ionization temperature only after their envelope of gas had dissipated, forming consequently a pure source of Lyman continuum (Ly-c) photons. However, the description of the various phases of evolution of PAGB in \citet{2016Bertolami} is slightly more complex. About the ``detectability'' of PN, Miller Bertolami explained that this depends critically on the relationship between the evolutionary timescales of the CSPN, which is the source of ionizing photons, and on the dynamical timescales of the circumstellar material ejected at the end of the AGB phase: if the CSPN evolves too fast, the PN is ionized only for a short time, and thus has a low detection probability, however, if the star evolves too slowly---which would be the case of lazy-PAGB---the ionization of the nebula would take place when the ejected material is already dispersed, the ionized envelope being too faint then to be detected. In both cases, therefore, the envelope is ionized and thus these are not free sources of Ly-c photons. On the other hand, in some cases, the envelope might not have enough time to be ionized but that would be due to the rapid transformation phase into a white dwarf, since once the transformation is complete, although the temperature of the central star is very high, the number of ionizing photons emitted is very low, due to the small surface of the white dwarf; the number of photons emitted being proportional to the area of the star, a typical white dwarf with typical mass 0.5 M$_\odot$ and radius R$_{WD} \sim 0.01$ R$_\odot$ \citep{2010Parsons} only emits $10^{-6}$ the number of photons emitted by the Sun (even smaller for a 1 M$_\odot$ white-dwarf, since the radius decrease with the mass). 

There are consequently two difficulties with the lazy-PAGB hypothesis: 1) the process by which a PAGB becomes lazy does not depend solely on the initial mass of the AGB stars, but also on the formation and evolution of the AGB envelope, a messy and not fully understood process, which implies that the Lazy-PAGB phase cannot be considered typical or even common to all PAGB with an initial mass lower than 1 M$_\odot$, 2) the fraction of ionizing-photon escaping to space is expected to be low, since the gas of its ejected envelope (the most nearby gas) is ionized, although too dispersed to be detected. These two difficulties reduce not only the potential number of lazy-PAGB but also the probability they could ionize the gas over kpc distance. Consequently, one would need an enormous amount of lazy-PAGB dispersed over kpc regions to reproduce the luminosity of the emission lines we measured in the LINER in our sample (more specifically [OIII]$\lambda5007$, see next section below). Two questions then seem relevant: how many of these stars must we expect and what would be their typical luminosity? 

Thanks to the study made by \citet{2015Vickers}, who determined the SEDs of 209 PAGB in our galaxy from the Toru\'{n} survey, we do not have to guess, but can directly search for lazy-PAGB: those are stars that show neither an IR envelop in their SED nor an ionization spectrum. Of the 209 likely PAGB in the Toru\'{n} list determined by \citet{2015Vickers}, only 18 (8.6\%) agrees with this definition. In terms of flux, L$_\odot$ kpc$^{-1}$ (which is the luminosity at distance of 1 kpc), 13 have on average $1.1 \times 10^{35}$ erg s$^{-1}$ and 5 have $2.6 \times 10^{36}$ erg s$^{-1}$, which considered together represent an average luminosity of only $8.1 \times 10^{35}$ erg s$^{-1}$. Therefore, only a few \% of PAGB are lazy, and their luminosity varies between $10^{35}$ and $10^{36}$ erg s$^{-1}$. Compared to the relatively high luminosity in [OIII]$\lambda5007$ we observe for the LINER in our sample (cf. figure~\ref{fig10:OsterbrockTEST}), the number of co-evolved lazy-PAGB required to ionize such quantity of gas over kpc regions would need consequently to be unrealistically large. In conclusion, the idea lazy-PAGB can be the source of ionization of gas in LINER seems extremely improbable. 

\subsection{AGN luminosity in Sy2, LINER and RG} 
\label{Subsec2.2}

\begin{deluxetable*}{lcccc}
\tablecaption{Linear regressions in Figure~\ref{fig10:OsterbrockTEST} and Figure~\ref{fig12:OIIIvsLNVSS}\label{Table1:L5100vsLOIII}}
\tablewidth{0pt}
\tablehead{
\colhead{Sample}  
& \colhead{$\alpha$}
& \colhead{$\beta$}  
& \colhead{r(Pearson) } 
& \colhead{p value} 
}
\startdata
\multicolumn{5}{c}{L[OIII] vs. $\lambda{\rm L[5100]}$ uncorrected}\\
\hline
Sy2         & $1.457 \pm 0.022$ & $-22.698 \pm 0.572$ & $0.999$ & $<0.001$  \\
LINER       & $1.224 \pm 0.008$ & $-13.439 \pm 0.212$ & $0.996$ & $<0.001$  \\
RG-Sy2      & $0.451 \pm 0.034$ & $ 21.311 \pm 1.584$ & $0.969$ & $<0.001$  \\
RG-LINER    & $0.450 \pm 0.028$ & $ 20.489 \pm 1.363$ & $0.985$ & $<0.001$  \\
\hline
\multicolumn{5}{c}{L[OIII] vs. $\lambda{\rm L[5100]}$ corrected}\\
\hline
Sy2        & $1.018 \pm 0.009$ & $-2.149 \pm 0.423$ & $0.999$ & $<0.001$  \\
LINER      & $1.003 \pm 0.008$ & $-1.539 \pm 0.371$ & $0.999$ & $<0.001$  \\
RG-Sy2     & $0.451 \pm 0.034$ & $20.673 \pm 1.584$ & $0.969$ & $<0.001$  \\
RG-LINER   & $0.450 \pm 0.028$ & $19.452 \pm 1.363$ & $0.985$ & $<0.001$  \\
\hline
\multicolumn{5}{c}{L[OIII] vs. L$_{NVSS}$ }\\
\hline
RG-Sy2     & $0.305 \pm 0.006$ & $-0.562 \pm 0.183$ & $0.976$ & $<0.001$  \\
RG-LINER   & $0.320 \pm 0.007$ & $-1.403 \pm 0.082$ & $0.986$ & $<0.001$  \\
\enddata
\end{deluxetable*}

The lack of evidence in favor of the PAGB hypothesis for LINER led us to consider the problem from the opposite point of view: what makes LINER an AGN? The answer is the ``Osterbrock's test'' \citep{1989Osterbrock}, which consists in comparing the luminosity of the ionized gas in the NLR with the luminosity of the adjacent ``featureless'' continuum and verifying that the energy of ionization is consistent with the energy expected from the accretion of matter onto a SMBH. By featureless continuum, one means the continuum radiation emitted by the accretion disk of an active SMBH. In \citet{1989Osterbrock}, this test was applied on different AGN types, Sy1, Sy2 and RG, comparing the luminosity of H$\alpha$ with the luminosity of the monochromatic continuum at 4800 \AA, finding that the relation between the luminosities seems to be unique, compatible with a power law, $F_\nu \propto \nu^{-\alpha}$ with $\alpha \sim 1.05$. 
\begin{figure*}[ht!]
\gridline{\fig{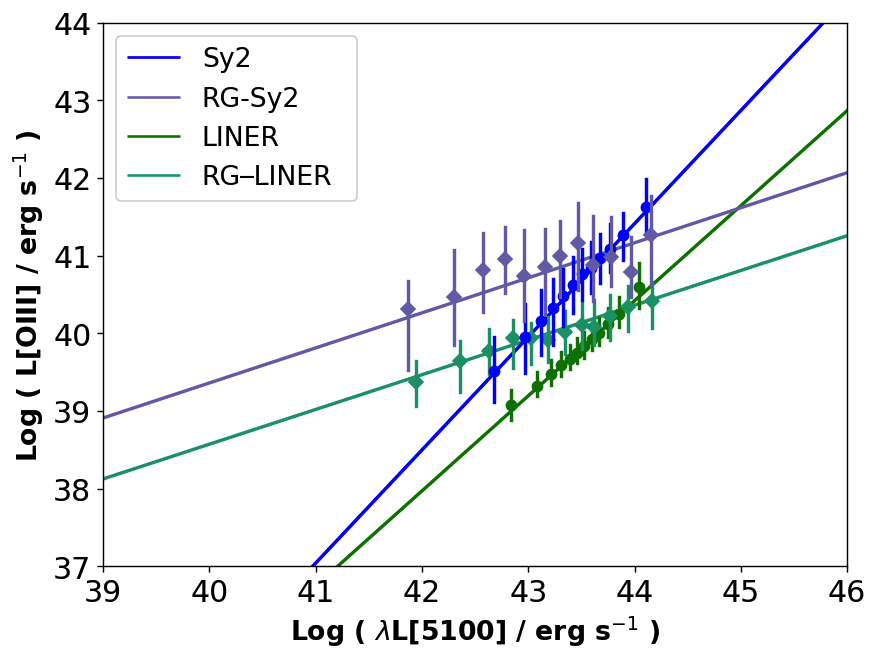}{0.4\textwidth}{(a)}
\fig{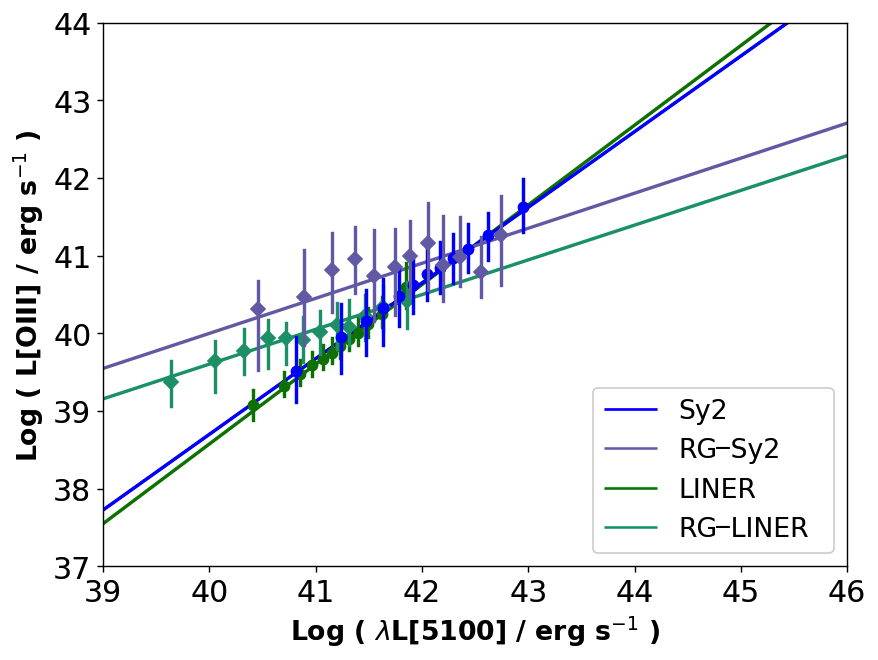}{0.4\textwidth}{(b)}  
}
\caption{Comparing the fluxes in [OIII] with the total fluxes of the continuum at 5100 \AA\, (a) before correcting for the contribution of the stellar populations, (b) after correction. 
The corresponding parameters for the linear relations ${\rm Log}({\rm L[OIII]})  = \alpha \times {\rm Log} (\lambda{\rm L[5100]}) + \beta$ are given in Table~\ref{Table1:L5100vsLOIII}.\label{fig10:OsterbrockTEST}
}
\end{figure*}
In Figure~\ref{fig10:OsterbrockTEST}a, we compare the luminosity of the [OIII]$\lambda5007$ line (excluding resolved OFs) with the luminosity of the continuum at 5100~\AA, for the four samples, Sy2, LINER, and RG with emission either classified as Sy2 (RG-Sy2) or LINER (RG-LINER). Using the cumulative percentiles as determined in bins containing equal numbers of galaxies, we fitted a linear regression on the medians obtaining four relations of the form ${\rm Log}({\rm L[OIII]})  = \alpha \times {\rm Log} (\lambda{\rm L[5100]}) + \beta$. As can be appreciated in Table~\ref{Table1:L5100vsLOIII}, the regressions have very significant $p$ values and relatively high Spearman correlation coefficients. Three important facts about the sources of ionization of the gas in these galaxies can be inferred from these correlations: 1- L[OIII] is relatively high (ranging from $10^{39}$ erg s$^{-1}$ to $10^{42}$ erg s$^{-1}$) and well correlated with the L[5100] (Table~\ref{Table1:L5100vsLOIII}). This suggests that the sources of ionization of the gas in all these galaxies are intense. 2- the slopes of the correlations for the Sy2 and LINER are steeper than the slope of the relation obtained by \citet{1989Osterbrock}. Note that although the [OIII] luminosity was obtained after subtracting the underlying stellar population (SP) contribution using STARLIGHT, the luminosity of the continuum still includes this component. Consequently, the stellar contribution to the continuum increases the luminosity explaining the steeper slopes. Finally, 3- at low continuum luminosities, the RG-Sy2 and RG-LINER show a relative excess of ionizing photons compared to their radio-quiet counterparts, explaining the shallower slopes. This ionization excess is remarkable considering that in the alternative diagram, RG-Sy2 and RG-LINER have clearly lower star formation (based on the higher D$_n(4000)$) than in the RG-SFG and RG-TO.
\begin{figure*}[ht!]
\gridline{\fig{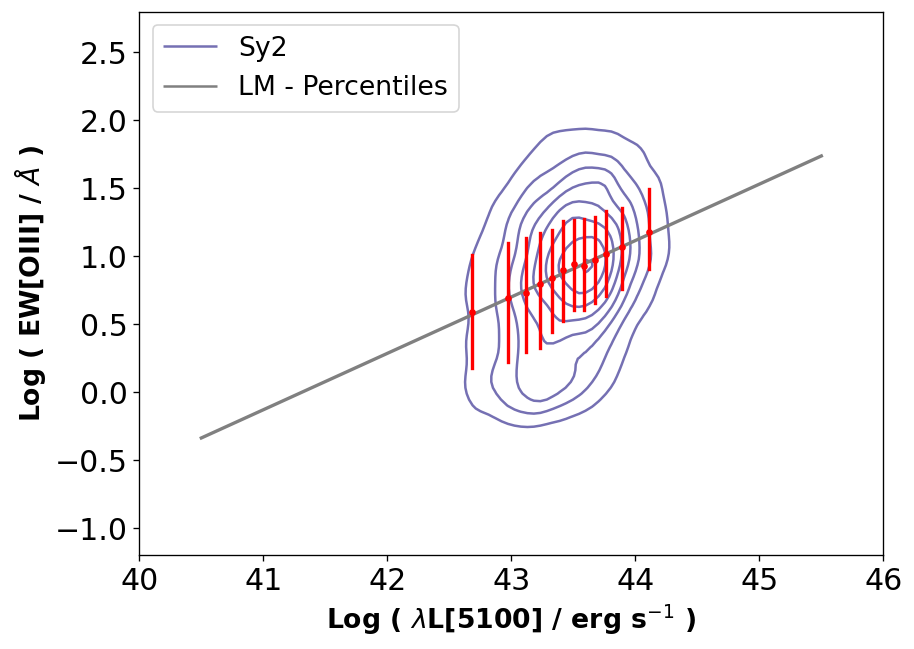}{0.35\textwidth}{(a)}
\fig{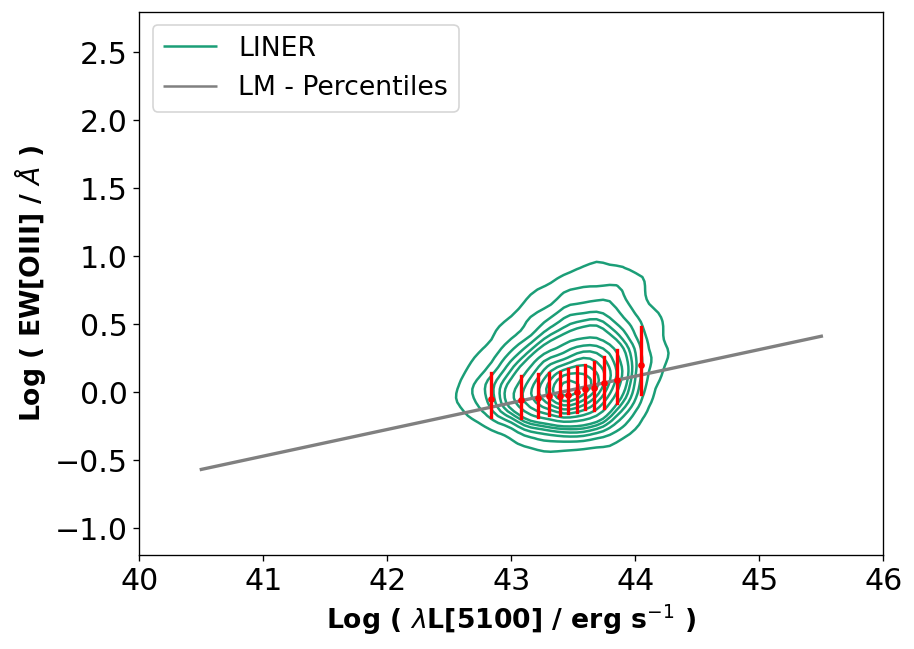}{0.35\textwidth}{(b)}  
}
 \gridline{\fig{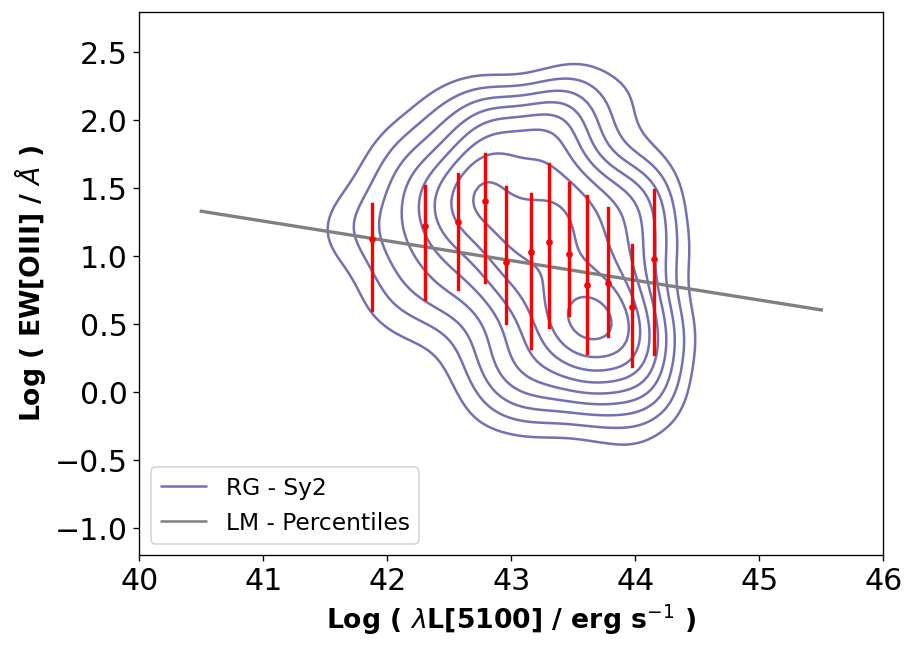}{0.35\textwidth}{(c)}
          \fig{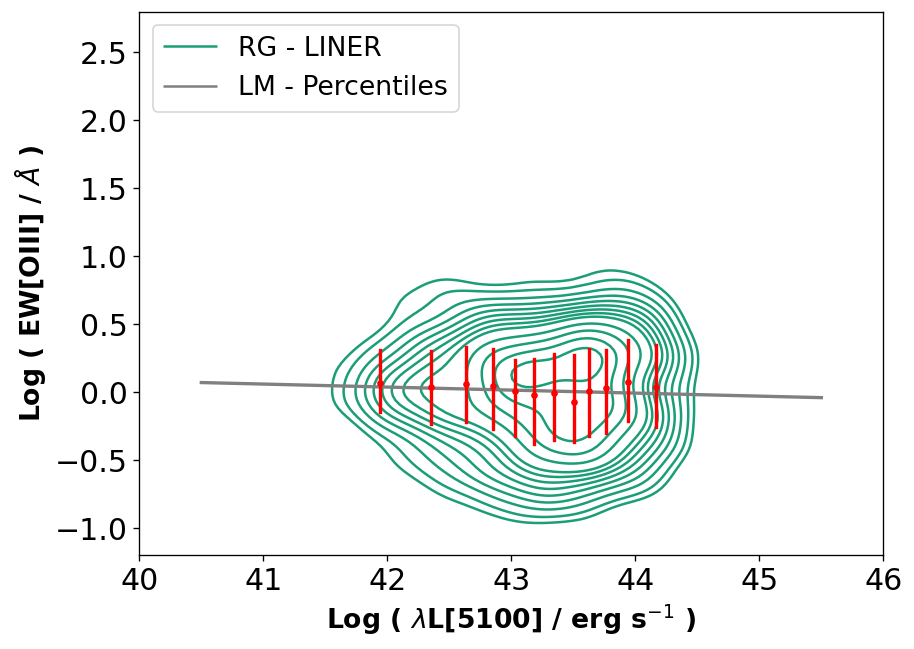}{0.35\textwidth}{(d)} 
          }
\caption{Comparing EW$([{\rm OIII}])$ with $\lambda$L[5100] \textcolor{red}{as observed} in (a) Sy2, (b) LINER, (c) RG-Sy2 and (d) RG-LINER. For each distribution we calculated two linear regressions, one on the whole sample (contour levels) and the other on the medians of the cumulative percentiles (red bars), reducing the dispersion.\label{fig11:EO3vsLC}
}
\end{figure*}

From point two, we conclude that the total continuum at 5100 \AA\ is composed of different sources, which we simplified as a sum of two fluxes, $F_{5100} = A \times F_{SP} + B \times F_{AGN}$ \citep{1981Shuder}, where $A$ is the fraction contributed by the stellar population (SP) and $B$ the fraction contributed by the AGN. To get L$_{AGN}$ we must thus first quantify these components. In practice, this is not a simple task, since it implies knowing what forms the SEDs of these galaxies have. However, based on the equivalent width of the emission line, EW$([{\rm OIII}])$, one can get a relatively good estimate for B. This is how it is done. According to the definition of the equivalent bwidth, EW$([{\rm OIII}])= F_{line}/F_{cont}$, where $F_{line}$ is the ionizing flux and $F_{cont}$ is the total flux in the continuum.  Therefore, assuming EW$([{\rm OIII}]) \approx F_{AGN}/ (F_{SP}+F_{AGN})$ (star formation is negligible) and comparing EW([OIII]) with L[5100], a higher contribution from $F_{SP}$ would produce lower EW$([{\rm OIII}])$ while a lower contribution would produce the inverse (the increase or decrease of $F_{AGN}$ in the numerator and denominator cancelling themselves). 

In Figure~\ref{fig11:EO3vsLC}a and~\ref{fig11:EO3vsLC}b, we compare EW$([{\rm OIII}])$ with $\lambda$L[5100] for the Sy2 and LINER, respectively. In both sample, ${\rm EW}([{\rm OIII}])$ increases with the luminosity, which suggests, as expected, that the contribution of the stellar populations to the continuum decreases as the luminosity of the AGN increases. To quantify this decrement, we first calculate two linear regressions, ${\rm Log}({\rm EW([OIII])})  = a \times {\rm Log} (\lambda{\rm L[5100]})_{obs} + b$, where $a = 0.415\pm0.008$ and $b = -17.145\pm 0.356$ for Sy2 and $a = 0.196\pm0.008$ and $b = -8.507\pm0.262$ for LINER. For both regression the $p$ values are highly significant, $p <0.001$, and the Pearson's correlation coefficients are high, $r=0.995$ and $r=0.892$, for Sy2 and LINER, respectively. Now, assuming the contribution of the AGN in Sy2 is maximum at the highest continuum luminosity, $B = B_{max}$, and EW$([{\rm OIII}])$ is consequently also maximum, EW$_{max}$, the ratios EW/EW$_{max}$ at lower luminosity could be taken as a quantitative estimate how the contribution of the AGN decreases with the luminosity. To calibrate our corrections, we used the SED model for type~2 AGN produced by \citet{2016CalistroRivera}, determining that the maximum contribution by the AGN to the continuum luminosity at 5100 \AA\ is $B_{max} = 0.1$. 

In practice, starting with the Sy2, we separated the range in continuum luminosity in 5 bins \textcolor{red}{(in erg $s^{-1}$)}, ${\rm Log} (\lambda{\rm L[5100]}) = 42.5, 43.0, 43.5, 44.0, 44.5$ and use the EW-Luminosity relation traced in Figure~\ref{fig11:EO3vsLC}a to estimate the observed EW([OIII]) in each bin. Then we calculate the ratios EW/EW$_{max}$, multiplying by $B_{max} = 0.1$ for calibration. Finally, we adopt these ratios as corrections on the continuum luminosity, and performed a linear regression to get the luminosity correction function: ${\rm Log}(\lambda{\rm L[5100]})_{cor}  = 0.415 \times {\rm Log} (\lambda{\rm L[5100]})_{obs} - 19.47$. In \citet{2016CalistroRivera}, the authors also determined an average bolometric luminosity of $10^{43.25}$ erg s$^{-1}$, which, after applying our correction to the continuum luminosity of the Sy2 implies an average bolometric correction factor of  order $f = 10$ must be applied to get L$_{Bol}$, that is,  L$_{Bol} = {\rm Log}(\lambda{\rm L[5100]})_{cor} + 1$ \citep[][]{2000Kaspi}. This bolometric correction factor is within the range of values for different types of AGN as determined by \citet[][]{2006Richards}. In Figure~\ref{fig10:OsterbrockTEST}b we trace ${\rm Log}({\rm L[OIII]})$  as a function of ${\rm Log} (\lambda{\rm L[5100]})_{cor}$ for the Sy2. The parameters of the linear regression are reported in Table~\ref{Table1:L5100vsLOIII}. After correction, the Sy2 in Figure~\ref{fig10:OsterbrockTEST}b trace a linear relation with a slope consistent with a power law with exponent $\alpha = 1.018 \pm 0.016$, that is, nearly reproducing the results of \citet[][]{1989Osterbrock} and \citet[][using a different method]{1981Shuder}.  

Adopting the view that LINER are the low luminosity counterparts of Sy2 \citep{Veilleux1987}, we determine another luminosity correction function, ${\rm Log}(\lambda{\rm L[5100]})_{cor}  = 0.196 \times {\rm Log} (\lambda{\rm L[5100]})_{obs} - 10.83$, by comparing the ${\rm EW}([{\rm OIII}])$ in LINER with those of the Sy2 in each corresponding luminosity bin. This correction is now congruent with an average bolometric luminosity of $10^{42.25}$ erg s$^{-1}$ for LINER (consistent with an intrinsic lower luminosity) and an average bolometric correction also of order $f=10$. In Figure~\ref{fig10:OsterbrockTEST}b we trace ${\rm Log}({\rm L[OIII]})$ as a function of ${\rm Log} (\lambda{\rm L[5100]}{cor})$ for LINER, reporting the parameters of the linear regression in Table~\ref{Table1:L5100vsLOIII}. After correction, the LINER follow the same relation as the Sy2, with a slope consistent with a exponent $\alpha = 1.003 \pm 0.034$, which again is in good agreement with the results obtained by \citet{1989Osterbrock} and the conclusion that AGN in different classes follow the same relation. 
\begin{figure*}[ht!]
\gridline{\fig{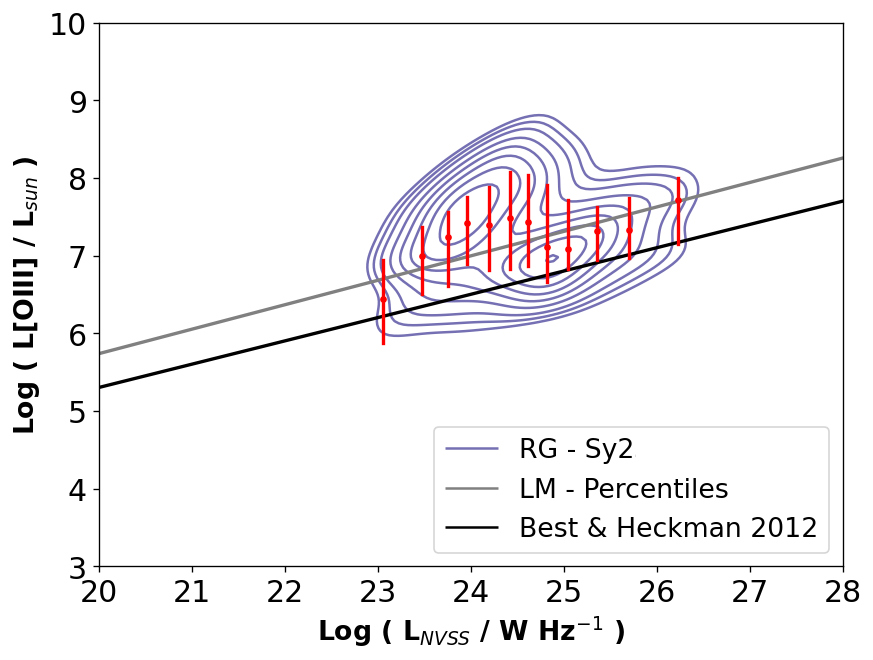}{0.4\textwidth}{(a)}
\fig{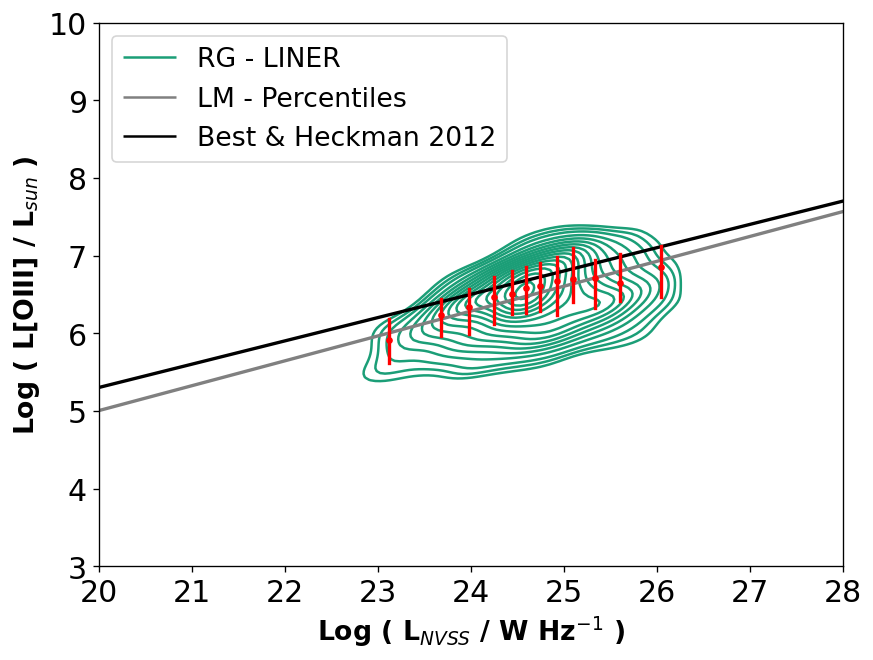}{0.4\textwidth}{(b)}  
}
\caption{Comparing the [OIII] luminosities with the radio luminosities in (a) the RG-Sy2 and (b) the RG-LINER. The black line is the division between high-excitation RG (HERG) and low-excitation RG (ERG) as proposed by \citet{2012Best}: \textcolor{red}{in general, HERG have higher [OIII] luminosities than LERG.} Note that while this separation criterion suggests most RG-Sy2 are HERG, the classification of RG-LINER has LERG is less certain, since the bulk of their distribution (traced by the cumulative percentiles) falls almost right on the separation line. \label{fig12:OIIIvsLNVSS}
}
\end{figure*}

Except for the excess of ionization, the SP correction functions for the RG-Sy2 and RG-LINER are expected to be similar to those of their radio quiet counterparts. This excess can either be due to an UV bump \citep{1999Koratkar,1994Elvis}, which appear at low luminosity in RG host galaxies depleted in gas in their center (the UV escaping to the NLR), or to the radio structures themselves, in particular, collimated radio jets \citep{2000Axon,2008Privon}. Supporting this last interpretation, we show in Figure~\ref{fig12:OIIIvsLNVSS}a and~\ref{fig12:OIIIvsLNVSS}b that the radio luminosities are correlated with the [OIII] luminosity, with log-log linear relations, as reported in Table~\ref{Table1:L5100vsLOIII}, that have highly significant $p$ values and high Pearson's correlation coefficients.

In Figure~\ref{fig11:EO3vsLC}c and Figure~\ref{fig11:EO3vsLC}d, the excess of ionization make the two EW-luminosity relations flat, ${\rm EW}([{\rm OIII}])$ being constant, tending to 10 \AA\ in RG-Sy2 and 1 \AA\ in RG-LINER. Consequently, we apply two constant SP corrections equal to the values applied to their counterparts Sy2 and LINER at the pivot luminosity $\lambda{\rm L[5100]}=10^{43.5}$ erg s$^{-1}$, obtaining two correction functions ${\rm Log}(\lambda{\rm L[5100]})_{cor}  = {\rm Log} (\lambda{\rm L[5100]})_{obs} - C$ where $C= -1.415$ for the RG-Sy2 and $C= -2.305$ for the RG-LINER. Applying the bolometric correction factor $f = 10$ like for the other AGN yields bolometric luminosities $\lambda{\rm L[5100]}=10^{43.1}$ erg s$^{-1}$ for RG-Sy2 and $\lambda{\rm L[5100]}=10^{42.2}$ erg s$^{-1}$ for RG-LINER, similar to their radio quiet counterparts. These bolometric luminosities are in good agreement with the luminosities estimated by \citet{2022Azadi} RG. In Figure~\ref{fig10:OsterbrockTEST}c and~\ref{fig10:OsterbrockTEST}d we added the corrected luminosity-luminosity relations for the RG-Sy2 and RG-LINER, where the excesses of ionization are preserved.

Since RG with weak or no emission occupy the same locus in the alternative diagram as the RG-LINER, but tend to be more luminous in radio, then one could expect their SMBH to be has active. Consequently, the reason why they show no emission lines in the optical is probably due to these galaxies being deficient in gas \citep[although many do show weak emission lines typical of LLAGN][]{Coziol1998}. On the other hand, the fact that they have similar Dn(4000) values as the RG-LINER suggests their stellar components should be similar. Consequently, we tentatively apply the same SP correction to their continuum as the RG-LINER.  

\subsection{Peculiar redshift distributions}
\label{subsec2.3}

\begin{figure*}[ht!]
\gridline{\fig{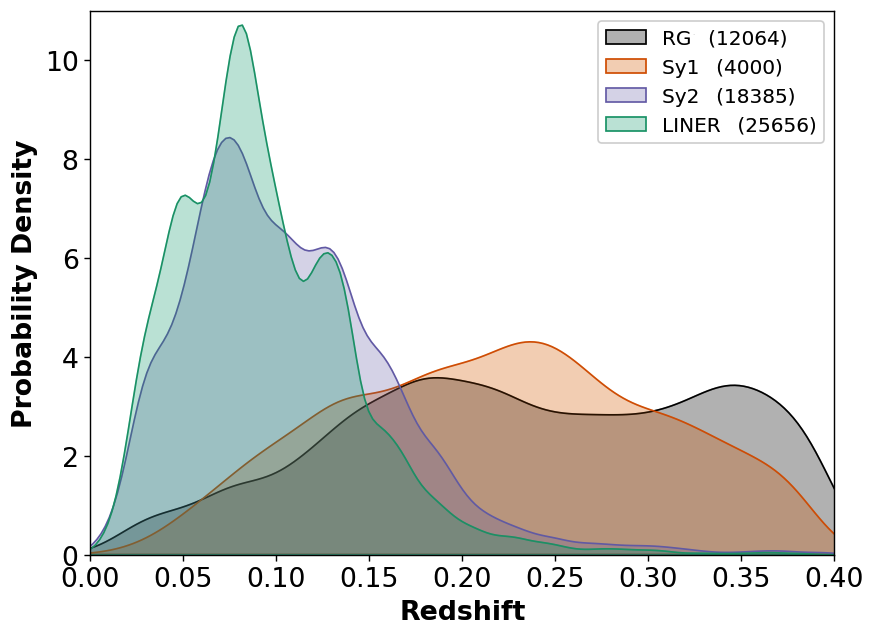}{0.48\textwidth}{(a)}
\fig{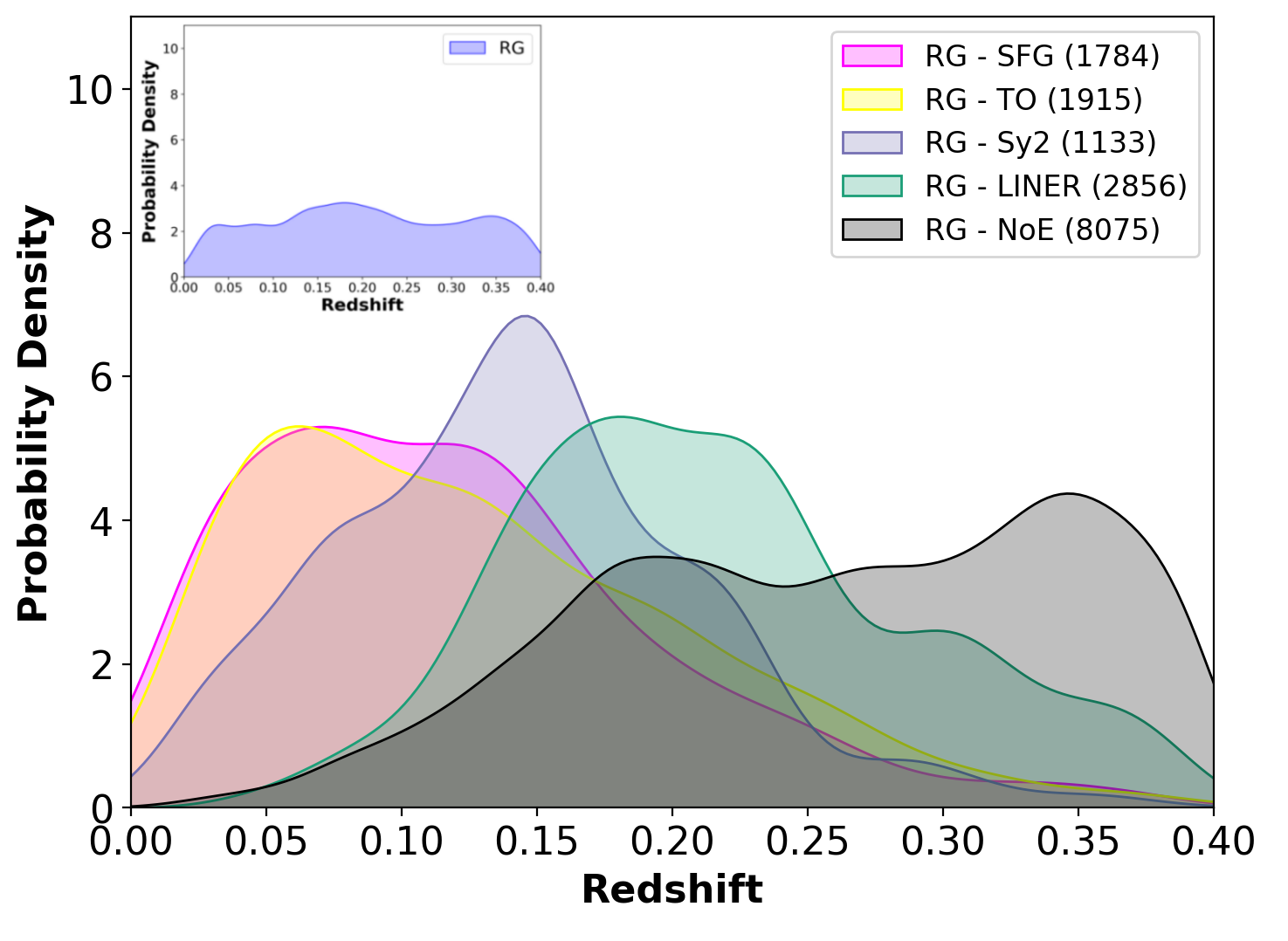}{0.48\textwidth}{(b)}  
}
\caption{Density distribution of redshifts, \textcolor{red}{a) for our sample of galaxies with different AGN classes (omitting RG classified as SFG and TO), b) showing the distributions only for the RG with different AGN classes. The inset shows (on same scale) a flat distribution for all the RG without class distinction.}
\label{Fig6:z}}
\end{figure*}


In Figure~\ref{Fig6:z}a we trace the density distributions in redshift of our samples with different AGN spectral classes. As mentioned earlier, RG-SFG and RG-TO were excluded to avoid the ambiguity as to the source of the radio emission in these galaxies (star formation vs. AGN). For comparison sake, we have also included our previous sample of Sy1 \citep[][adding 14 galaxies from SDSS DR8]{2020Torres-Papaqui}. Already we can see an important difference: while LINER and Sy2 appear mostly at low redshifts (below $z \sim 0.15$), Sy1 and RG tend to be at higher redshifts. For the Sy1 this could be due to a bias in luminosity: because Sy1 are more luminous than Sy2 and LINER, they are more easily detected at higher redshifts. However, based on the way the Sy1 were selected---showing typical broad lines in their spectra, independently from their luminosity \citep{2020Torres-Papaqui}---we see no reason why we should not detect more of them below $z \sim 0.15$. 

For RG, on the other hand, the skewed redshift distribution is easy to explain: adding back in Figure~\ref{Fig6:z}b those RG that we classified as SFG and TO does produce a flat redshift distribution for the whole sample (see inset in Figure~\ref{Fig6:z}b).  Since the detection criterion of RG is in radio, therefore, there are no observational bias that can explain the different redshift distribution.  Consequently, in Figure~\ref{Fig6:z}b one can also see that in the RG class there is a systematic change in activity type with the redshift: \textcolor{red}{RG with weak or without emission} are preferably at high reshifts, followed by RG-LINER and RG-Sy2, the former being at higher redshifts than the latter, and finally RG-SFG and RG-TO appearing only at low redshifts, below $z \sim 0.15$.Actually, this distribution in spectral type follows the variations shown in the alternative diagnostic diagram (cf.~Figure~\ref{Fig5:AltDiag_RG}), connecting the disappearance of emission lines to an increase of D$_n(4000)$ and L$_{\rm NVSS}/$M$_*$ with the redshift. 

Coincidentally, the fact that both probability densities for RG classified as AGN and Sy1 decrease in exactly the same way in in Figure~\ref{Fig6:z}a might also suggest a parallel change of spectral class in radio quiet AGN: Sy1 at higher redshifts being followed by Sy2/LINER at low z. However, identifying the physical causes for these differences is not obvious. In particular, considering the diversity of spectral classes and how their distributions vary with the redshifts an explanation in terms of observational biases may not be as simple as it seems. 

Alternatively, a difference in redshift distributions could indicate different evolutionary phases (like evolution between spectral class), while a difference in environment could allude to different formation processes of the host galaxies. For example, note that there are no Sy1 in our RG sample at high z. This could suggest these two types of AGN form in different density environments: RG forming in denser environments than Sy1. Actually, a deficit of Sy1 compared to Sy2 and LINER has already been reported in compact groups \citep{2008Martinez}, suggesting the same difference would also exists for these AGN types. Therefore, assuming Sy1 are typical of galaxies in the field while Sy2 and LINER are typical of groups, one would naturally expect the probability of detecting a Sy1 to increase with the redshift as the volume surveyed increases. Moreover, such a bias becoming less effective in radio survey would also explain why RG-Sy2/LINER in Figure~\ref{Fig6:z} are found at higher redshifts than their radio-quiet counterparts. Note that if the environment of galaxies plays a role in explaining the different AGN types, then differences in terms of galaxy morphology and mass, for both the SMBH and their hosts may also be expected to be observable. 


On the other hand, assuming some kind of evolution exists between the AGN spectral classes, for example a Sy1 eventually transforming into a Sy2 or a LINER, and assuming OF plays some role in this transformation \citep[e.g.,][]{2024Silk}, a relatively short time gap $\Delta t \sim 1.2$ Gyr, according to the respective peaks in redshifts, might fit the idea of delayed OF feedback effects, and such hypothesis possibly could be falsified. 

\subsection{Characterization of galaxy hosts and SMBHs in different AGN spectral classes}
\label{subsec2.4} 

In order to be able to distinguish feedback effects due to AGN winds from transformations related to secular evolution or different formation processes, it is important to get more information about the galaxy hosts and their SMBHs. For the narrow-line AGN, running STARLIGHT allowed us to retrieved data about the stellar mass, M$_*$, the velocity dispersion of the stars in the bulge, $\sigma$, the age of the older stellar populations, t$_{old}$, and the presence of younger stellar populations through the SFR \citep[for a detailed description how this is done, see][]{2013Torres-Papaqui}. 

Other relevant data retrieved from SDSS, like the \textit{ugriz} photometric magnitudes, the colors, $u-g$, $g-r$, $r-i$, $i-z$, the effective radius, $R_e$, and inverse concentration index, $R_{50}(r)$/$R_{90}(r)$, which is the ratio of the Petrosian radii (Petrosian 1976) containing 50\% and 90\% of the total flux in the $r$ band, were used to determined the morphological types of the host galaxies. Here is a brief description how this was done \citep[more details in][]{2020Torres-Papaqui}. First, following Blanton \& Roweis (2007), a K-correction is applied on the magnitudes, then the equivalent of the Hubble morphological types are deduced for all the AGN hosts by applying the correlations established by \citet{Shimasaku2001} and \citet{Fukugita2007}. The morphologies are finally classified using a numeral code, T, following an equivalence relation with the Hubble types: (0, E), (1, S0), (2, Sa), (3, Sb), (4, Sc), (5, Sd) and (6, Im), with intermediate mixed types, for example, (0.5, E/S0), (1.5, S0/Sa), etc. All this was done automatically using well constructed and thoroughly tested scripts written in IDL. 

In Table~\ref{RES_AVhost}, we report the average characteristics of the host galaxies in each AGN class. For comparison sake, we also included the corresponding data for the Sy1 class, as previously analysed before in \citet{2020Torres-Papaqui}. Note that since we cannot run STARLIGHT on a broad-line AGN (because the absorption lines of the stellar populations are flooded by the AGN continuum), $\sigma$ and t$_{old}$ for Sy1 remain unknown. To estimate M$_*$ for the Sy1 host galaxies, the masses of their SMBH, M$_{BH}$, as determined using L$_{AGN}$ and the FWHM of the broad H$\beta$ line measured in the SDSS spectra \citep[as done in][]{2020Torres-Papaqui}, were used in combination with the relation M$_{BH}-$M$_{*}$ as established by \citet{2015Reines} for nearby AGN. As for their SFRs, a rough estimate was obtained by applying the WISE color vs. STARLIGHT-SFR correlation calculated by \cite[][]{2020Torres-Papaqui} using the narrow-line AGN. However, note that depending on the varying contributions of the AGN components in MIR, these SFRs for type~1 AGN might be underestimated \citep[see][]{2023CutivaAlvarez}.

Completing our knowledge about our AGN sample, we added in Table~\ref{RES_AVhost} the frequency (percentage) of radio detection for Sy1, Sy2 and LINER in non-RG galaxies. Since how we got these last numbers is not trivial, a brief explanation is given here. One first preoccupation in assembling our sample was to make sure that the AGN selected in radio were different from those we selected based solely on their optical spectra. First, we verified that no galaxy in the RG sample was duplicated in the other non-RG samples, by cross-correlating their positions. This test yielding no match, we concluded that none of the non-RG Sy2, LINER and Sy1 were included in the original sample of \citet{2012Best}. However, before concluding that all these galaxies are radio-quiet, we asked a specialist of radio galaxy working at our university (Dr. Heinz Andernach) to verify our results. Having performed a thorough search in NVSS and FIRST, he was able to find a few extra sources, but only detected in FIRST. These are the results we reported in Table~\ref{RES_AVhost}. Since these extra sources were not included in the sample of \citet{2012Best}, we concluded that they did not comply with their selection criteria; these radio sources tend to be more compact and weaker in intensity than in RG. Since the number of radio detection of these weak sources is small, we feel safe qualifying our whole sample of non-RG AGN as radio quiet. 
\begin{deluxetable*}{lcccccccc}
\tablecaption{Average characteristics of AGN hosts \label{RES_AVhost}}
\tablewidth{0pt}
\tablehead{
\colhead{AGN}  & \colhead{Total} & \colhead{Radio} &  \colhead{z} &  \colhead{T} 
& \colhead{Log (t$_{old}$)}
& \colhead{$\sigma$ }  
& \colhead{Log (M$_{*}$) } 
& \colhead{Log (SFR)} 
 \\
\colhead{}  
& \colhead{} 
& \colhead{(\%)} 
&\colhead{}  
& \colhead{}
& \colhead{(yr)}
& \colhead{(km s$^{-1}$)}  
& \colhead{(M$_{\odot}$)}
& \colhead{(M$_\odot$ yr$^{-1}$)} 
}
\startdata
RG     & 12064  & 100 & 0.23 & 1.7 & 9.78 & 238 & 11.5 & -0.42\\
LINER  & 25656  &   3 & 0.09 & 2.0 & 9.81 & 160 & 11.0 & -0.45\\
Sy2    & 18385  &  10 & 0.11 & 2.6 & 9.38 & 134 & 10.8 & -0.01\\
Sy1    &  4000  &   5 & 0.22 & 2.3 &  -   & -   & 10.9 & -0.04\\
\enddata
\end{deluxetable*}
\begin{deluxetable*}{lccccccccc}
\tablecaption{Average characteristics of SMBHs \label{RES_AVBH}}
\tablewidth{0pt}
\tablehead{
\colhead{AGN}  
& \colhead{Log(L$_{AGN}$)}
& \colhead{Log(L$_{Bol}$)}  
& \colhead{Log(M$_{BH}$) } 
& \colhead{N$_{Edd}$}
& \colhead{Log(BHAR)} 
 \\
\colhead{}  
& \colhead{(erg s$^{-1}$)} 
& \colhead{(erg s$^{-1}$)} 
& \colhead{(M$_{\odot}$)}
&\colhead{}
& \colhead{(M$_\odot$ yr$^{-1}$)} 
}
\startdata
RG      & 41.4 & 42.3 & 8.6 & -4.3 & -3.4 \\
LINER   & 41.7 & 42.1 & 7.9 & -3.0 & -2.7 \\
Sy2     & 41.9 & 42.8 & 7.6 & -2.0 & -2.0 \\
Sy1     & 43.9 & 44.8 & 8.0 & -1.2 & -0.9 \\
\enddata
\end{deluxetable*}

The average characteristics for the SMBHs are reported in Table~\ref{RES_AVBH}. These consist in the AGN luminosity, L$_{AGN} = \lambda$L$_\lambda(5100)$, where L$_\lambda(5100)$  is the corrected (featureless) continuum at 5100 \AA, and the bolometric luminosity, L$_{bol}$, estimated by multiplying L$_{AGN}$ by a factor 10 (as explained in Section~\ref{Subsec2.2}). For M$_{BH}$, we used for the narrow-line AGN the velocity dispersion of the stars in their bulges, $\sigma$, as determined by STARLIGHT, and applied the M$_{BH}$-$\sigma$ relation determined by \citet{Gultekin2009}, while for Sy1, we use the virial relation that combines the FWHM of the broad line component of H$\beta$ and L$_{AGN}$ \citep[see][]{2017Coziol}. From these values, the Eddington ratios, N$_{Edd} =$ L$_{bol}$/L$_{Edd}$, were calculated, allowing the BH accretion rates, BHAR$\ = \dot m\ = $\ L$_{bol}$/$\eta$c$^2$, to be estimated (note that by definition BHAR is independent from $\eta$). 

From the results reported in Table~\ref{RES_AVhost} and Table~\ref{RES_AVBH}, we can already draw a global picture of how the characteristics of the AGN hosts and their SMBHs vary in the different spectral classes. The average stellar mass, M$_*$, increases along the sequence Sy2$\to$Sy1$\to$LINER$\to$RG. Consequently, both the velocity dispersion and the age of the stellar populations in the narrow-line AGN increase along the same sequence: Sy2$\to$LINER$\to$RG. In principle, the mass of a SMBH should follow the mass of its host galaxy, which is almost what we observe, M$_{BH}$ increasing along the sequence: Sy2$\to$LINER$\to$Sy1$\to$RG. Note that due to the different ways the galaxy masses for Sy1 were estimated, we cannot put too much emphasis on the small difference (a factor 1.2 on average) between LINER and Sy1 in the two sequences. In general, therefore, we can securely state that in AGN the mass of the SMBH tends to grow with the mass of its galaxy host, Sy2 being the less massive and RG being the most massive, while LINER and Sy1 have comparable, intermediate masses. 

Maybe more significant, we also find the morphological type of the galaxies to follow the stellar mass: T decreasing from late types (high values) to early types (low values) along the same sequence, Sy2$\to$Sy1$\to$LINER$\to$RG. Consequently, and as we expect for normal (non-AGN) galaxies, SFR also decreases along this sequence, except that it is slightly higher in RG than in LINER: Sy2$\to$Sy1$\to$RG$\to$LINER. The other way for LINER and RG was expected, because since the RG host galaxies turned out to have earlier morphological types than in LINER, lower SFR would have seems more normal. However, the difference in SFR is small, and there is a non-zero probability that due to the radio emission (the effect of a jet) the SED of RG is slightly flatter in the blue than in LINER (as we reported in sub-section~\ref{Subsec2.2}). If this is true, then STARLIGHT could have misinterpreted this feature as evidence of a younger SSP, explaining the higher SFR. Consequently, without a more thorough study of the SEDs of these galaxies (in preparation), this difference of SFR should not be taken at face value. 

In conclusion, except for slight apparent peculiarities, the variations we observe for the host galaxies and their SMBHs seem to be consistent with what we could have expected for non-AGN galaxies. However, since the level of AGN activity is a phenomenon that is assumed to be transient, and thus expected to fade with time, that is, with the redshift, our samples could be seen as characteristics of two different epochs, where one AGN is more evolved than the other: Sy1 being less evolved than RG at high redshifts (peaking at $z \sim$ 0.22-0.23), and Sy2 being less evolved than LINER at low redshifts ($z \sim$ 0.09-0.11). However, Whether these epochs are connected by evolution through the effects of OFs is an hypothesis that still needs to be established.  

\subsection{Detection and measurement of resolved OF in AGN} 
\label{subsec2.5}
In \citep{2020Torres-Papaqui}, a method to automatically detect and measure OFs in the line [OIII]$\lambda$5007 was developed using an IDL script. Our method only consider OFs with S/N~$>3$ and that are spectroscopically resolved, that is, appearing as a broad Doppler components blue-shifted from the core of the main emission line by a few~\AA, corresponding to $\Delta v \ge - 69$~km~s$^{-1}$, which is equal to the resolution of the SDSS spectra \textcolor{red}{(see the caption of Figure~\ref{fig14:OFmethod2} for how  $\Delta v$ is measured)}. Applying this resolution criterion allows to disentangled the OF component from the narrow component associated with the gas at rest in the galaxy \citep[e.g.,][]{Woo2016,Perna2017}.

First, the script subtracts from each spectrum the component of the stellar template produced by STARLIGHT around the [OIII] line, obtaining a redressed spectrum where the fluxes can be more precisely measured. This step also allows to simplify the automatic detection and measurements of OFs.
To retrieved the total flux, the script fits two Gaussian functions using the Levenberg-Marquardt fitting algorithm, as implemented with the MPFIT library in IDL \citep[][]{Markwardt2009}. Together with the positions of the cores, the routine gives the flux intensities, with uncertainties of the order of 15\%, as well as their FWHMs \citep{2020Torres-Papaqui}. 
\begin{figure*}[ht!]
\gridline{
\fig{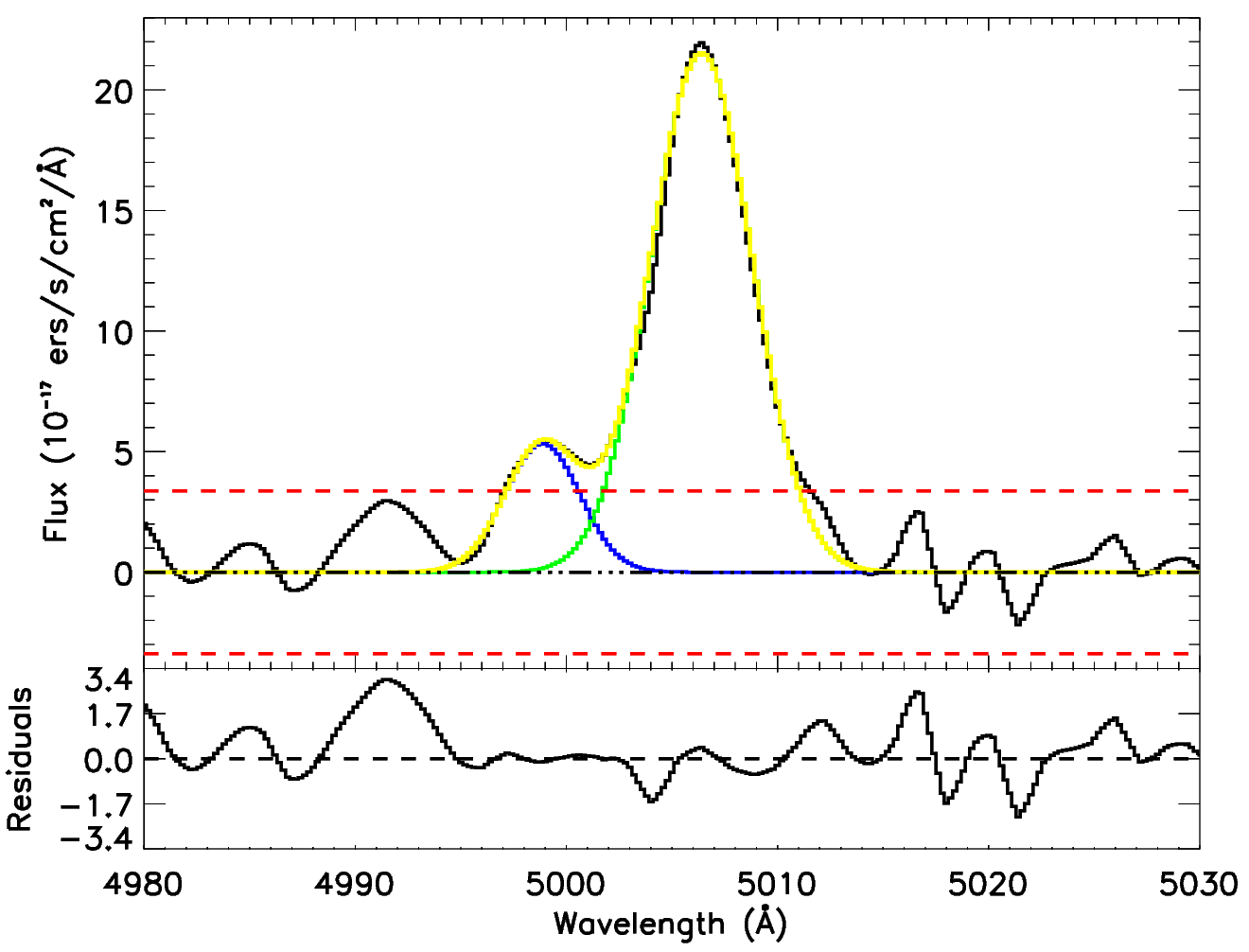}{0.45\textwidth}{(a)}
\fig{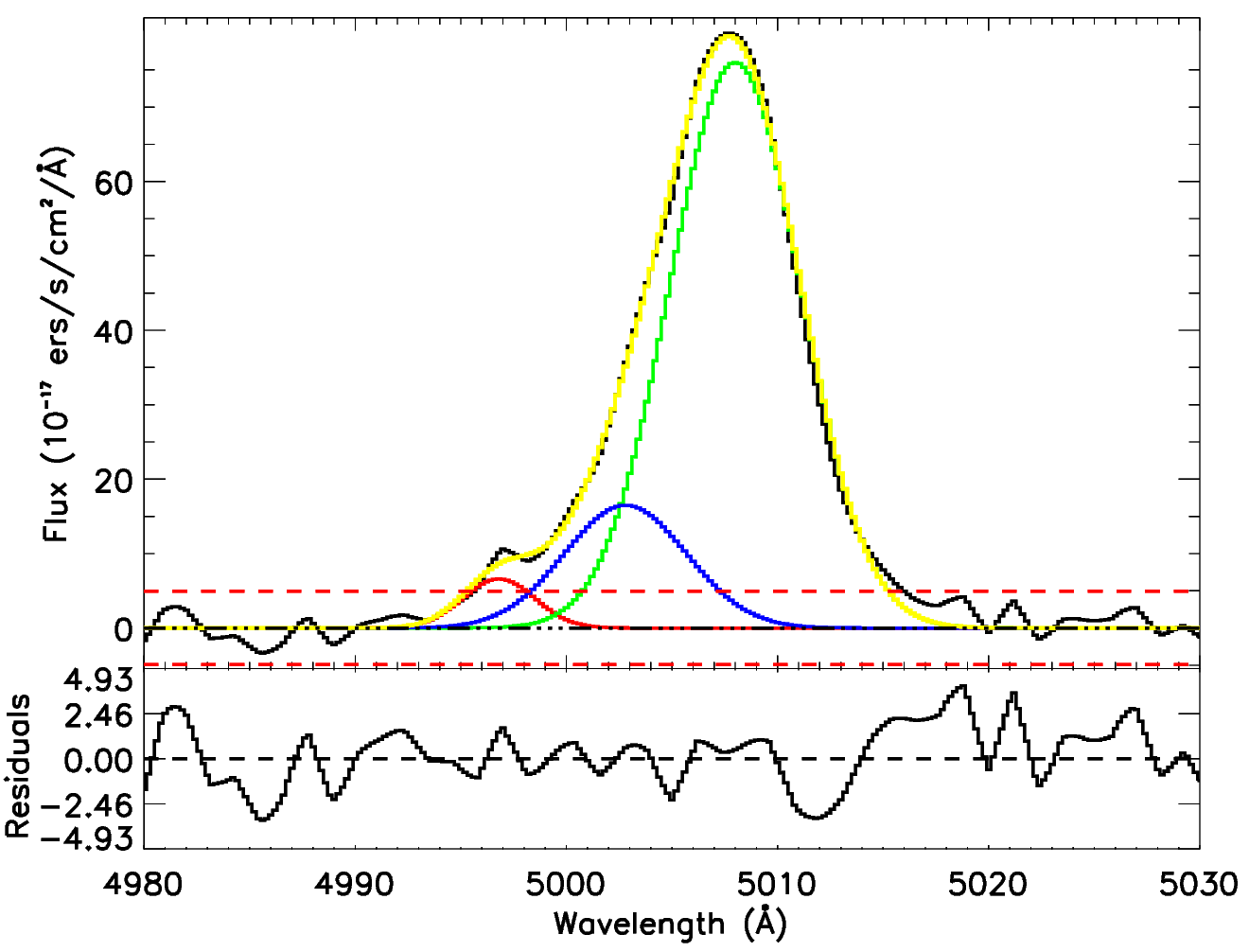}{0.45\textwidth}{(b)}
}
\caption{(a) Example of OF fitting using Gaussians. The green component is the main line, the blue is the OF, and the yellow is the sum of both (the red wing is not included). The red dash lines are the rms consistent with  S/N = 3. Note that the peak in OF is resolved, blueshifted by more than 69 km s$^{-1}$. (b) Example of double OF: the second component in red is resolved and has S/N $> 3$. 
\label{fig13:OFmethod1}
}
\end{figure*}
 
\begin{figure}[ht!]
\epsscale{0.5}
\plotone{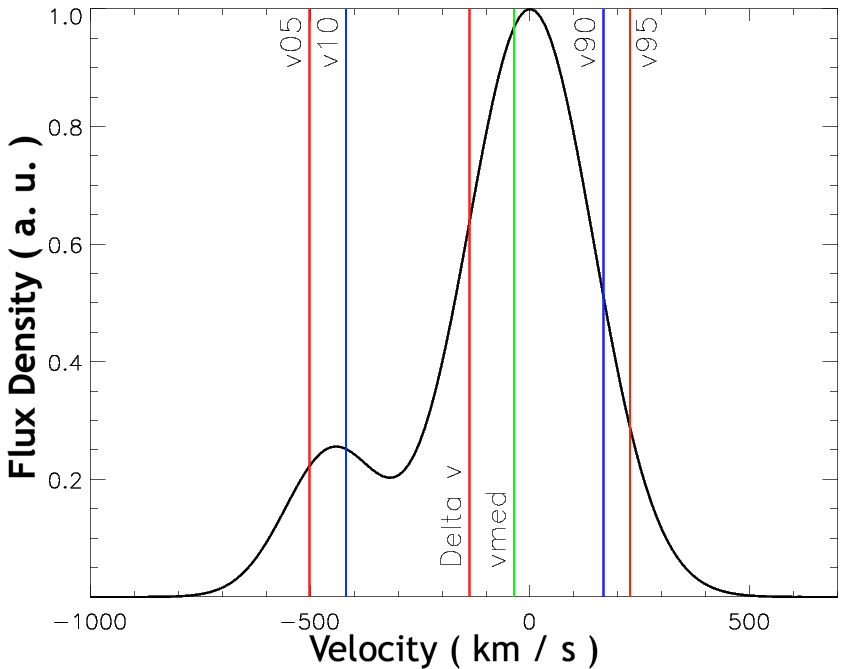}
\caption{Reconstruction of the velocity profile of the line showed in (a): v05, v10, v90 and v95, are the velocities in km/s, at the 5, 10, 90 and 95 flux percentiles, respectively. These percentiles are determined by tracing the total flux cumulative function, where the line width is defined as $W80 = v90 - v10$ and the velocity offset as $\Delta v = (v95 + v0.5)/2$ \citep{2014Harrison}. In absence of OF, $\Delta v =0$ and $W80 \sim {\rm FWHM}$, where FWHM is the full width half maximum of the main Gaussian at 0 velocity. 
\label{fig14:OFmethod2}}
\end{figure}

One example of resolved OF is shown in Figure~\ref{fig13:OFmethod1}a. Note in this example that the two Gaussians do not include the weak wing to the red of the line, because the signal it produces in the residual has S/N~$< 3$. This is different in the blue, where sometimes a second signal appears in the residual which is resolved and has S/N~$>3$, consistent with a double OF component. One example is shown in Figure~\ref{fig13:OFmethod1}b. When this happens, the script automatically fits a third Gaussian to retrieve the total flux within this wavelength range. 

Double OFs are very few, appearing in only 1\% of the Sy2 with OFs and 3\% in the other radio quiet AGN. They do not appear in RG. Generally, the second OF component has a higher blueshift and lower intensity than the first component. The origin of such structures is not obvious. They could either be evidence of recurrent events, as if the accretion of matter in the SMBH was intermittent, or evidence the two feedback modes happen at the same time, the secondary OF being created by a weak radio jet, briefly developing in compact radio sources. Supporting the second possibility, when weak radio sources are detected in radio quiet AGN (in Table~2) the probability of finding a double OF increases to 38\% in Sy2, 33\% in LINER but only 15\% in Sy1. In our analysis, we will consider only the first OF component, since double OFs are few in number and we do not have enough information to study this phenomenon thoroughly.

Using the sum of fluxes of the two most important Gaussians (double OFs are not considered and the flux in the red wing is negligible, representing less than 1\% of total flux), the velocity profile of the whole [OIII] line, in Figure~\ref{fig14:OFmethod2}, can easily be reconstructed and the percentiles estimated using a cumulative function. For comparison with other studies in the literature, the width velocity at 80\% (containing 80\% of the fluxes), $W80 = v90 - v10$ was estimated and used to characterize the intensity of the OF \citep{2014Harrison}. In principle, our Gaussian fitting method  (compared to directly integrating the fluxes) could sub-estimate $W80$. However, this uncertainty is mitigated by the fact that all the OFs in our study are well resolved, resulting in an average ratio $W80/{\rm FWHM} \sim 2$ \citep[for an explanation about the significance of this ratio, see][and references therein]{2021Aitor}. The main advantage of considering only OFs that are resolved is that are estimation of their intensity is free of contamination by the dynamics of the underlying stellar populations of the galaxy hosts \citep[which is not the case in][]{Woo2016,Perna2017}.  

\section{Results and analysis} \label{sec3:RES}

In Table~\ref{tab:OF}, we summarize the results for our search of resolved OFs in our three narrow-line AGN samples, also adding for comparison the results for our updated Sy1 sample. While the detection rate in Sy2 is comparable (although slightly lower) to what we observe in Sy1, the detection rates are significantly lower in LINER and RG. Because there are 4.6 times more Sy2 than Sy1 and only 1.3 times more LINER than Sy2, we can safely state that the probability to detect resolved OF decreases along the sequence Sy1$\to$Sy2$\to$LINER. On the other hand, because there is almost two times more LINER than RG, the detection rates are comparable. These results suggest that the detection of an OF strongly depends on the AGN class, Sy1/Sy2 vs. LINER/RG. Moreover, because LINER, Sy2 and Sy1 are mostly radio quiet compared to RG, our results also imply that the presence of a resolved OF does not depend on the capacity of an AGN to emit in radio. 
\begin{deluxetable*}{lcccccc}[ht!]
\tablecaption{OF detection\label{tab:OF}}
\tablewidth{0pt}
\tablehead{
\colhead{AGN class}  & \colhead{Total} 
& \colhead{OF} & \multicolumn{4}{c}{OF detection in each redshift bin}\\
\cline{4-7}
\colhead{} & \colhead{} 
&\colhead{} &\colhead{(0.0-0.1]} &\colhead{(0.1-0.2]} &\colhead{(0.2-0.3]} &\colhead{(0.3-0.4]} 
}
\startdata
RG-AGN   & 12064 
& 1292(11\%)  & 329(25.5\%) &  412(31.9\%) &  313(24.2\%)   & 238(18.4\%) \\
RG-SFG/TO   & 3729 
& 1132(30\%)  & 958(84.6\%) &  168(14.8\%) &  6(0.5\%)   & 0 \\
LINER & 25656 
& 4779(19\%)  & 2735(17.1\%) & 1807(20.1\%) &  194(29.4\%)  &  43(39.8\%) \\
Sy2   & 18385 
& 9683(53\%)  & 4850(51.6\%) & 4240(52.9\%) &  481(59.8\%)  & 112(68.3\%)\\
Sy1   &  4000 
& 2641(66\%)  &  254(69.0\%) &  831(63.9\%) & 1019(66.6\%)  & 537(67.0\%)\\
\enddata
\end{deluxetable*}
\begin{figure}[ht!]
\epsscale{0.7}
\plotone{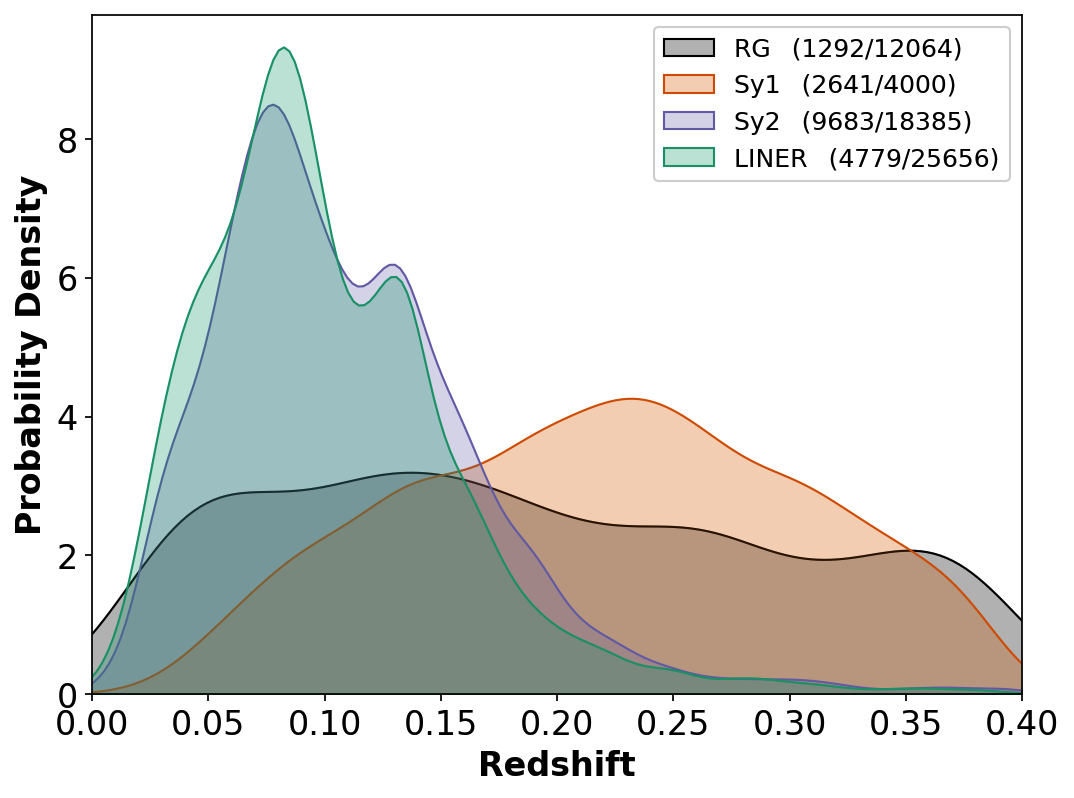}
\caption{Density distribution in redshift of galaxies with different AGN spectral classes and where OF are detected. 
\label{fig8:DD_OF_z}}
\end{figure} 

\subsection{Presence of outflows in different AGN spectral classes }
\label{Res:OF}
In Figure~\ref{fig8:DD_OF_z}, we trace the redshift density distributions of the AGN with detected OFs. Compared with Figure~\ref{Fig6:z}, only the distribution for RG is different, being almost flat over the whole range of redshifts. To understand why this happens, we have extended our search for OF in RG classified as SFG or TO, finding that 37\%  of them (three times more than in RG-AGN) show the same kind of outflows. 
Then, in Table~\ref{tab:OF} we calculated the fractions of OFs in different redshift bins. Although these fractions stay almost constant for RG-AGN, up to the last bin, they decrease rapidly for RG-SFG/TO which are late-type galaxies at low redshifts. 
This suggests that in the whole sample of RG (AGN and SFG/TO), OFs appear more frequently at low redshifts due to a change in morphology of the host galaxies, from early-type at high redshifts to late-type at low redshifts (c.f. Figure~\ref{Fig5:AltDiag_RG} and Figure~\ref{Fig6:BPT_RG_model}).

\begin{figure*}[ht!]
\gridline{\fig{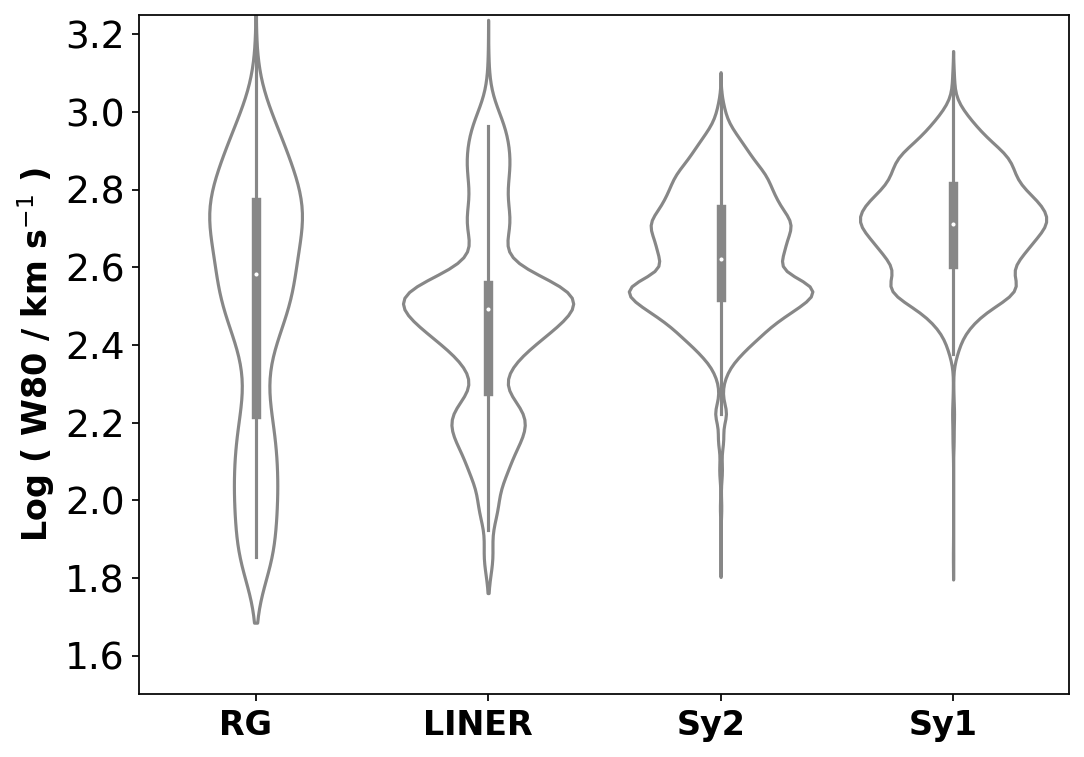}{0.45\textwidth}{(a)}
                \fig{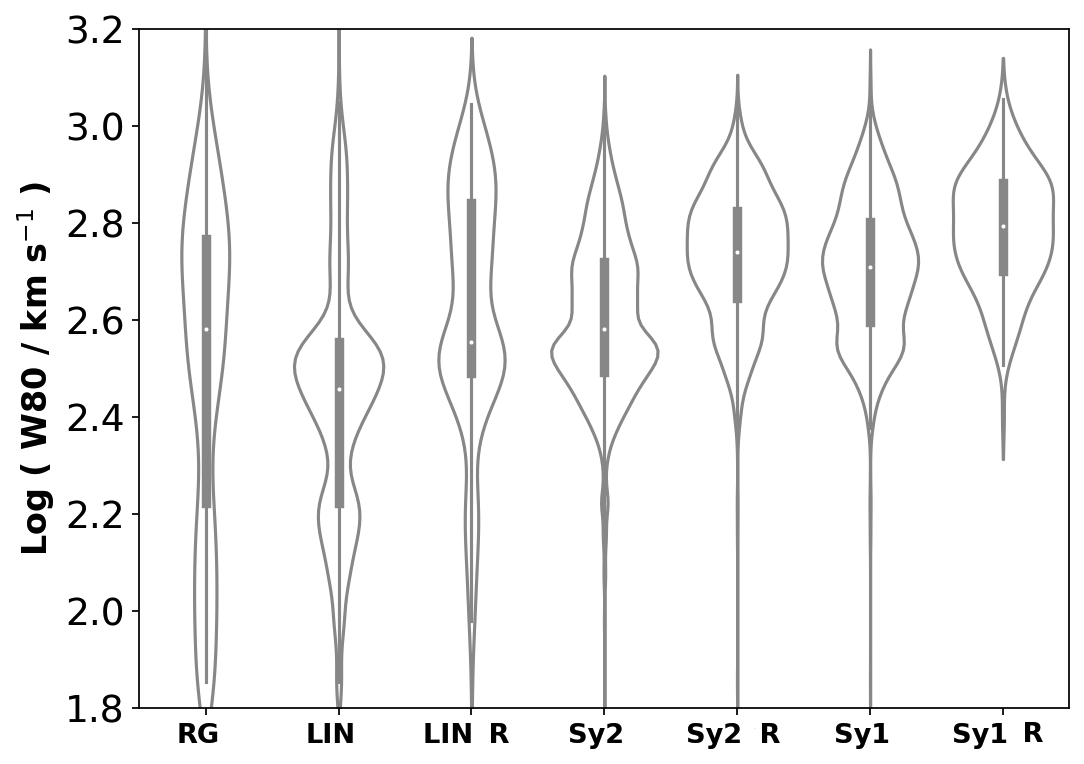}{0.45\textwidth}{(b)}}
\caption{(a) Violin plots comparing the intensity of the outflows in the different AGN types; (b) Same comparison distinguishing between AGN not detected in radio and those detected (adding R to their class).
\label{fig15:w80}
}
\end{figure*}

In the case of the radio-quiet AGN, the fraction of detected OFs in any redshift bin in Table~\ref{tab:OF} remains almost constant. Actually, the fraction slightly increases at high redshifts, which implies that the probability of observational biases affecting the detection of OF at high redshifts is nil. Consequently, the fact that at any redshift the rates of detection in Sy2 and Sy1 remain higher than in LINER and RG-AGN, implies that their detection depends not only on the different morphological types but also on the AGN luminosity (or level of AGN activity). Moreover, in Figure~\ref{fig15:w80}a there is a clear trend for the intensity of the AGN winds, based on W80, to increase along the luminosity sequence LINER$\to$Sy2$\to$Sy1. This is in good agreement with the conclusion in \citet{2020Torres-Papaqui} that OFs are intrinsic features related to the accretion process of matter onto a SMBH.  

However, in RG, which are also low-luminosity AGN, W80 tends to be higher than in LINER, in fact, getting as high as what we observe in Sy2 and Sy1. Since LINER are mostly radio quiet, we wondered whether this difference observed in Figure~\ref{fig15:w80}a could be due to the jet mode being dominant in RG \citep{2019Molyneux,2020Santoro}. To verify this hypothesis, in Figure~\ref{fig15:w80}b, we once again trace the violin plots for W80 but this time distinguishing between those AGN with and without radio detection. As one can see, W80 is systematically higher in AGN detected in radio. This suggests that there could be a difference in energy, the jet mode injecting more energy to the OF than the radiative mode (perhaps because jets are collimated). Note that this observation is fully consistent with our explanation for the excess of ionization observed in the RG-Sy2 and RG-LINER compared to their radio-quiet counterparts.

On the other hand, noting that even for the radio detected AGN W80 increases along the sequence RG/LINER$\to$Sy2$\to$Sy1, it seems possible that in radio-quiet AGN both modes, radiative and radio jet, contribute simultaneously in triggering OFs \citep[][]{2022Ayubinia,2022Silpa,2023Venturi,2023aSingha,2023bSingha}. Indeed, one can easily imagine a scenario where these two modes are not exclusive: at high luminosity (QSO/Sy1/Sy2), the radiative mode, being primordial, would be dominant---radio jets being weak and compact and thus more difficult to detect---while at lower luminosity (LINER/RG) the jet mode would dominate, but only when the conditions are favorable \citep[like suggested in][]{2017Coziol}, explaining why in general very few AGN are radio loud. 

\begin{figure*}[ht!]
\gridline{\fig{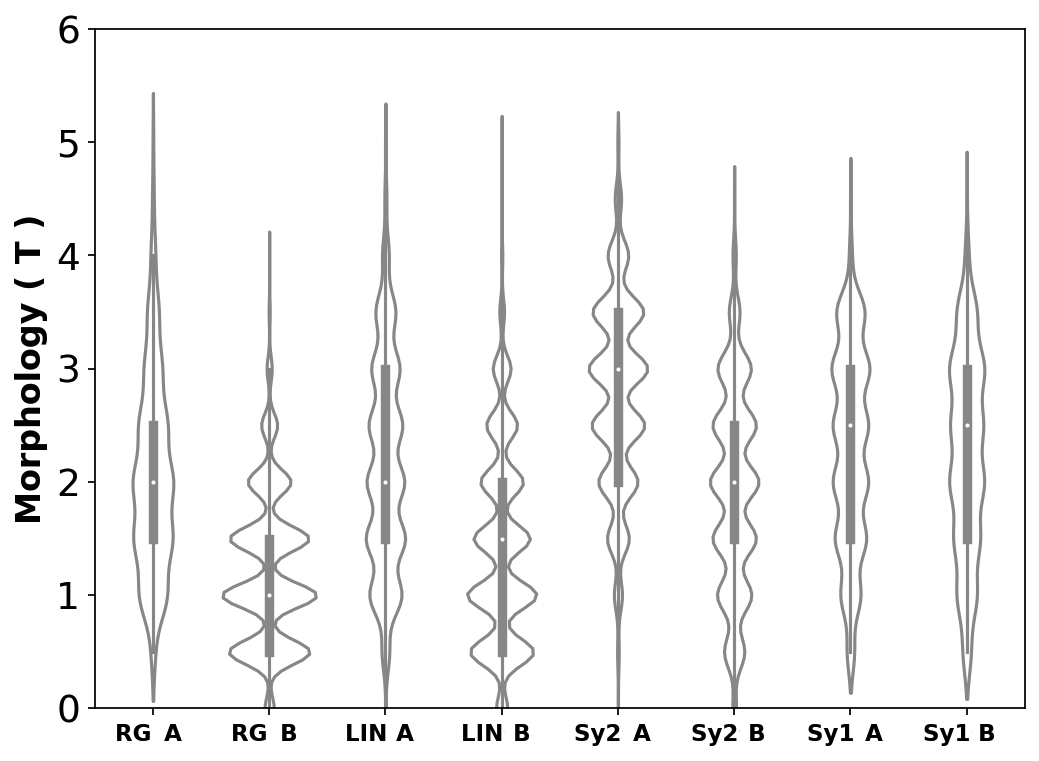}{0.45\textwidth}{(a)}
          \fig{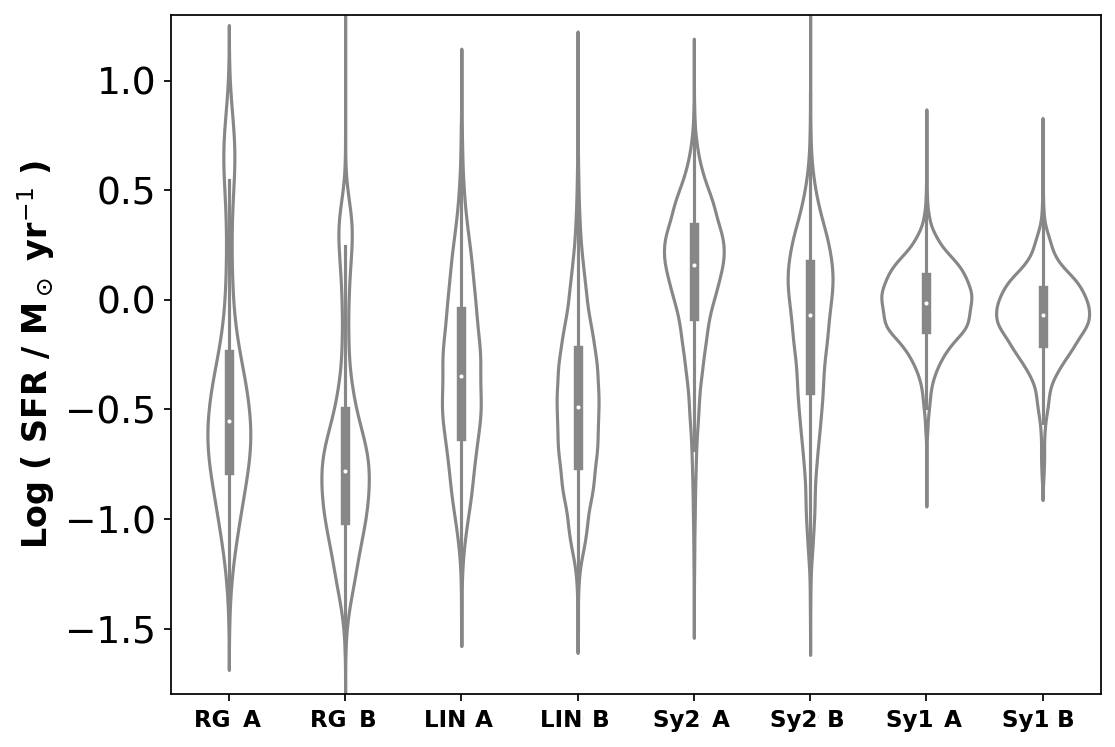}{0.48\textwidth}{(b)}
 }
\caption{(a) Violin plots comparing the morphologies of AGN with OFs (identified as A) and without OF (identified as B). The equivalent in terms of Hubble types are: (0, E), (1, S0), (2, Sa), (3, Sb), (4, Sc), (5, Sd) and (6, Im), with intermediate mixed types, for example, (0.5, E/S0), (1.5, S0/Sa), etc.; (b) comparison of SFR. 
\label{fig16:OF_T+SFR}
}
\end{figure*}

Additionally, a change of dominant morphology at low redshifts, from early- to late-type, seems to significantly increase the detection of OFs in RG. Normally, late-type spiral galaxies have disks that are rich in gas, and assuming a SMBH is at the source of the wind in all AGN, having a huge amount of gas would increase the chance for OFs to form in their narrow-line regions. If this interpretation is correct that may also apply to other AGN classes. Comparing in Figure~\ref{fig16:OF_T+SFR}a the morphological types of the AGN host galaxies with OFs (identified as A) with their counterpart without OF (identified as B), one can see a clear trend for the former to have a later-type morphology than the latter. This suggests that one extra condition to observe an OF in any class of AGN is a galaxy host particularly rich in gas \citep[][]{2019Luo,2020Jarvis,2023Harrison}. 

Incidentally, the fact that AGN with OFs are late-type spirals that are rich in gas also makes them ideal candidates to search for direct feedback effects, most specifically, evidence in their disks of quenching or triggering star formation. Searching for such effects, in Figure~\ref{fig16:OF_T+SFR}b we now compare the SFR in galaxies with and without OF. This shows that in any AGN class, galaxies with OFs have actually higher SFR than those without OF \citep{2020Woo,2021Luo}. Apparently this result contradicts the quenching hypothesis. However, at the same time one can see that SFR increases along the sequence RG$\to$LINER$\to$Sy1$\to$Sy2, which roughly points, assuming evolution, towards the right direction for quenching, that is, high-luminosity, high-SFR AGN evolve into low-luminosity, low-SFR AGN \citep[e.g.,][]{2016Wylezalek}. But whether that evolution necessarily involves OFs is unclear (this is circumstantial evidence).\footnote{Here we use the jurist definition: ``A circumstantial evidence is an indirect evidence that does not, on its face, prove a fact in issue but gives rise to a logical inference that the fact exists. Circumstantial evidence requires drawing additional reasonable inferences in order to support the claim''; Definition of LII legal information institute Cornell Law School, Ithaca, NY.} Conversely, the fact galaxies with OFs have higher SFR than their counterpart without OF might also be taken as evidence for positive feedback \citep{2019Shin,2021Yao,2023Mercedes-Feliz,2023Mahoro}. Actually, based on our data, the increase in SFR between galaxies with and without OF is relatively small, which suggests that since AGN with OFs have later-type morphologies than their counterpart without OF, finding a slightly higher SFR in these galaxies would rather seem natural. Normally, late-type spiral galaxies have disks that are rich in gas and, consequently, have higher star formation rates than early type galaxies \citep{Kennicutt2012}. This is a natural trait that is usually interpreted as due to different formation processes, late-type spirals forming their stars less rapidly than early-type spirals \citep{Sandage1986}. This is well explained by the hierarchical galaxy formation scenario \citep{1998Coziol_a}---the  higher the number of mergers of proto-galaxies, the faster their astration rates, that is, how fast the galaxies transform their gas into stars; in this model spiral galaxies form in less dense regions where the number of proto-galaxies that merge are fewer.

Consequently, we tentatively conclude that there are no clear evidence of quenching or triggering of star formation in our AGN samples. This would imply that we observe OFs in AGN only when there is still a huge amount of gas in the narrow-line regions of the galaxies. In AGN with OFs the intensity of the winds (W80) increases with the AGN luminosity---or accretion rate consistent with radiative mode---and are amplified when the jet mode is present or become dominant like in RG. What happens to these OFs over a longer period of time and what role could they play in the evolution of galaxies in the different AGN classes is still something that needs to be investigated further. 

\subsection{In search for OF feedback in nearby AGN}
\label{res:COMP_SMBH_host}

Understanding how AGN winds contribute to the formation/evolution of galaxies is a much more complex subject than searching for direct evidence of OF effects like quenching or triggering of star formation. After they are produced, the gas forming the OFs is expected to eventually mix with the gas in the NLRs \citep{2023Jones} and for the OFs to slowly disappear over a certain period of time as their effects take place. Therefore, assuming a sufficiently long delay, not only an evolution from high to low SFR might be expected \citep[e.g.][]{2016Wylezalek} but at the same time, an evolution from high to low AGN luminosity, as the accretion activity also diminishes \citep[e.g.,][]{2023Temple}. 

As it is, what we observe in Figure~\ref{fig16:OF_T+SFR}b might be interpreted as evidence of quenching of SFR. However, since the behavior is general, appearing in each AGN class, connecting this evolution to the effect of OFs would imply such evolution only happens within galaxies in each class. However, for RG, LINER and Sy2, such evolution does not fit the differences in morphology observed between galaxies with and without OFs, since quenching of SFR, even after a delay, would not explain the more massive bulges (and higher BH masses) of galaxies without OFs. Only Sy1, where in Figure~\ref{fig16:OF_T+SFR}a we do not observe a difference in morphology, might fit such simple scenario. 

\begin{figure*}[ht!]
\gridline{\fig{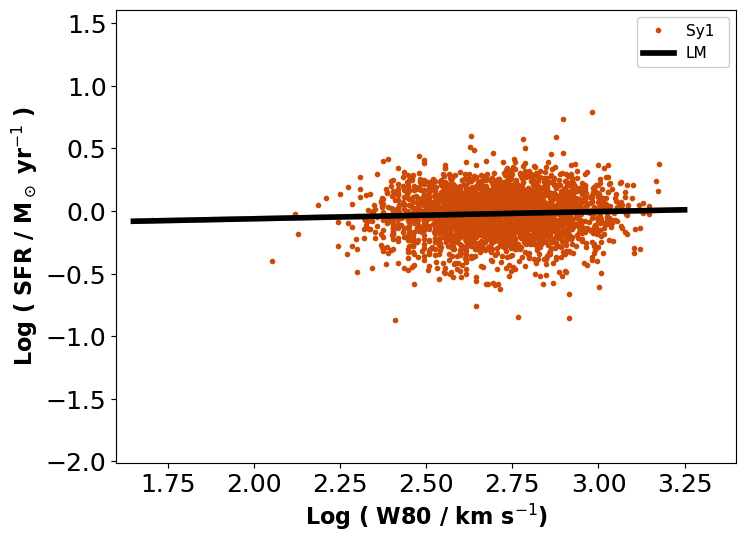}{0.35\textwidth}{(a)}
          \fig{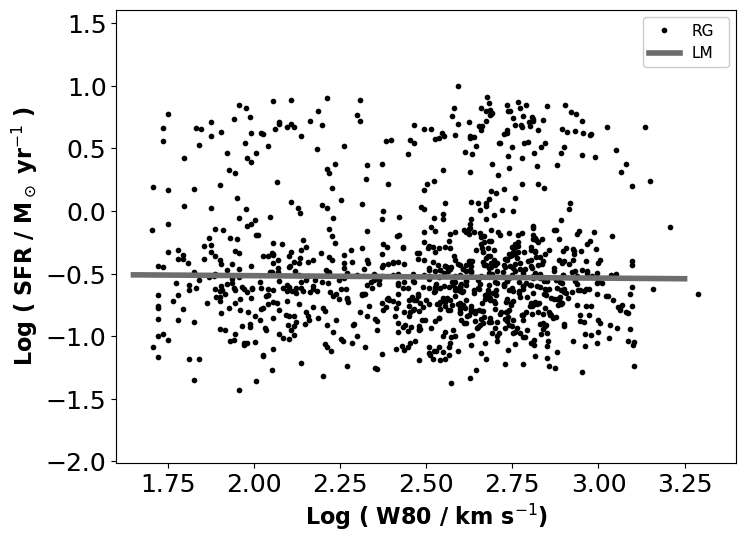}{0.35\textwidth}{(b)}
 }
 \gridline{\fig{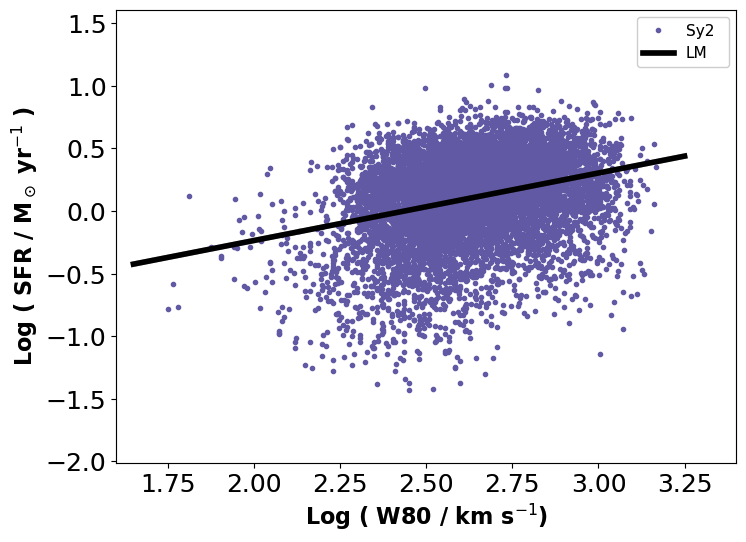}{0.35\textwidth}{(c)}
          \fig{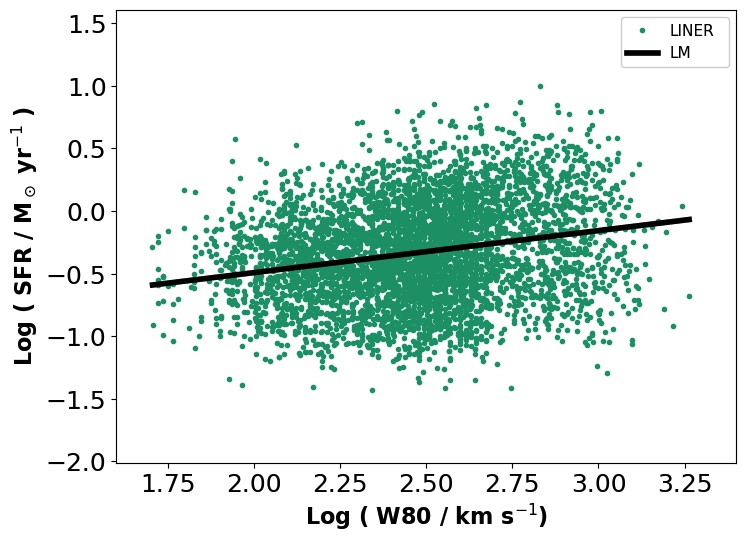}{0.35\textwidth}{(d)} }
\caption{Comparing the SFR with the intensity of the wind, W80, in (a) Sy1, (b) RG, (c) Sy2 and (d) LINER. 
\label{fig17:SFRvsW80}
}
\end{figure*}

This is easy to check. In \citet[e.g.,][]{2016Wylezalek} the authors compared in a inhomogeneous sample of 132 radio-quiet AGN with OFs the SFR with W80, concluding that due to AGN winds SFR decreases as W80 increases. Doing the same comparison for the Sy1 with OFs in Figure~\ref{fig17:SFRvsW80}a, the SFR tend to be independent from the wind intensity. Coincidentally, the fact SFR is constant in Sy1 is as expected in galaxies having the same morphology, which is what we see in Figure~\ref{fig16:OF_T+SFR}a. Consequently, despite W80 being highest in Sy1 host galaxies, their SFRs continue to be consistent with their morphological types. 

However, SFR in RG also seems, in Figure~\ref{fig17:SFRvsW80}b, to stay constant as W80 increases. This is important because, contrary to the SFRs in Sy1, those in RG were more directly determined using STARLIGHT. Moreover, there is also a positive trend for SFR to increase with W80 in Sy2 in Figure~\ref{fig17:SFRvsW80}c and LINER in Figure~\ref{fig17:SFRvsW80}d. Considering that W80 increases along the sequence LINER$\to$Sy2/RG$\to$Sy1 in Figure~\ref{fig15:w80}b, maybe feedback effects only happens in optically the optically luminous Sy1 and radio luminous RG, where the OF intensity crosses a certain threshold that makes them quenching efficient \citep[e.g.,][]{2019Smith,2020Voit,2022Zhu,2023Bollati}.
However, considering that SFR in Figure~\ref{fig16:OF_T+SFR}b decreases along a different sequence, Sy2$\to$Sy1$\to$LINER$\to$RG, it is not clear why LINER, assumed to be below the threshold would have lower SFR than Sy1 (remembering also that SFR in Sy1 could be underestimated). Moreover, this would neither explain why the detection of OFs in LINER is much lower than in Sy2 or Sy1, which both have high detection rates. Indeed, based on the OF detection rates and SFRs, one could have expected RG and LINER to be those AGN more affected by quenching. 

Alternatively, evolution night connect two different AGN classes. More specifically, one could imagine Sy2 or Sy1 transforming into LINER (assuming the missing OFs in LINER already did their deeds before disappearing). But such evolution would already goes beyond the quenching effect due to OFs, suggesting that the higher SFR in Sy2 and Sy1 would also have something to do with the transformation of morphology, from late- in Sy2/Sy1 to early-type in LINER, also implying a grow in mass in parallel of their SMBHs. Without any clear evidence to distinguish between OF effects and processes related to the formation of galaxies, that hypothesis would be highly speculative! The putative role of OFs in RG is possibly even more complex, since these would be triggered by radio jets, and, consequently their formation and evolution would depend on conditions that might not exist in the radio-quiet AGN classes, albeit producing the same morphological type transformations. 

Therefore, due to the difference in morphological types the galaxies with and without OF in each AGN class do not seem to be connected by any evolution pattern. Similarly, the same differences seem to eliminate the possibility of an evolution of galaxies connecting the different AGN classes. In fact, the dichotomy in redshifts could suggest we observed two different levels of activity, high/low, at two different epochs, Sy1/RG above $z \sim 0.15$ and Sy2/LINER below, where galaxies with OFs in each class only appear because of their different distributions in morphology due to different formation processes; different AGN classes producing different fractions of late-type spirals riches in gas. Consequently, how will these OFs evolve with time and whether this evolution has any impact on their galaxy hosts is a question that only involves those galaxies that form as late-type spirals, not the whole sample of galaxies in their class. 

Considering the whole samples, what we should expect is a multi-scenario of local and individual formation/evolutionary paths where the Eddington ratio decreases as the BH mass increases, the AGN luminosity decreases as the accretion decreases, and the SFR decreases with the astration rate as the the reservoir of gas decreases. Therefore, assuming the AGN phase to be a special phase in the formation galaxies and recognizing that AGN winds are intrinsic to the AGN activity, and as such could be recurrent in time with cumulative effects on their hosts \citep{2023Harrison}, one can easily postulates they might have some important effects. However, most of these effects imputed to OFs would have happened in the past, making AGN winds a part of the standard formation/evolutionary process of galaxies. Consequently, discerning their effects from the effects of other intrinsic processes of formation/evolution of galaxies, in particular, astration---how galaxies normally exhaust their reservoir of gas by forming stars---should be especially difficult.

This suggests that in order to assess correctly what can be the effects of OFs on the evolution of their host galaxies, we need first to establish what characterizes the galaxies in each spectral class and determine how these characteristics are connected with the formation/evolution of the SMBHs at their centers. 
\subsection{Characteristics of galaxies and SMBH in different AGN spectral classes}

\begin{figure*}[ht!]
\gridline{\fig{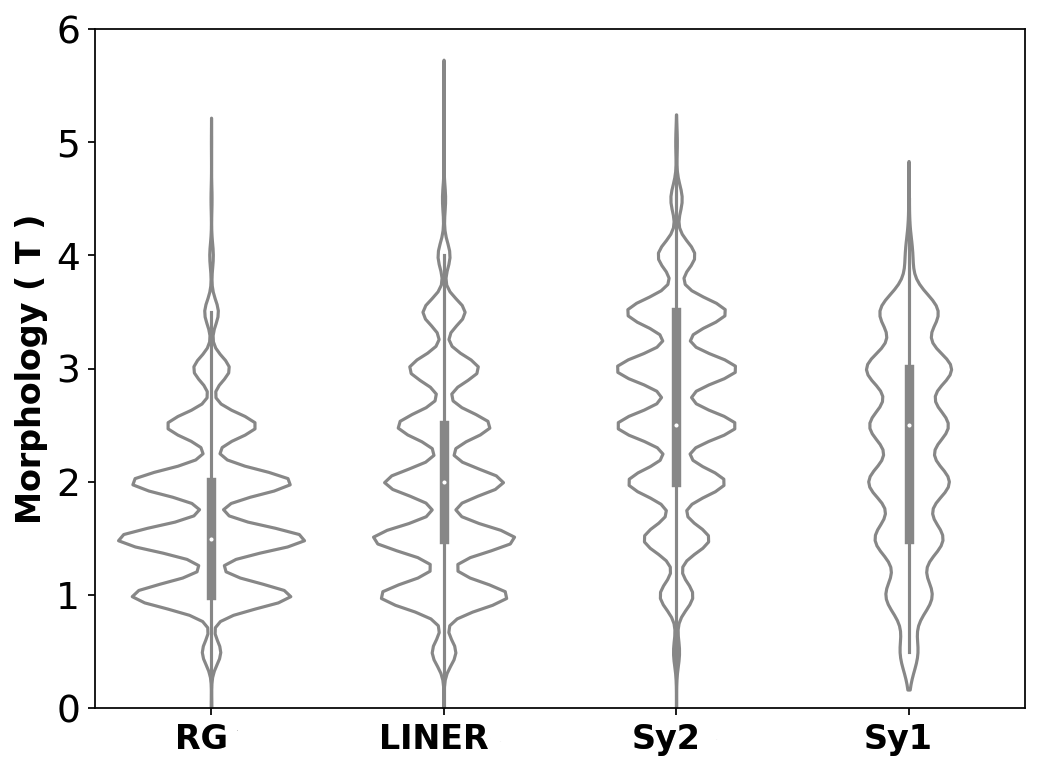}{0.38\textwidth}{(a)}
\fig{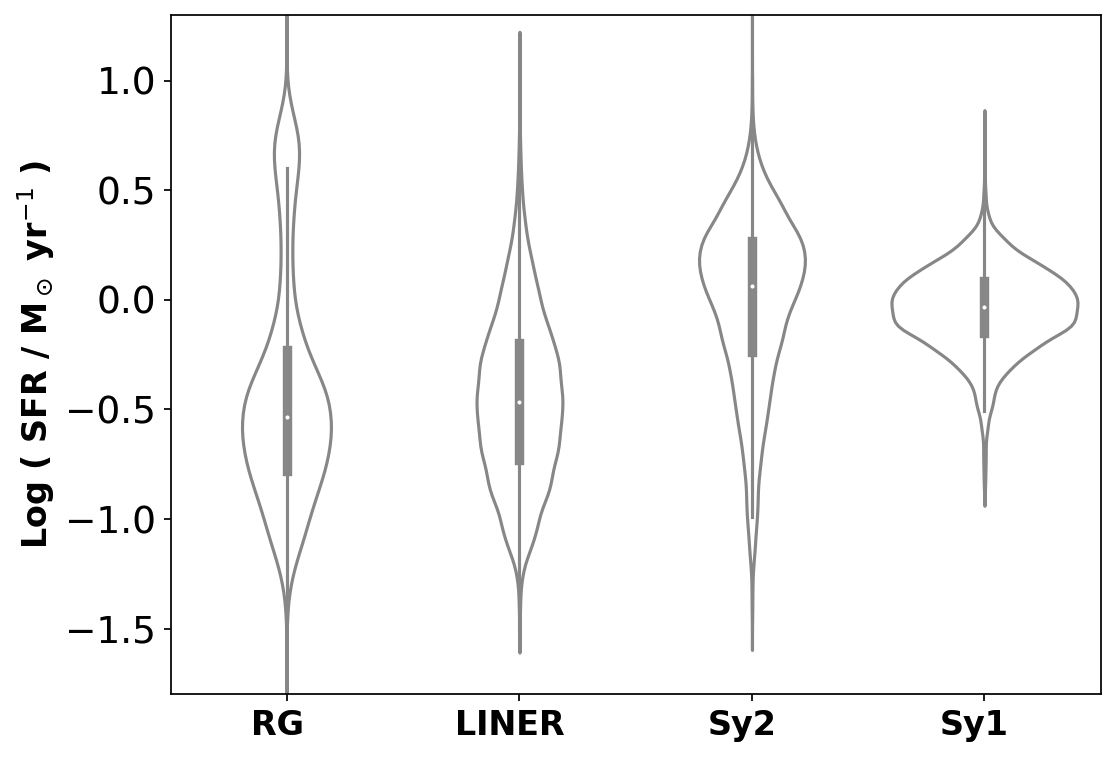}{0.4\textwidth}{(b)}  
}
 \gridline{\fig{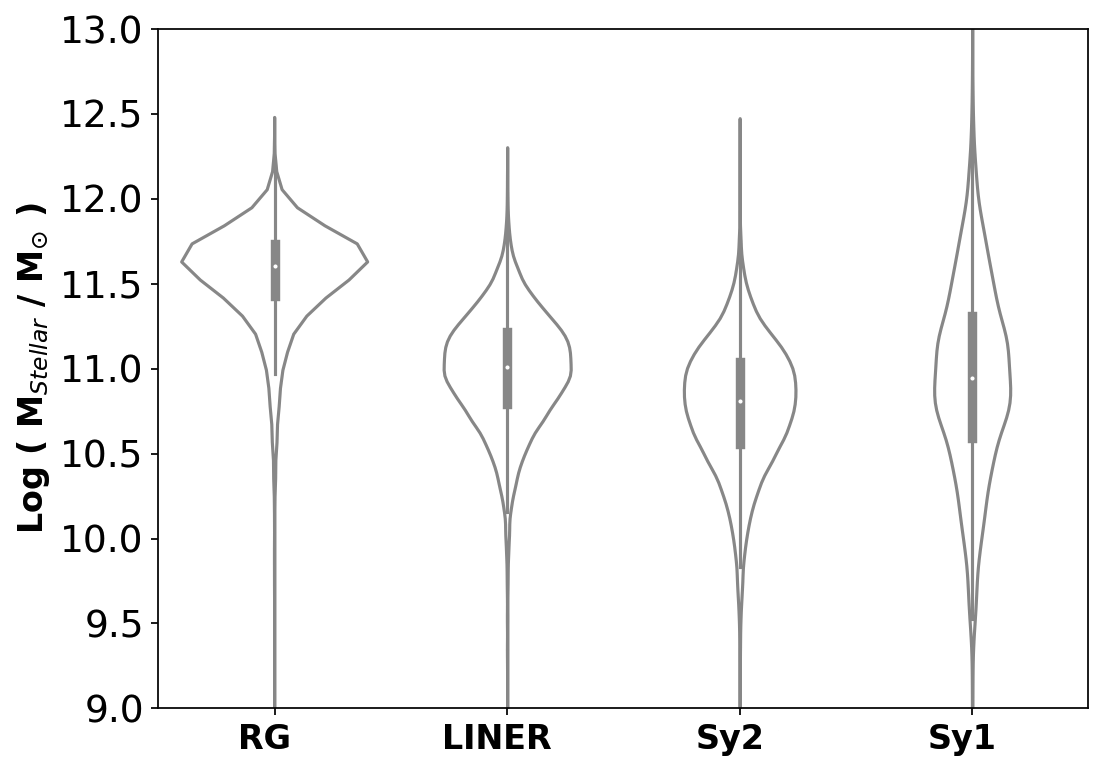}{0.4\textwidth}{(c)}
          \fig{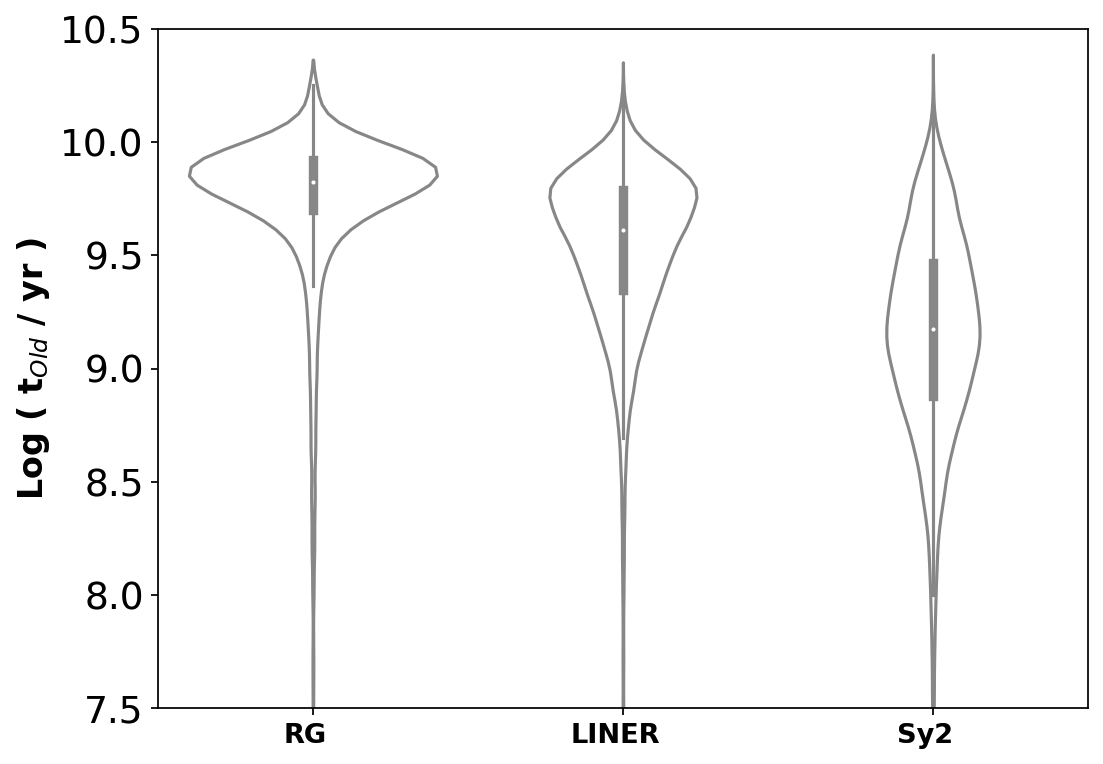}{0.4\textwidth}{(d)} }
\caption{(a) Violin plots comparing the morphology of the host in the different AGN types. The equivalent in terms of Hubble types are: (0, E), (1, S0), (2, Sa), (3, Sb), (4, Sc), (5, Sd) and (6, Im), with mixed types, for example, (0.5, E/S0), (1.5, S0/Sa), etc.; (b) comparison of SFRs, (c) stellar masses, M$_*$ and (d) ages of the oldest stellar populations, $t_{old}$.\label{fig18:Host}
}
\end{figure*}

Comparing the host characteristics of the AGN with different spectral classes in Figure~\ref{fig18:Host}, there are not much new results compared to what is already known. In terms of morphology, Figure~\ref{fig18:Host}a shows that Sy1 and Sy2 are hosted by early-type spirals (Sab/Sbc), with a trend for Sy2 hosts to have slightly later morphological types than Sy1 hosts (which have smoother density distribution in violin plot), while LINER and RG hosts have significantly earlier types (S0/Sa) than in the Seyfert spectral classes. Note that since LINER and Sy2 are located on average at the same redshift, their morphological difference cannot be attributed to observational biases. Similarly, Sy1 and RG-AGN being at high redshifts, their differences must also be physical; the only bias is RG are radio loud while Sy1 are radio quiet. Consequently, at their respective redshifts, the hosts of RG and LINER tend to have earlier morphological types than the hosts of the two Seyfert galaxies. In Figure~\ref{fig18:Host}b we can also verify, as we noted before, that the differences in SFR follow the morphological type differences. This is a common trait of all galaxies, not only AGN. 

\begin{figure*}[ht!]
\gridline{\fig{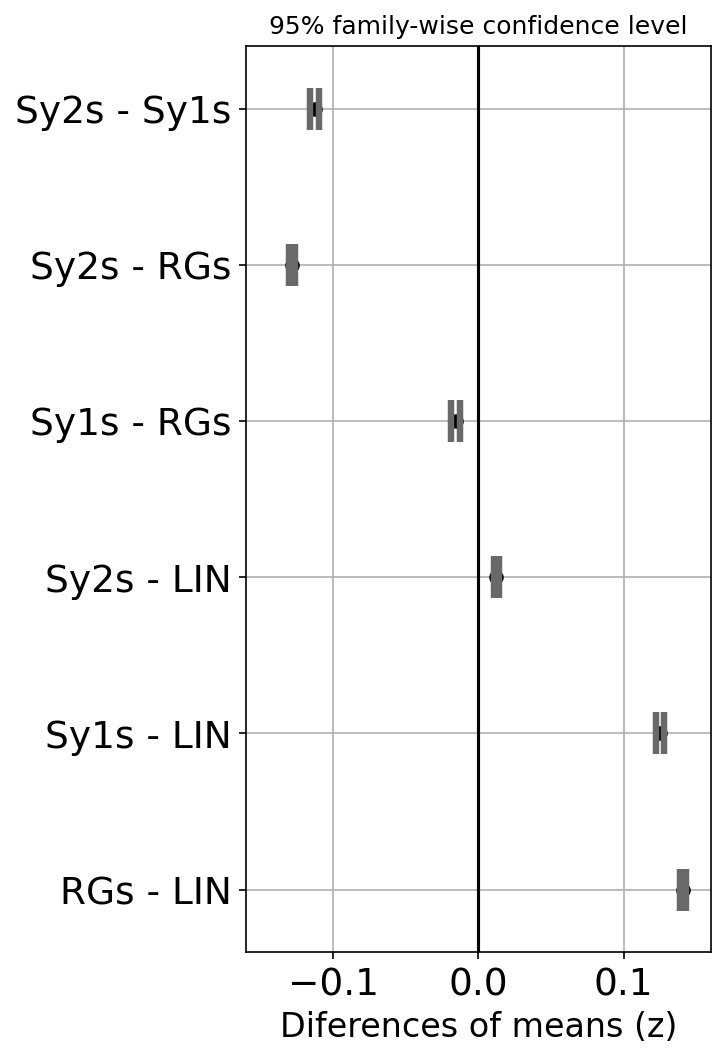}{0.25\textwidth}{(a)}
               \fig{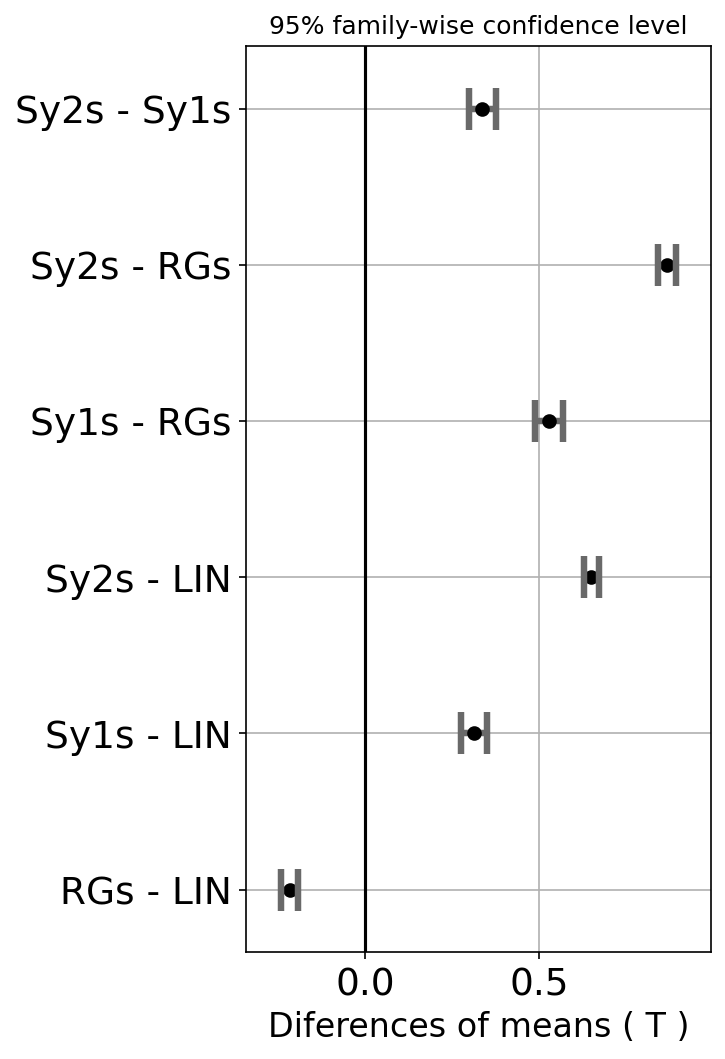}{0.25\textwidth}{(b)}
               \fig{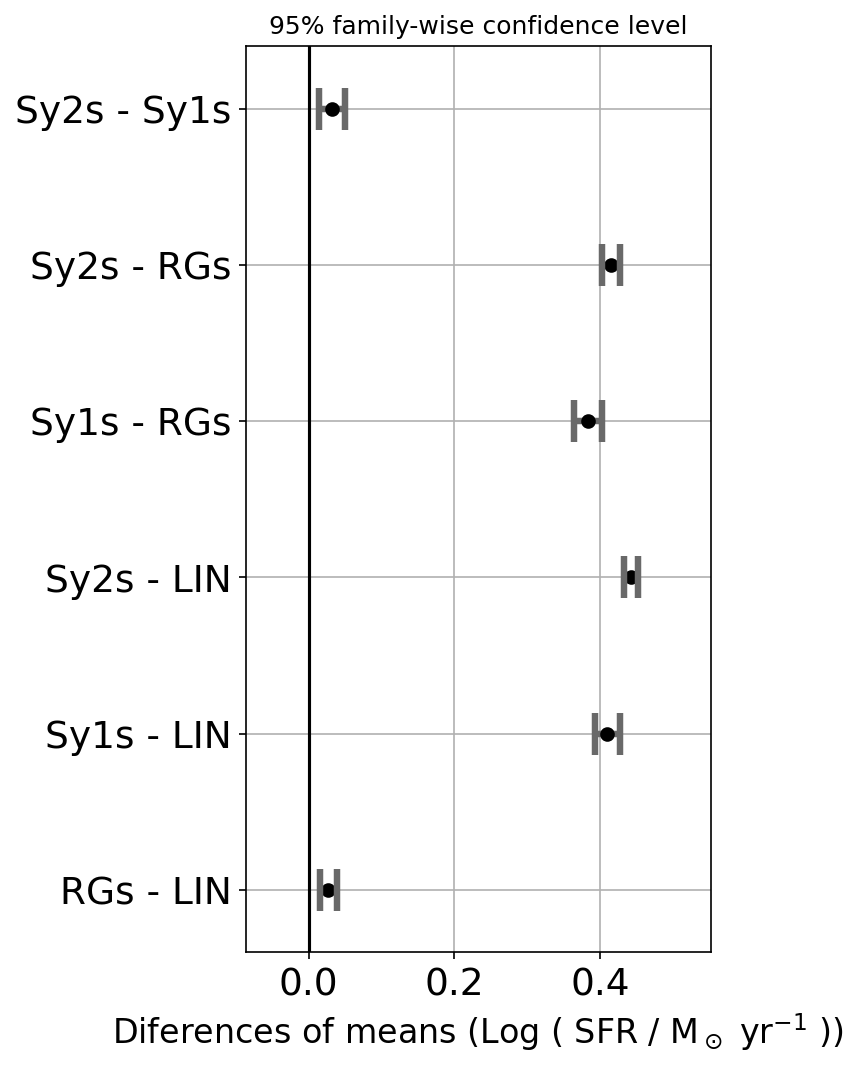}{0.29\textwidth}{(c)}
    }
\caption{Max-t test comparing the host characteristics of galaxies with different AGN types: (a) Redshift (z), (b) Morphology (T), (c) Star formation rate (SFR).  \label{fig19:host1}
}
\end{figure*}

To test the statistical significance of these differences, we applied the max-t~test, which is a parametric test based on simultaneous pairwise comparisons of means \citep[see explanations in][]{2012aJPTP}. This multi-comparison test, which was performed using the R statistics programming language,\footnote{https://www.r-project.org} has the advantage that it is insensitive to the differences in size of the samples compared, although becoming more effective as the sizes increase, allowing, consequently, to detect smaller differences. In Figure~\ref{fig19:host1} we show one example; for clarity sake, all the other test results will be presented in the Appendix. The results are presented under the form of simultaneous confidence intervals for all pairwise comparisons of the means, $(p_1 - p_2)$,  in the four AGN classes. A positive difference implies that the first member of the pair, $p_1$, has higher mean than the second member (the inverse when negative). Confidence intervals including zero indicate no statistically significant difference and the farther from zero the more statistically significant the difference. 

In Figure~\ref{fig19:host1}a, the max-t test confirms, at a level of confidence of 95\%, the close similitude in redshift of Sy2 with LINER and of Sy1 with RG. The small positive differences imply that Sy1 are at slightly higher redshifts than RG, and that Sy2 are at slightly higher redshifts than LINER. In Figure~\ref{fig19:host1}b, the test confirms the similitude in morphology of RG and LINER, the difference being slightly negative showing a weak trend for LINER to have later morphological types. The max-t test also shows that Sy2 have later morphological types than Sy1, and both have later morphological types than LINER and RG. In general, the test confirms the sequence along which T decreases: Sy2$\to$Sy1$\to$LINER$\to$RG. Finally, in Figure~\ref{fig19:host1}c the max-t test confirms that the differences in SFRs are small between RG and LINER and between Sy1 and Sy2. Moreover, as expected for any galaxy (except perhaps starbursts) there is a strong connection between SFR and T. In other words, the different AGN spectral classes do not change the relationships between the characteristics usually found in galaxies with different morphological types \citep[e.g.,][]{2011Coziol}. 

In Table~\ref{RES_AVhost}, we already saw that the velocity dispersion increases along the sequence, Sy2$\to$LINER$\to$RG, which implies, consistent with their earlier morphological types, that RG have more massive bulges, on average, than LINER. In Figure~\ref{fig18:Host}c we also saw that the stellar masses are also significantly different, RG having the highest masses, followed by LINER, then Sy1 and Sy2. Therefore, in terms of stellar mass in Figure~\ref{fig18:Host}c, the variation from high to low is the inverse of the morphology sequence from early- to late-type, RG$\to$LINER$\to$Sy1$\to$Sy2. Interestingly, LINER are almost as massive as Sy1. This is confirmed by the max-t test in Figure~\ref{figA1:maxt_host}a. In general, therefore, as the mass decreases, galaxies at low and high redshifts tend to have later morphological types. This is also a trait common to all galaxies. 
Consequently, in Figure~\ref{fig18:Host}d, the variation in age of the older stellar populations reflects what is seen in any normal galaxies: as the mass of a galaxy increases, its stellar population gets older.  

\begin{figure*}[ht!]
\gridline{\fig{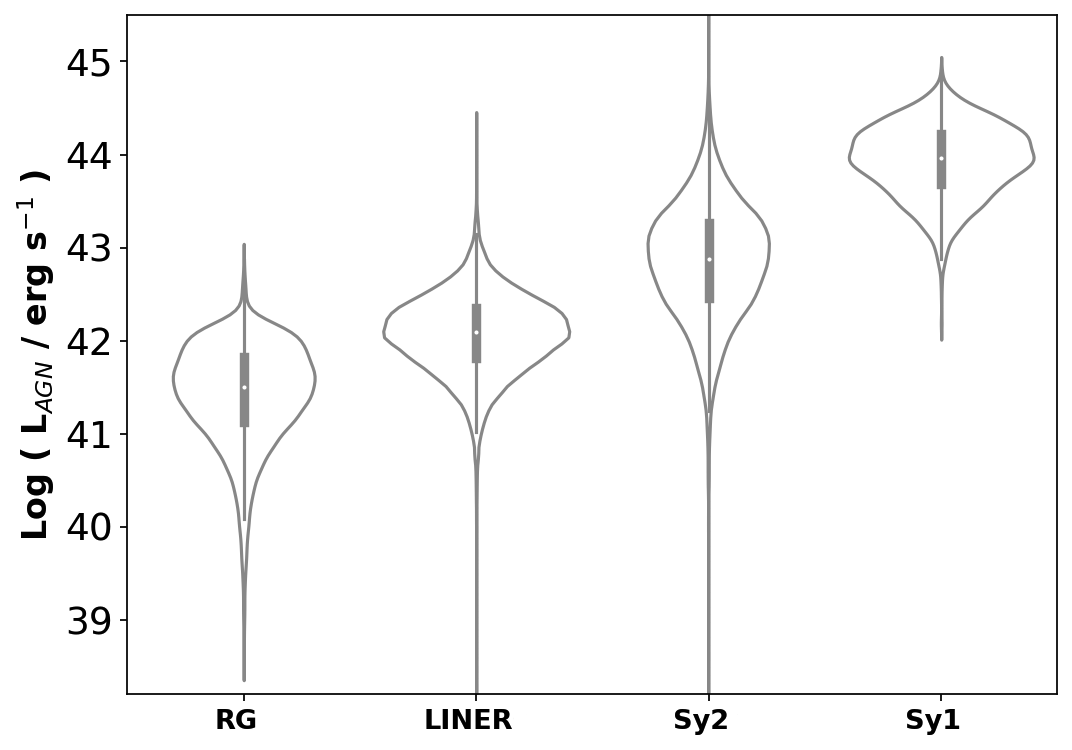}{0.4\textwidth}{(a)}
                \fig{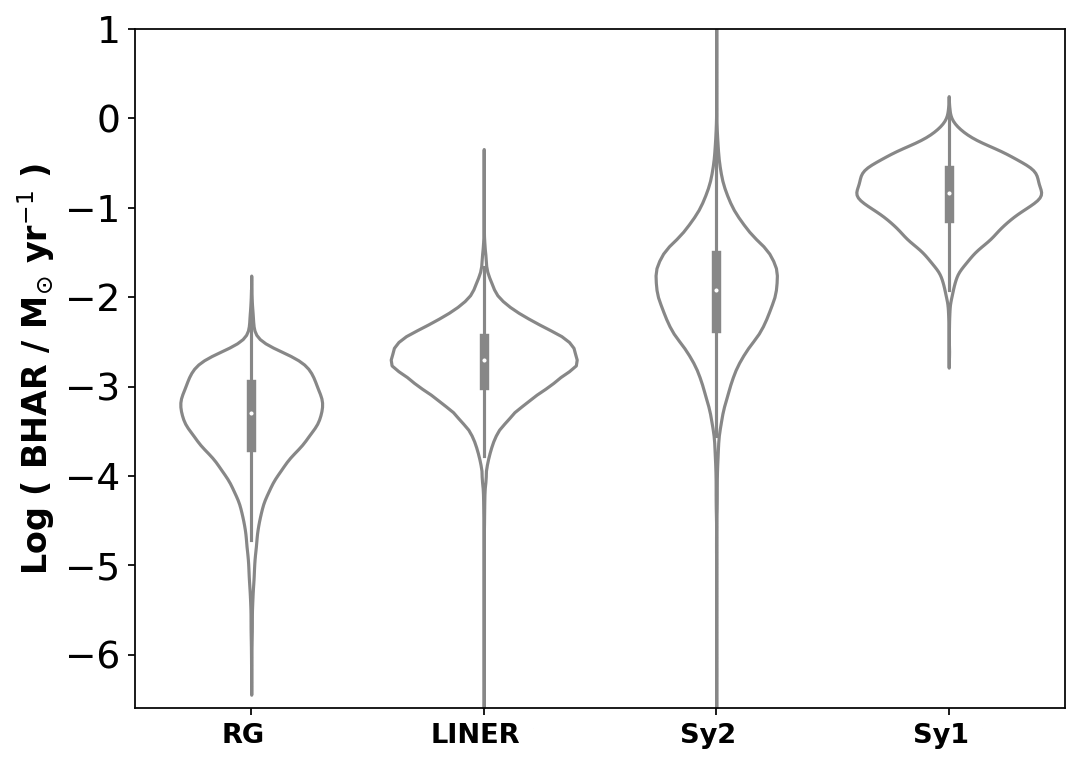}{0.4\textwidth}{(b)}}
 \gridline{\fig{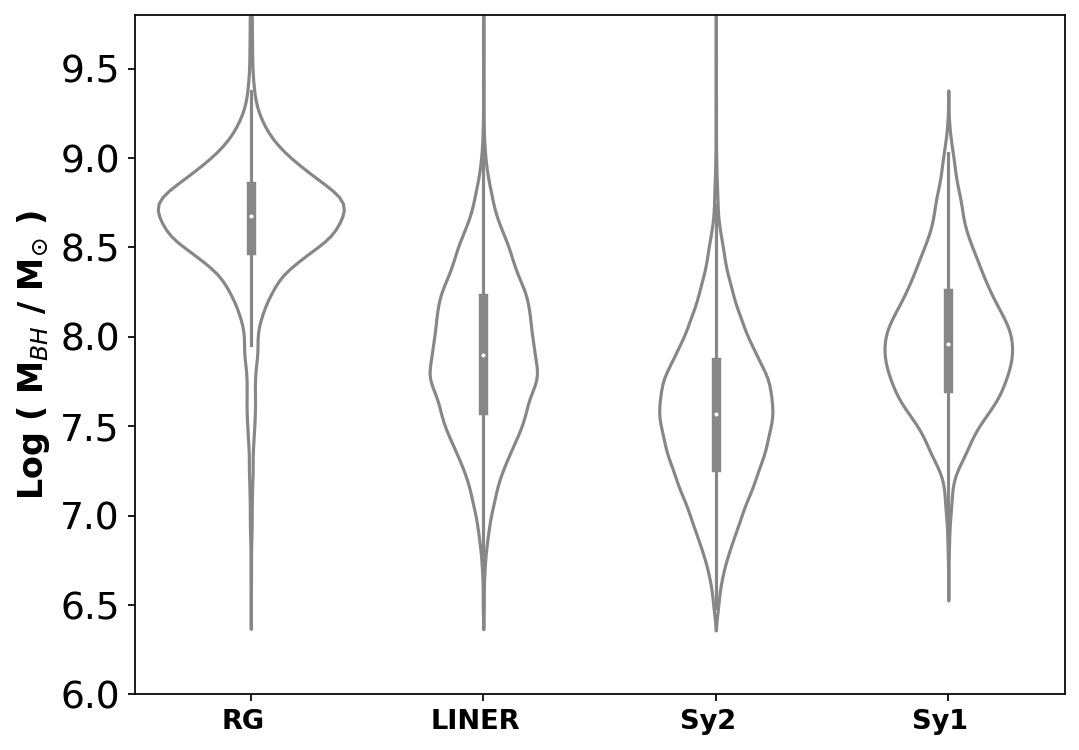}{0.4\textwidth}{(c)}
          \fig{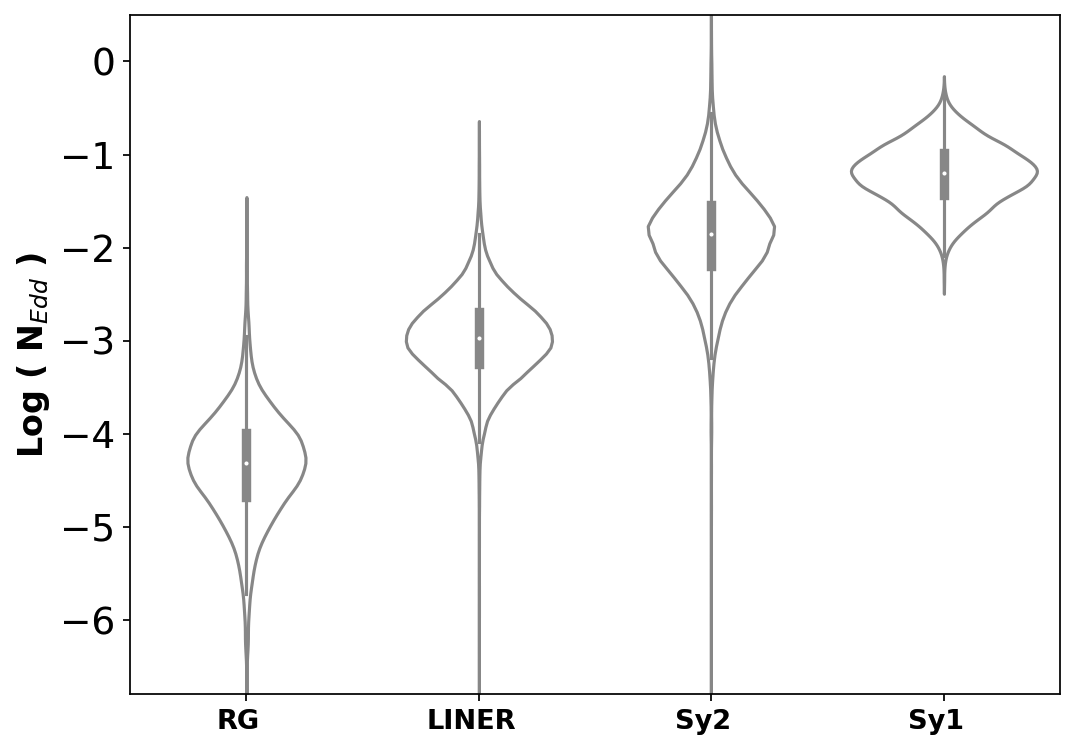}{0.4\textwidth}{(d)} }
\caption{Violin plots comparing for the different AGN types: (a) the AGN luminosity, (b) BH accretion rate, (c) BH mass, and (d) Eddington  ratio.\label{fig20:BH}
}
\end{figure*}

Considered as a whole, the AGN in our sample can be separated in two groups: group-E, formed by RG and LINER, where galaxies are massive, have an early morphological type (with massive bulge), and are old and evolved in terms of stellar population and star formation, and group-L, formed by Sy2 and Sy1, where galaxies are less massive, have later morphological types (spirals with massive disks), younger (at least the Sy2) and less evolved in terms of SFR. However, there is one important difference that transcends these groups, which is that LINER, Sy2 and Sy1 are mostly radio-quiet. In fact, LINER, which are most similar to RG, register the lowest detection frequency in radio (only 3\% in Table~\ref{RES_AVhost}, despite their large number). In terms of radio emission, they should thus be associated to group-L rather than group-E. In addition to the difference in radio emission, the most significant difference between LINER and RG is the stellar mass: according to the max-t test, Figure~\ref{figA1:maxt_host}a, the latter are almost three times more massive than the former. In models of galaxy formation, a difference in mass and stellar populations usually implies a difference in formation process, more massive galaxies forming their stars (exhausting their gas reservoir) more rapidly, form more massive bulges \citep{Sandage1986} and thus more massive SMBHs. In general, therefore, the ages of the stellar populations in Figure~\ref{figA1:maxt_host}b are consistent with a process of formation of galaxies varying with the mass: RG formed their stars more rapidly than LINER, and LINER formed their stars more rapidly than Sy2 and Sy1.

In Figure~\ref{fig20:BH}, we now trace the violin plots describing the characteristics of the SMBHs in AGN with different spectral classes, using the max-t test in Figure~\ref{figA2:maxt_BH} and~\ref{figA3:maxt_BH} to confirm the statistical significance of any difference observed. In Figure~\ref{fig20:BH}a Sy1 have the highest AGN luminosities, L$_{AGN}$. In Figure~\ref{figA2:maxt_BH}a, the max-t test confirms this specificity of the Sy1 class, confirming at the same time the higher luminosity of Sy2 compared to LINER and RG, and of LINER compared to RG. Following its definition, BHAR in Figure~\ref{fig20:BH}b naturally follows the AGN luminosity, increasing along the same sequence RG$\to$LINER$\to$Sy2$\to$Sy1. This is also confirmed statistically in Figure~\ref{figA2:maxt_BH}b. 

Comparing M$_{BH}$ in Figure~\ref{fig20:BH}c, we observe that the differences are almost the same as for the stellar masses in Figure~\ref{fig18:Host}c. This appears more clearly using the max-t tests, comparing Figure~\ref{figA3:maxt_BH}a with Figure~\ref{figA1:maxt_host}a. Note that, contrary to the transformation of L$_{AGN}$ into BHAR, there is no trivial relation between M$_*$ and M$_{BH}$ for the narrow-line AGN: practically, M$_*$ were estimated using an SSP synthesis code (STARLIGHT) while M$_{BH}$ were estimated using the velocity dispersion of the stars in the bulge, $\sigma$ (also produced by STARLIGHT) and applying an independent M$_{BH}$-$\sigma_*$ relation, based on reverberation observations. Consequently, there is a priori no simple relation expected between the stellar velocity dispersion of the stars in the bulge and the stellar mass of a galaxy, except through its morphology, that is, the importance of its bulge, since this is also one criterion entering in the definition of the different Hubble morphological types. 
\begin{figure*}[ht!]
\gridline{\fig{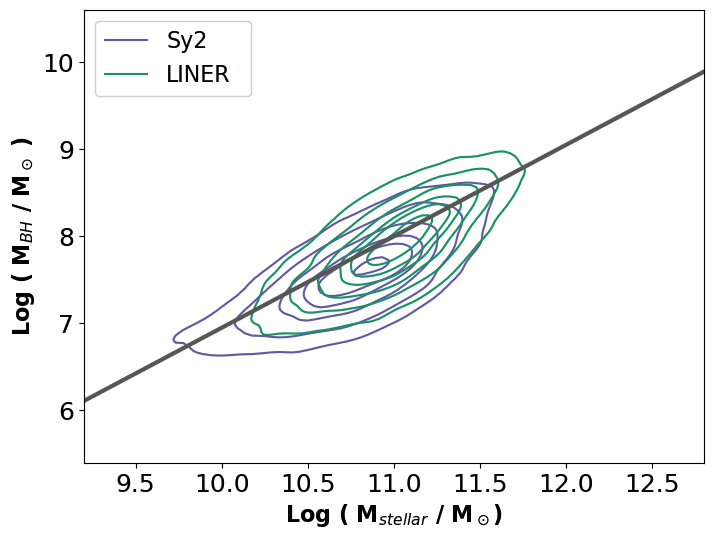}{0.4\textwidth}{(a)}
 \fig{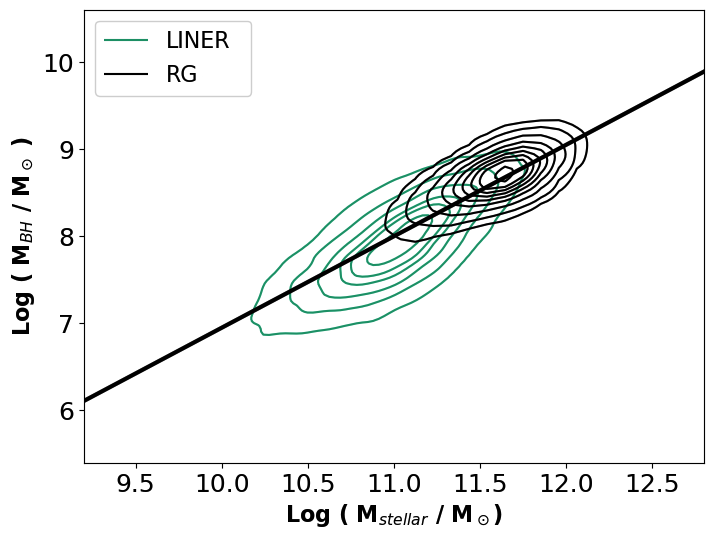}{0.4\textwidth}{(b)}
                }
\caption{Comparing the stellar masses with the SMBH masses in (a) LINER vs. Sy2, and (b) LINER vs. RG. The solid line is the relation established by  \citet{2015Reines}.\label{fig16:Massses}
}
\end{figure*}

In Figure~\ref{fig16:Massses}, we trace the relation between M$_{BH}$ and M$_{*}$ for the narrow-line AGN, comparing our data with the relation obtained by \citet{2015Reines} for local AGN, which we used to estimate the stellar masses of the Sy1 galaxy hosts based on the masses of their SMBHs. We find an excellent agreement for RG with the empirical relation, while our BH masses for Sy2 and LINER fall slightly below it. This concordance suggests that the formation processes of the galaxies and their SMBHs are tightly connected \citep[e.g.,][]{2017Yang}, consistent with the M$_{BH}-\sigma$ relation for AGN. Usually this relation is interpreted as evidence that the formation of a SMBH at the center of a galaxy determines the mass of the bulge of its host through AGN feedback \citep[e.g.,][]{2007Schawinski}. However, this relation could also be interpreted in reverse, the formation of the bulge of galaxies (through multiple mergers of proto-galaxies and fusion of BH seeds) is what primordially determines the masses of the galaxies and their SMBHs. 

\subsection{Characteristics of galaxies with OF in different AGN spectral classes}

In general, independent of their spectral class, the AGN galaxy hosts seem to show the same characteristics as normal (non-AGN) galaxies. Assuming feedback, however, we might expect something different. Continuing our comparison of AGN with and without OF, we first confirm, using the max-t test in Figure~\ref{figA4:maxt_OF1}a, that RG-AGN with OFs have significantly lower redshifts than their counterpart without OF. However, due to the high number of galaxies in our sample, the max-t test also allows to detect that Sy2 and LINER with OFs tend to be at slightly higher redshifts that Sy2 and LINER without OF, something that does not happen for the Sy1. The max-t tests similarly confirms that Sy2, LINER and RG with OFs have later morphological types, Figure~\ref{figA4:maxt_OF1}b, and higher SFR, Figure~\ref{figA4:maxt_OF1}c, than their counterparts without OF. Again, Sy1 show no such difference. This trait of AGN with OFs in the RG class explains the differences in redshifts in Figure~\ref{figA4:maxt_OF1}a, since late-type galaxies in this class only appear at low redshifts. This is the inverse in the LINER and Sy2 classes, where the number of late-type galaxies increases at high redshifts. Consistent with the later-type morphologies and higher SFR in AGN with OFs, in Figure~\ref{figA5:maxt_OF2}b t$_{old}$ is found to be significantly lower, because galaxies with active star formation tend to have on average younger stellar populations. 

\begin{figure*}[ht!]
\gridline{\fig{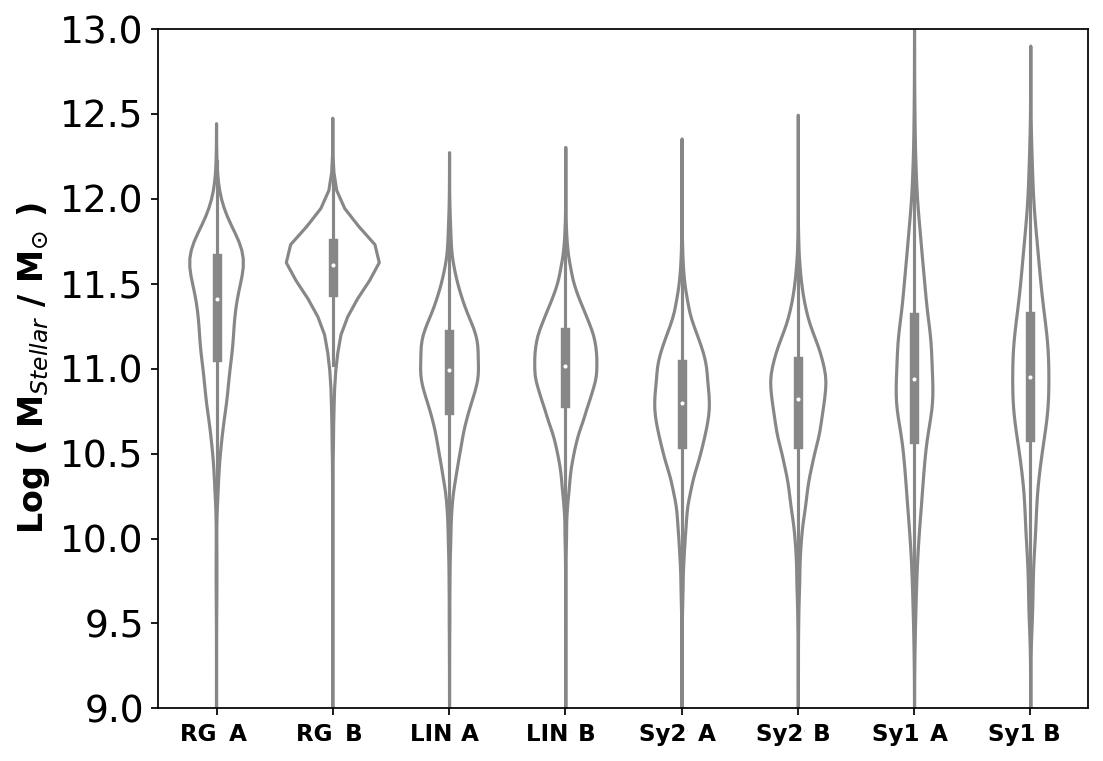}{0.4\textwidth}{(a)}
 \fig{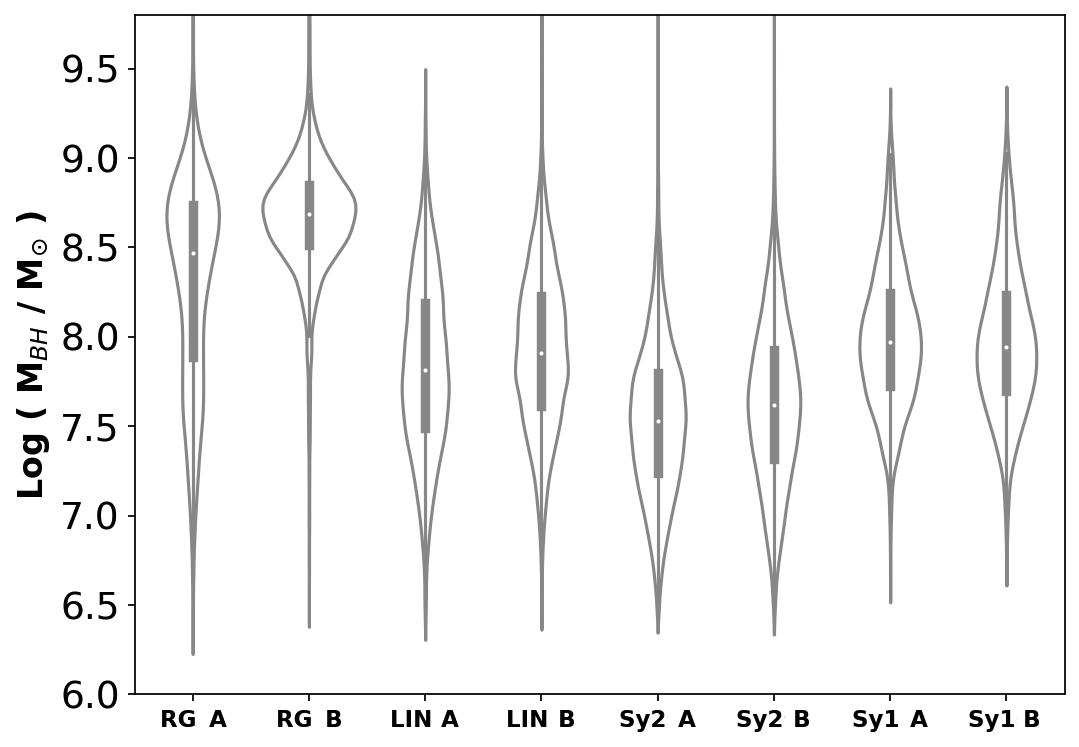}{0.4\textwidth}{(b)}
                }
\gridline{\fig{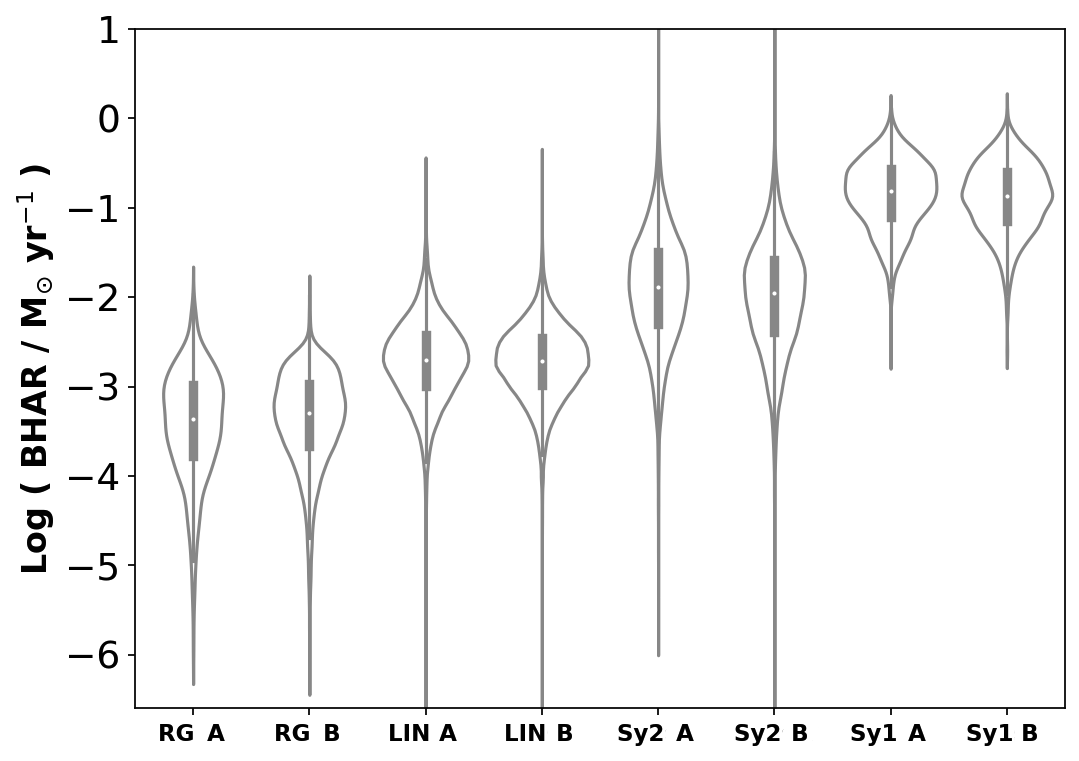}{0.4\textwidth}{(c)}
 \fig{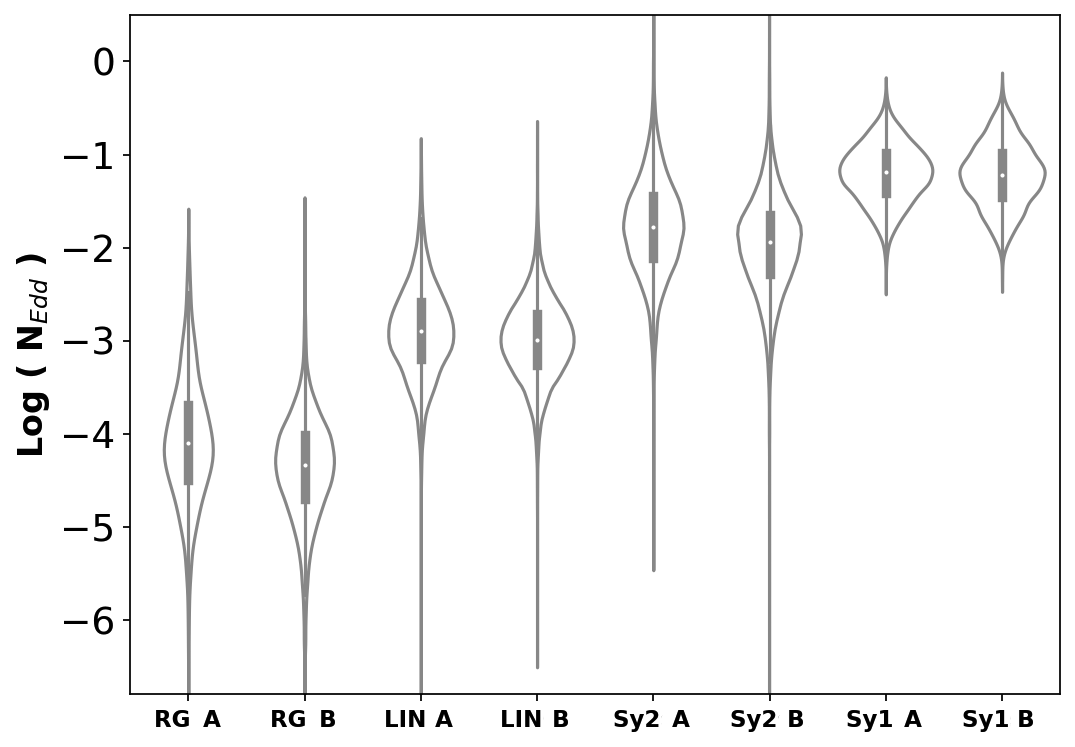}{0.4\textwidth}{(d)}
           } 
\caption{Violin plots comparing in AGN with OF (A) and without OF (B): (a) Stellar mass of host, (b) BH accretion rate, (c) BH mass and (d) Eddington ratios. \label{fig22:Mst_BHM_SFR_BHAR_Ledd}
}
\end{figure*}

In Figure~\ref{fig22:Mst_BHM_SFR_BHAR_Ledd}a we show the violin plots for the stellar masses. The only difference related to the presence of OFs is seen in the RG class: RG with OFs have lower stellar masses than those without OF. The same is true comparing the SMBH masses in Figure~\ref{fig22:Mst_BHM_SFR_BHAR_Ledd}b: the SMBH mass following the stellar masses, RG with OFs have less massive SMBHs than those without OF. 

In comparison, the differences for the other AGN spectral classes are very small and not always consistent. To summarize our analysis, we reproduced in Table~\ref{RES_maxt} the results of the max-t tests presented in Figure~\ref{figA4:maxt_OF1}, Figure~\ref{figA5:maxt_OF2} and Figure~\ref{figA6:maxt_OF_BH}, comparing galaxies with OFs ($A$) with galaxies without OF ($B$). In general, galaxies with OFs have later  morphological types (column~3), higher SFRs (column~4), younger stellar populations (column~6) and less massive SMBHs (column~7) than their counterpart without OF. The Sy1 are the only one showing almost no differences due to the presence of OF, except for the slightly higher SFR and W80. The max-t test also shows no difference (a value of 0) in terms of stellar mass for the Sy2, in terms of BHAR for the LINER and RG, and N$_{Edd}$ for the RG. In general, the results of the max-t test confirm that AGN with OFs, independently of their AGN class, are late-type spirals with disks that are particularly rich in gas, explaining why they have higher SFR.   

Finally, in Figure~\ref{figA7:maxt_W80}a the max-t test confirms that Sy1/RG have higher W80 than Sy2/LINER, also confirming that the intensity of AGN winds increases with the level of activity of the AGN, while in Figure~\ref{figA7:maxt_W80}b the max-t test confirms W80 is higher when the AGN are detected in radio, confirming that radio jets boost the energy of OFs. 
\begin{deluxetable*}{lcccccccc}
\tablecaption{Results of max-t tests \label{RES_maxt}}
\tablewidth{0pt}
\tablehead{
\colhead{AGN}  
& \colhead{$z$} 
& \colhead{T} 
& \colhead{SFR} 
& \colhead{M$_*$} 
& \colhead{t$_{old}$} 
& \colhead{M$_{BH}$} 
& \colhead{BHAR} 
& \colhead{N$_{Edd}$} 
\\
\colhead{(1)}  
& \colhead{(2)}
& \colhead{(3)}
& \colhead{(4)}
& \colhead{(5)}
& \colhead{(6)}
& \colhead{(7)}
& \colhead{(8)} 
& \colhead{(9)} 
}
\startdata
Sy2   & $A>B$ & $A>B$ & $A>B$ &   0   & $A<B$ & $A<B$ &$A>B$ & $A>B$  \\
Sy1   & 0     & 0     & $A>B$ &   0   &       & 0     & 0    & 0      \\   
LINER & $A>B$ & $A>B$ & $A>B$ & $A<B$ & $A<B$ & $A<B$ & 0    & $A>B$  \\ 
RG    & $A<B$ & $A>B$ & $A>B$ & $A<B$ & $A<B$ & $A<B$ & 0    & $A>B$  \\
\enddata
\end{deluxetable*}

\subsection{Multi-correlations analysis}

Due to the large number of galaxies in our sample, the max-t test is extremely sensitive, capable of detecting minute differences. Taking this point into consideration, the most obvious differences comparing AGN with and without OF are all related to the characteristics of their galaxy hosts: galaxies with OFs have later morphological types in Figure~\ref{figA4:maxt_OF1}b, higher SFR in Figure~\ref{figA4:maxt_OF1}c, lower stellar masses in RG and LINER (but not in Sy2 and Sy1), and thus smaller mass BH (including the Sy2) in Figure~\ref{figA6:maxt_OF_BH}a and younger stellar ages in Figure~\ref{figA5:maxt_OF2}b (except Sy1). This suggests that in each AGN class, galaxies with OFs are possibly at a younger phase of formation/evolution than galaxies without OF. Presented in this way, the idea that galaxies with OFs could transform with time into galaxies without OF in the same AGN class does seem possible and the idea that OFs could play an active role in this transformation over a longer interval of time (consistent with the delay hypothesis) looks like a plausible conjecture. 

Consequently, as an ultimate tentative to detect feedback effects on the host galaxies, we applied a multi-correlation analysis using R, to test in each AGN class how the characteristics of the hosts could be evolutionary connected with the characteristics of their SMBHs and whether the presence of an OF changes in some way the correlations. This test is first applied on the whole sample, then only on the respective sub-sample with OFs. The results of multi-correlations are shown in Table~\ref{Corr_Sy2} for Sy2, Table~\ref{Corr_LNR} for LINER,  Table~\ref{Corr_Sy1} for Sy1 and finally Table~\ref{Corr_RG} for RG. The correlations significance in terms of probability, $p$, are shown using a color code: black, $ p < 0.001$, blue, $ 0.01 > p \ge 0.001$, and red, $ 0.05 > p \ge 0.01$, while correlations with $ p > 0.05$ are not reported. The smaller the probability, the most significant the correlation: strong in black, intermediate in blue and weak in red. To simplify the analyses, we rearranged the correlation matrix in 3 sections: the first section occupied one column for the redshift, $z$, the second regroups the parameters of the hosts, star formation rate, SFR, stellar mass, M$_*$, morphology, T, and age of older stellar population, t$_{old}$, while the third regroups the parameters proper to the SMBH, mass of BH, M$_{BH}$, luminosity of AGN, L$_{AGN}$ and radio,  L$_{NVSS}$, and Eddington ratio, N$_{Edd}$. 
\begin{deluxetable*}{lr|rrrr|rrr}
\decimalcolnumbers
\tablecaption{Spearman correlation matrices for the Sy2\label{Corr_Sy2}}
\tablewidth{0pt}
\tablehead{\colhead{\textbf{Sy2}} &\colhead{z} 
&\colhead{SFR} &\colhead{M$_*$} &\colhead{T} &\colhead{t$_{old}$} 
&\colhead{M$_{BH}$} &\colhead{L$_{AGN}$} &\colhead{N$_{Edd}$}
} 
\startdata
SFR            & 0.23   & \nodata    &	\nodata  &	\nodata   &	\nodata	& \nodata &	\nodata	& \nodata\\
M$_*$          & 0.55	&{\red 0.05} & \nodata	& \nodata &\nodata & \nodata	& \nodata & \nodata \\
T              & 0.28	& 0.53	     &{\red $-0.06$}	& \nodata & \nodata & \nodata & \nodata & \nodata \\
t$_{old}$	   & $-0.39$ &$-0.60$	 &{\red 0.05} &$-0.54$	& \nodata & \nodata & \nodata & \nodata	\\ 
M$_{BH}$   & 0.38	&$-0.24$  & 0.67    &$-0.36$	& {\blue$0.17$} & \nodata & \nodata & \nodata \\	
L$_{AGN}$  & 0.87	&{\blue$0.18$} & 0.67 & {\blue$0.16$} & $-0.33$ &0.54	& \nodata & \nodata \\
N$_{Edd}$  & 0.69	&0.41	  & 0.23	& 0.48	& $-0.52$ & {\blue$-0.19$}	&0.35	& \nodata  \\
\hline 
Sy2 + OF       & z & SFR & M$_*$ & T & t$_{old}$ &	 M$_{BH}$ & L$_{AGN}$ & N$_{Edd}$  \\
\hline 
SFR            &{\blue 0.14}	& \nodata &	\nodata  &	\nodata   &	\nodata	& \nodata &	\nodata	& \nodata\\
M$_*$          &0.53	&{\red0.03	}&	\nodata  &	\nodata   &	\nodata	& \nodata &	\nodata	& \nodata\\
T              &0.26	&0.48	&{\red $-0.04$}	&	\nodata   &	\nodata	& \nodata &	\nodata	& \nodata\\
t$_{old}$	   &$-0.39$	&$-0.54$ & {\red 0.04}	&$-0.52$	& \nodata& \nodata&	\nodata & \nodata \\
M$_{BH}$   &0.40	&$-0.20$ & 0.64	& {$-0.32$}	& {\red 0.08}	& \nodata &	\nodata	& \nodata\\	
L$_{AGN}$  &0.87	&{\blue 0.14} & 0.65	& {\blue$0.16$} & $-0.36$	 &0.54	& \nodata &	\nodata	\\
N$_{Edd}$  &0.69	& 0.33	&	0.24 & 0.44	& $-0.49$ &{\blue$-0.17$}	&0.73	&	\nodata	\\
\hline
W80	           &0.26	&0.30	&0.27	&{\red 0.09}	&$-0.22$ &0.22	&0.34	&{\blue 0.12}	\\
\enddata			
\tablecomments {Color code: black, $ p < 0.001$, blue, $ 0.01 > p \ge 0.001$, red, $ 0.05 > p \ge 0.01$; correlations with $ p > 0.05$ are not reported.}
\end{deluxetable*}

We start our analysis with the two classes at low redshifts (Sy2 and LINER). In Table~\ref{Corr_Sy2} for the Sy2 class, col.~2 shows that all the parameters are strongly positively correlated with the z, except for t$_{old}$, the stellar populations getting younger as the galaxies get more massive and later in morphological types at high z. The strongest correlation is with L$_{AGN}$, which is trivial, since by definition the luminosity increases with z. What is more significant physically is that both the mass and SFR increase with z, a behavior consistent with downsizing. In col.~3 we can also see that SFR is more strongly correlated to T (positively) and t$_{old}$ (negatively) than to M$_*$. Also highly significant, in col.~4 we find that M$_*$ is strongly correlated with the BH parameters, M$_{BH}$, L$_{AGN}$ and N$_{Edd}$, which suggests that the formation process of the BH is tightly connected with the formation process of its host galaxy \citep{ 2015Reines,2017Yang}. This explains the strong anti-correlation between M$_{BH}$ and T in col.~5: as expected in the hierarchical structure formation theory (HSFT), black holes are less massive in late-type galaxies than in early-type galaxies. Also obvious, all the parameters related to the AGN (col.~7, 8 and 9) are correlated together. However, what is less obvious is the fact that N$_{Edd}$ is anti-correlated (at intermediate level) with M$_{BH}$, although  correlated with L$_{AGN}$. To understand what this means, we need to remember how the Eddington ratio is calculated. By definition, N$_{Edd} \propto$ L$_{AGN}/$M$_{BH}$, which means that since L$_{AGN}$ increases with M$_{BH}$, the anti-correlation implies the mass of the BH is bigger than what the luminosity would imply. This suggests that the growth of the SMBH was possibly more rapid in the past due to higher accretion rates. As should be well known now, the formation of a SMBH is a very fast process \citep[see][and reference therein]{2023CutivaAlvarez}.

Searching next for the effects of the observed OFs, in the second part of Table~\ref{Corr_Sy2} for the Sy2 with OFs we find almost exactly the same correlations than for their counterpart without OF. Taken at face value, this suggests that the presence of an OF has (or had) no impact in the formation/evolution processes of the galaxy hosts. Considering the intensity of the wind, W80 is found to increase at higher redshifts and with the SFR. However, since W80 is more tightly correlated to M$_*$ than T while SFR is more tightly correlated to T than M$_*$, and both W80 and SFR are strongly anti-correlated with t$_{old}$, we suspect that the strong correlation of W80 with SFR is indirect: W80 increases with the amount of gas (increasing with M$_*$), while SFR increases due to the usual mechanisms (like density waves, bars or stochastic self-propagation) favoring star formation in late-type spirals, thus increasing with T. One thing is very clear though, which is that the correlations with the BH parameters are congruent with the radiative mode: the wind intensity increases as the accretion rate (L$_{AGN}$) increases, in good agreement with the positive correlation with N$_{Edd}$. 
\begin{deluxetable*}{lr|rrrr|rrr}
\tablecaption{Spearman correlation matrices for LINER\label{Corr_LNR}}
\tablewidth{0pt}
\tablehead{ \colhead{\textbf{LINER}} &\colhead{z} &\colhead{SFR} &\colhead{M$_*$} &\colhead{T} &\colhead{t$_{old}$} &\colhead{M$_{BH}$} &\colhead{L$_{AGN}$} &\colhead{N$_{Edd}$} 
} 
\decimalcolnumbers
\startdata
SFR            &{\blue 0.18}	& \nodata        &	\nodata  &	\nodata     &\nodata& \nodata &	\nodata	& \nodata\\
M$_*$          &0.38	        &{\blue $-0.20$} &	\nodata  &	\nodata     &\nodata& \nodata &	\nodata	& \nodata\\
T              &{\blue 0.19}	& 0.60	         &{\blue$-0.22$}	     &	\nodata     &\nodata& \nodata &	\nodata	& \nodata\\
t$_{old}$	   &$-0.32$	        &$-0.58$	     &0.30	     &$-0.48$       & \nodata & \nodata & \nodata& \nodata \\
M$_{BH}$   &{\blue$0.22$}   &$-0.50$	     &0.71	     &$-0.53$       & 0.42 & \nodata &	\nodata	& \nodata\\
L$_{AGN}$  &0.83	        &{\red $-0.04$}  &0.57	     &{\red $-0.01$}& {\blue$-0.17$} & 0.47	&	\nodata	& \nodata\\
N$_{Edd}$  &0.58	        &0.46	         &{\blue$-0.16$}     &0.51	        & $-0.58 $ &$-0.55$	&0.48	  &	\nodata	\\
\hline 
LINER + OF & z & SFR & M$_*$ & T & t$_{old}$ &	 M$_{BH}$ & L$_{AGN}$ & N$_{Edd}$  \\
\hline 
SFR            &{\blue 0.13} & \nodata        &	\nodata  &	\nodata     &\nodata& \nodata &	\nodata	& \nodata\\
M$_*$          &0.38 	&$-0.26$	&	\nodata  &	\nodata     &\nodata& \nodata &	\nodata	& \nodata\\
T              &{\blue$0.21$} &0.60	&$-0.26$ &	\nodata     &\nodata& \nodata &	\nodata	& \nodata\\
t$_{old}$	   &$-0.37$ &$-0.60$	&{\blue 0.29} &$-0.54$& \nodata& \nodata & \nodata & \nodata \\
M$_{BH}$   &{\blue$0.22$}	&$-0.53$	&0.71	&$-0.54$ & 0.41 & \nodata & \nodata & \nodata\\	
L$_{AGN}$  &0.84	&{\red $-0.04$}	&0.54	&{\red$0.02$}	& {\blue$-0.23$}  &0.44	& \nodata & \nodata\\
N$_{Edd}$  &0.57	&0.46	&{\blue $-0.17$}&0.54	& $-0.61$ &$-0.64$& 0.39	 & \nodata\\
\hline
W80	& {\red$0.05$} &{\blue 0.20} & {\red 0.05} &{\blue 0.08} &{\blue $-0.22$}& \nodata &{\red 0.08} &{\red 0.06} \\
\enddata			
\tablecomments {Color code: black, $ p < 0.001$, blue, $ 0.01 > p \ge 0.001$, red, $ 0.05 > p \ge 0.01$; correlations with $ p > 0.05$ are not reported.}
\end{deluxetable*}

Examining in second the LINER class in Table~\ref{Corr_LNR}, we find the same correlations with the redshift as in the Sy2 class, except that the levels of the correlations tend to be slightly lower (intermediate) for SFR, T and M$_{BH}$. One important difference is that although SFR is still strongly correlated to T, it decreases as M$_*$ increases, which is consistent with the idea that more massive galaxies form their stars more rapidly than less massive galaxies, exhausting their gas reservoir more rapidly; SFR decreases due to the astration rates, not by quenching effects (these galaxies has no OF). Actually, the correlation of SFR with L$_{AGN}$ is weakening in LINER, continuing the trend already observed in the Sy2 class---the level of  correlation being positively intermediate in Sy2 is now negatively weak in LINER. The same trend applies for SFR vs. M$_*$, the correlation being weak in Sy2 and anti-correlated (intermediate) in LINER. On the other hand, M$_*$ once again is strongly correlated with M$_{BH}$ and L$_{AGN}$, and strongly anti-correlated with N$_{Edd}$. This implies that SMBHs grow with the mass of their galaxies and in LINER the accretion rates in the past were higher than in Sy2. That interpretation would also explain their more massive SMBHs. All these differences in correlations are consistent with a difference in formation process, LINER forming their stars more rapidly than Sy2, explaining why they look``more evolved''; they now have lower SFR and BHAR, and thus lower L$_{AGN}$. Consequently, M$_*$ is anti-correlated with T, showing a strong correlation with t$_{old}$, because more massive galaxies, forming more massive bulges, have an early-type morphology and relatively older stellar populations than less massive galaxies. This difference in formation would also explain the low detection rate of OFs in these galaxies: high astration rates quickly drain the amount of gas, decreasing the chance for massive OFs to form. 

Reviewing the correlations in the sub-sample of LINER with OFs, we find once more the same correlations than in their counterpart without OFs, although with different levels of correlations for some parameters. More specifically, SFR is now strongly anti-correlated with M$_*$, while M$_*$ is strongly anti-correlated with T and mildly correlated with t$_{old}$, all these differences being due to the later morphological types of LINER with OFs. No significant differences appear between the parameters for the AGN, which suggests that the presence of OF in general has no effect on the AGN activity. Therefore, since LINER in general are more evolved and have less gas than Sy2, the intensity of the wind is expected to be generally lower, whence BHAR being lower the correlations of W80 with the AGN parameters decrease. In fact, the strongest correlations for W80 are with the host parameters, in particular, increasing with SFR and decreasing with t$_{old}$, both consistent with the later morphological types of LINER with OFs.
\begin{deluxetable*}{lr|rrr|rrrr}
\tablecaption{Spearman correlation matrices for Sy1\label{Corr_Sy1}}
\tablewidth{0pt}
\tablehead{ \colhead{\textbf{Sy1}} &\colhead{z} &\colhead{SFR} &\colhead{M$_*$} &\colhead{T} &\colhead{M$_{BH}$} &\colhead{L$_{AGN}$} &\colhead{N$_{Edd}$ }
} 
\decimalcolnumbers
\startdata
SFR            & {\blue 0.09} &	\nodata & \nodata & \nodata & \nodata &	\nodata	& \nodata\\
M$_*$          & 0.37 & {\red $-0.04$}	& \nodata & \nodata & \nodata &	\nodata	& \nodata\\
T              & {\blue 0.14} &	\nodata & 0.22	&\nodata & \nodata & \nodata & \nodata\\
M$_{BH}$   & 0.53 & {\red $-0.03$}	& 0.70	& {\blue 0.09} & \nodata & \nodata & \nodata\\	
L$_{AGN}$  & 0.85 & {\blue 0.14}	& 0.47	& {\blue 0.18} & 0.66 &	\nodata	& \nodata\\
N$_{Edd}$  & 0.37 & {\blue 0.20}& $-0.31$& {\blue 0.10}	& $-0.44$ & 0.39 & \nodata\\
\hline 
Sy1 + OF        & z & SFR & M$_*$ & T &	 M$_{BH}$ & L$_{AGN}$ & N$_{Edd}$ \\
\hline 
SFR             & {\blue 0.06}	&	\nodata & \nodata & \nodata & \nodata &	\nodata	& \nodata\\
M$_*$           & 0.36 & {\red $-0.07$} & \nodata & \nodata & \nodata &	\nodata	& \nodata\\
T               &{\blue 0.13} &	\nodata & {\blue$0.21$}	&\nodata & \nodata & \nodata & \nodata\\
M$_{BH}$    & 0.53 & {\red 0.06} & 0.71	& {\red 0.09}& \nodata & \nodata & \nodata\\	
L$_{AGN}$   & 0.85 & {\blue 0.10} & 0.47	& {\blue$0.16$}	& 0.68	&	\nodata	& \nodata\\
N$_{Edd}$   & 0.38 & {\blue 0.19} & $-0.31$& {\red 0.08}& $-0.42$& 0.39& \nodata\\
\hline
W80	       & 0.28 & {\red 0.06} & {\blue 0.19}	& {\red 0.03} & 0.26 & 0.39	& {\blue 0.15}\\
\enddata			
\tablecomments {Color code: black, $ p < 0.001$, blue, $ 0.01 > p \ge 0.001$, red, $ 0.05 > p \ge 0.01$; correlations with $ p > 0.05$ are not reported.}
\end{deluxetable*}

Passing now to the two classes of AGN at high redshifts, we first inspect the correlations for the Sy1, which represents the most active phase of AGN activity with higher evidence of OFs. In col.~2 of Table~\ref{Corr_Sy1}, we find once more all the parameters positively correlated with the redshift (there is no information about t$_{old}$), except that the correlations with SFR and T are only at an intermediate level. These weaker correlations is possibly due to the fact noted before that the Sy1 class is a more homogeneous sample in terms of morphology that the Sy2 class. Remarkably we find in col.~3 that all the correlations with SFR are weak or intermediate, the strongest being SFR with L$_{AGN}$ and N$_{Edd}$. On the other hand, the correlations are still strong between M$_*$ and the BH parameters, emphasizing once more the close connection between the formation of SMBH and their galaxy hosts. There is also a strong positive correlation between M$_*$ and T implying that as the mass increases the galaxies become increasingly late in morphology. Contrary to Sy2 and LINER, the correlation between T and the BH parameters is weak. Actually, the most important correlations are between the BH parameters: a stronger correlation between M$_{BH}$ and L$_{AGN}$ than in Sy2 and similar correlation than Sy2 between L$_{AGN}$ and N$_{Edd}$. The anti-correlation between M$_{BH}$ and N$_{Edd}$ is also slightly weaker although comparable than in LINER, implying that like these latter AGN the growth rate in mass of the SMBHs were higher in the past. Since Sy1 have later morphological types than LINER their galaxy hosts must have formed in different conditions. Compared to Sy2, which have similar morphological types, Sy1 have higher stellar masses and accretion rates, two characteristics that are typical at higher redshifts. Once more here we might have evidence of downsizing: more massive Sy1 at high redshifts forming before less massive Sy2 at lower redshifts.   
\begin{deluxetable*}{lr|rrrr|rrrr}
\tablecaption{Spearman correlation matrices for RG\label{Corr_RG}}
\tablewidth{0pt}
\tablehead{ \colhead{\textbf{RG}} &\colhead{z} 
&\colhead{SFR} &\colhead{M$_*$} &\colhead{T} &\colhead{t$_{old}$} 
&\colhead{M$_{BH}$} &\colhead{L$_{AGN}$} &\colhead{N$_{Edd}$} &\colhead{L$_{NVSS}$}
} 
\decimalcolnumbers
\startdata
M$_*$          & 0.51         & \nodata & \nodata          & \nodata  &\nodata  & \nodata  &\nodata  & \nodata & \nodata \\
T              &{\blue$-0.21$} & \nodata & {\blue$-0.20$}   & \nodata  &\nodata  & \nodata  &\nodata  & \nodata & \nodata \\
t$_{old}$	   &{\red$0.13$} & \nodata & 0.41	 & $-0.37$ & \nodata & \nodata & \nodata  &	\nodata	&  \nodata          \\
M$_{BH}$   & 0.40         & \nodata & 0.60	 & {\blue$-0.24$} & 0.43 & \nodata  &\nodata  & \nodata & \nodata \\
L$_{AGN}$  & 0.77	      & \nodata & 0.38	 & {\blue$0.21$} & {\red$0.04$} & {\blue$0.26$} & \nodata & \nodata &\nodata    \\
N$_{Edd}$  & 0.48	      & \nodata & {\red$-0.04$}	 & 0.35	& {\blue$-0.25$} & $-0.40$ & 0.78  & \nodata& \nodata  \\
L$_{NVSS}$ & 0.69	      & \nodata & 0.50	 & {\red$0.08$}	& {\blue$0.17$}   & 0.44      & 0.60  & {\blue$0.30$} & \nodata \\
\hline 
RG + OF & z & SFR & M$_*$ & T & t$_{old}$	&	 M$_{BH}$ & L$_{AGN}$ & N$_{Edd}$  & L$_{NVSS}$ \\
\hline 
M$_*$          & 0.67 & \nodata &\nodata  & \nodata  & \nodata  & \nodata  & \nodata  & \nodata & \nodata \\
T              &{\blue$-0.19$} & {\red$0.04$} & $-0.45$ & \nodata &\nodata &\nodata &\nodata &\nodata & \nodata\\
t$_{old}$	   & 0.42 & \nodata & 0.62	 & $-0.66$ & \nodata & \nodata	& \nodata	& \nodata	&\nodata  \\
M$_{BH}$   & 0.64 & \nodata & 0.78  & $-0.52$ & 0.62 &\nodata& \nodata  & \nodata & \nodata\\	
L$_{AGN}$  & 0.66 & {\red$0.05$} & 0.41  & \nodata & {\red$0.10$} & 0.33	&\nodata &\nodata &\nodata    \\
N$_{Edd}$  & {\red$0.06$} & {\red$0.07$} & {\blue$-0.27$} & 0.43 & $-0.42$ & $-0.52$ & 0.63	&\nodata &\nodata  \\	
L$_{NVSS}$ & 0.76 & \nodata & 0.61	& {\blue$-0.21$} & 0.35 & 0.62	& 0.62	& {\red$0.05$}	&  \nodata \\
\hline
W80	& $-0.39$ & \nodata & $-0.32$ & {\blue$0.21$} & $-0.36$ & $-0.34$ &{\blue $-0.17$} &{\red $0.13$} &{\blue $-0.29$}\\
\enddata			
\tablecomments {Color code: black, $ p < 0.001$, blue, $ 0.01 > p \ge 0.001$, red, $ 0.05 > p \ge 0.01$; correlations with $ p > 0.05$ are not reported.}
\end{deluxetable*}

Because both Seyfert galaxies have the highest frequencies of OFs, we would not expect differences in comparing the correlations in the sub-sample of Sy1 with OFs. Like in the other AGN examined so far, the presence of OFs seem to have had no effect on the formation/evolution of the host galaxies. For the intensity of the wind, we find W80 to be as strongly correlated with the redshift as in the Sy2 class. However the correlations are weak with SFR and T and intermediate with M$_*$, while they are as strong with the BH parameters as in Sy2, which is consistent with the radiative mode and a higher level of activity, explaining the slightly higher number of OFs detected. Therefore, downsizing, not evolution from Sy1 to Sy2 explains why the former are less frequent than the latter at low redshifts. Being more massive Sy1 also have more massive SMBHs than Sy2, suggesting that the two Seyfert classes might correspond to different kinds of galaxies with different cosmological formation/evolution processes. 

The last class of AGN at high redshifts is the RG-AGN. In Table~\ref{Corr_RG} col.~2 shows that all the parameters, except T, increase with z. As is always the case, the strongest positive correlations are the trivial correlations with the luminosities: L$_{AGN}$ and L$_{NVSS}$. Like in the other AGN classes, M$_*$ increases with the redshift but now T decreases as t$_{old}$ increases---as the redshift increases, the galaxies get more massive and the stellar populations get older as the morphology get earlier in morphological type. However, surely the most important difference is the entire lack of correlations with SFR, these galaxies being characterized by an almost complete absence of star formation. The correlations between M$_*$, T, t$_{old}$ and M$_{BH}$ are the same as in LINER, but slightly less strong for L$_{AGN}$ and weaker for N$_{Edd}$. The correlation between M$_*$ and L$_{NVSS}$ is very strong. As for the other AGN classes the correlations are relatively strong between the BH parameters: stronger between L$_{AGN}$ and N$_{Edd}$ than between L$_{AGN}$ and M$_{BH}$, while N$_{Edd}$ is still anti-correlated with M$_{BH}$, almost similar than what we observe in the Sy1. The correlations are also strong between L$_{NVSS}$, L$_{AGN}$ and M$_{BH}$ and mildly correlated with N$_{Edd}$. Based on these last correlations, there should be no doubt about the nature of the source of radio emission in RG being AGN.

Except for the weak correlations between SFR, T, L$_{AGN}$ and N$_{Edd}$, almost the same correlations appear for the sub-sample of RG with OFs, with slightly higher levels of significance. Since RG have almost no star formation and the frequency of OF detection is very small, the lack of  differences in the correlations suggests one cannot propose OFs in the past of these galaxies were responsible in quenching the formation of stars in these host. In general, this suggests that the differences observed between the AGN classes are related to different formation/evolution processes of the galaxy hosts. 

Examining the intensity of the wind, mostly only anti-correlations appear in the RG class: W80 decreases with the redshift and with M$_*$, M$_{BH}$, L$_{AGN}$ and L$_{NVSS}$, while increasing as the morphology gets later in type and the age of stellar population decreases. These last anti-correlations in RG-AGN are consistent with our general conclusion that OFs only appear in late-type galaxies that are rich in gas. 



\section{Discussion: connecting the different AGN classes to the different formation processes of their host galaxies} \label{sec4:Disc}

Based on the previous analyses, there is no evidence in our samples of AGN feedback affecting the formation/evolution of their host galaxies. Considered as a whole, our results seem to support the idea that the different AGN spectral classes correspond to different formation processes of their hosts, due to different masses, and that OFs appear to be specific in each AGN class to a ``special phase'' in spiral galaxies which have still profusion of gas extended over kpc regions in their disks. But what would be this special phase is not clear.

Consistent with the M$_{BH}-\sigma$ relation, more massive galaxies form more massive bulges and SMBHs, exhausting their reservoir of gas rapidly. The astration process being less efficient in smaller mass galaxies \citep{Sandage1986}, they form smaller bulges, less massive BHs and more massive spiral disks, where OFs can developed in their NLR. This difference in formation process explains why RG have the most massive BHs, suggesting that higher accretion rates in their past might also be the reason why they become radio loud \citep[e.g.,][]{2017Coziol}. A similar scenario applying to LINER would explain the lower levels of AGN activity of these two classes of AGN: by forming more massive bulges, LINER and RG exhausted most or all of their reservoir of gas, leading their SMBHs to starvation.

About the OFs, our analyses suggest that they are primordially energized by radiation (majority of AGN), their intensity increasing with the accretion activity of the SMBH (AGN luminosity), and when a radio-structure form, are boosted by the higher energy injected by the jets. One remaining question which we were not able to answer is what effects could these AGN winds have on the evolution of the SMBHs and/or on their host galaxies? Having found no evidence for feedback effects on the characteristics of the SMBHs and their host galaxies, we only see three remaining possibilities: 1- the effect is immediate, prolonging the conditions favorable to star formation in the disks and recurrent accretion on their SMBHs (like a ``fountain effect''), or 2- quenching of SFR happens over a  very long period in a cumulative way, assuming the AGN activity with OFs is recurrent \citep{2023Harrison}, 3) although AGN winds might have been more important during the formation of the galaxies \citep{Silk1998}, with time they would have much less impact on well formed galaxies at low redshifts (an OF affecting only the gas) and what we interpret today as quenching could simply be the results of different formation processes. 
\begin{figure*}[ht!]
\gridline{\fig{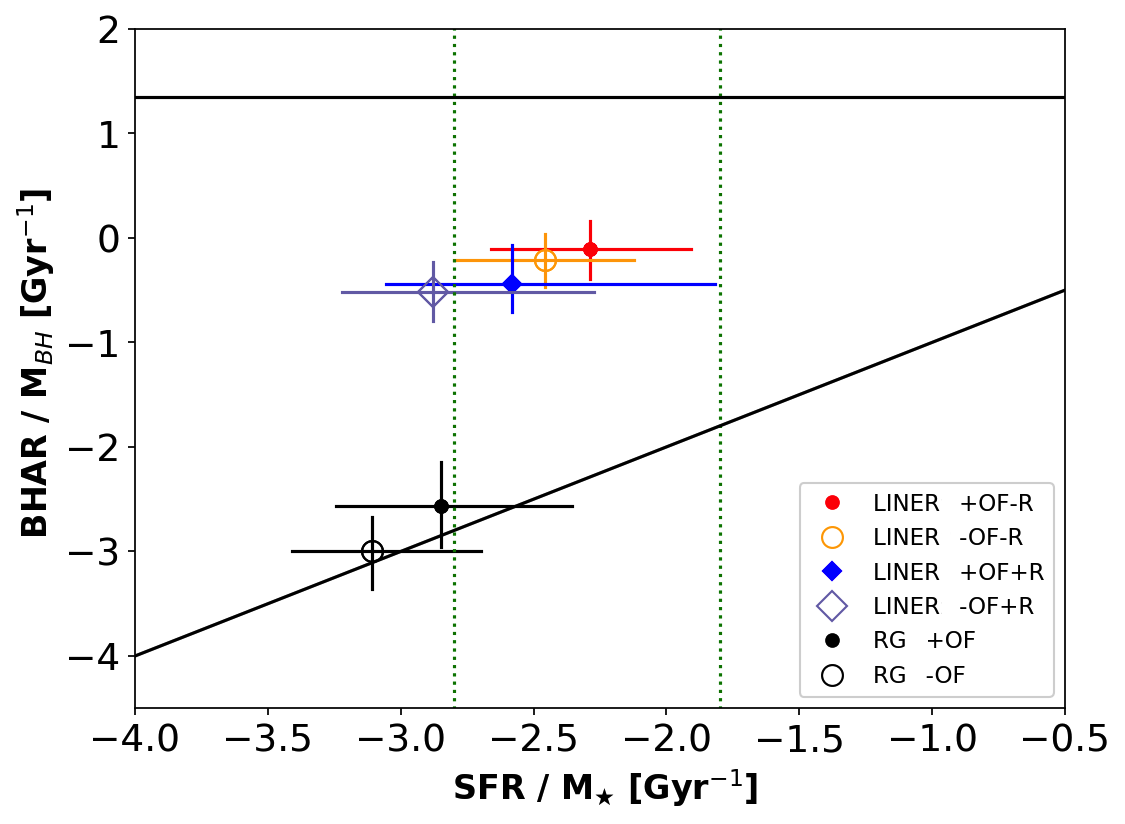}{0.4\textwidth}{(a)}
 \fig{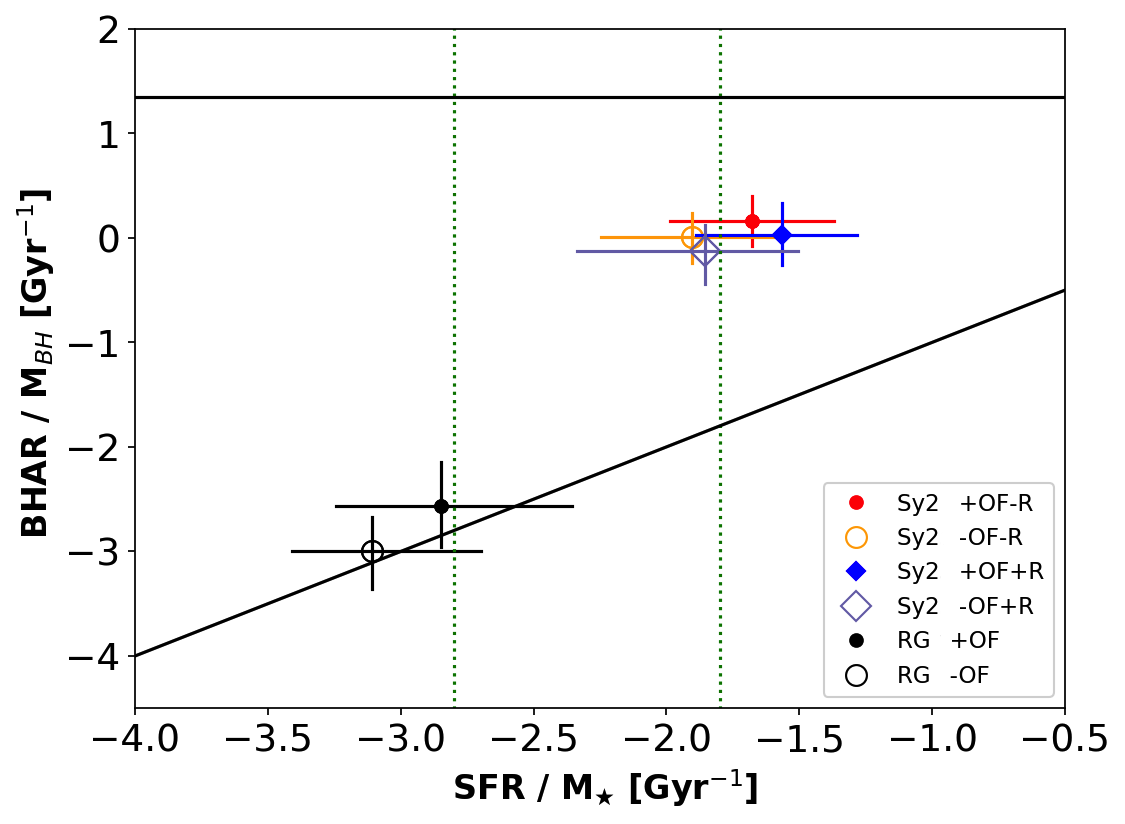}{0.4\textwidth}{(b)}
}
\gridline{\fig{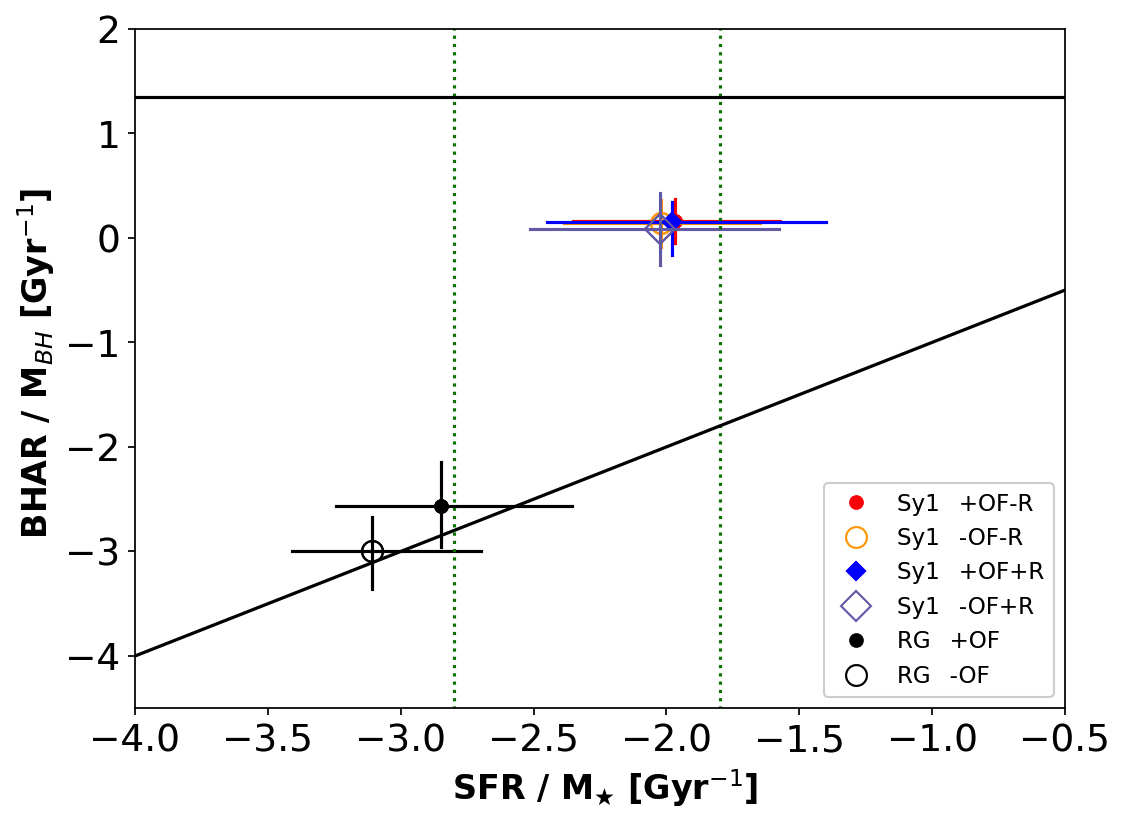}{0.4\textwidth}{(c)}}
\caption{Comparing the means and quartiles of sBHAR and sSFR of RG-AGN with (a) LINER, (b) Sy2 and (c) y1. All values are in logarithm exponents;  Different symbols distinguish radio quiet from radio loud, filled/open symbols are galaxies with OF/without OF. The three markers of galaxy evolution, horizontal, diagonal and vertical lines, are explained in the text. 
\label{fig23:sBHAR_sSFR}
}
\end{figure*}


To investigate which of the above possibilities is more probable, we trace in Figure~\ref{fig23:sBHAR_sSFR} the specific diagnostic diagrams for AGN with different spectral classes, which compare the growth rates of the SMBHs with the growth rates of their galaxy hosts, using the specific BH accretion rate, sBHAR = BHAR/M$_{BH}$, and specific star formation rate, sSFR = SFR/M$_{*}$ \citep[e.g.,][]{2021McDonald}. Over this diagram we have draw three markers that help us identify the present states of formation/evolution of the SMBHs and their galaxy hosts: 1) the horizontal bar at Log(sBHAR) $ = 1.35$ Gyr$^{-1}$ is the Eddington limit, which, in principle, is the highest accretion rate physically possible for AGN, 2) the diagonal is the one-to-one growth relation, which is the locus where the host galaxies and SMBHs grow in mass at equal rates, 3) the two vertical lines located at Log(sSFR)$ = -1.8$ and Log(sSFR)$ = -2.8$, which identify three important states in the star formation evolution of galaxies \citep{Bait2017}, the SFG state above -1.8 dex, where SFR is still high, the quenched state below -2.8 dex, where SFR is already very low, and the green valley in between, where SFR is going down.

Examining figure~\ref{fig23:sBHAR_sSFR} there are two general traits that are readily observable and that apply to the LINER, Sy2 and Sy1 classes. The first is that the galaxies are located well below the Eddington limit ratio. This is consistent with a mass growth for the SMBH limited by radiation pressure \citep[like for QSOs; see][]{2023CutivaAlvarez}. The other trait is that the AGN are located to the left of the one-to-one relation, implying that at the time of observation the masses of the galaxy hosts grow more slowly, by at least one or two orders, than the masses of their SMBHs. This is the standard interpretation. However, according to the results of our analyses, this could also be interpreted as evidence of a rapid formation of the galaxy hosts: the faster the formation process, the more massive the galaxy and its SMBH and the lower its gas reservoir, which lower the SFRs and BHARs, explaining the low sSFR and sBHAR of these galaxies at the time of observation. 

Locating these galaxies relative to the green valley, we found both LINER and Sy1 well within this zone, while Sy2 are in great number in the SGF zone. Note that this is also as expected based on the different formation processes of these galaxies: since LINER and Sy1 have formed more massive hosts with more massive SMBHs, their sSFR and sBHAR are naturally expected to be lower than in Sy2. Note that most of these AGN (except LINER without OF) have not reached a state consistent with the quenched state. Therefore, in agreement with our analyses there is no evidence of quenching related to OFs. In fact, AGN with OFs in LINER and Sy2 have generally higher sSFR and sBHAR than their counterparts without OF, while no difference appear in the Sy1. No difference in Sy1 can be distinguished either for those detected in radio, while Sy2 and LINER detected in radio have lower sBHAR, consistent with more massive SMBHs, but either higher sSFR in the former and lower in the latter. Note that since the two processes are not affected by OF quenching we do not expect sSFR and sBHAR to be synchronized.    

Only the RG are in the quenched zone also very close to the diagonal, suggesting that their SMBH now accrete at the same rate as they grow in mass. However, there is no contradiction with our interpretation based on different formation processes, since in RG both sSFR and sBHAR are extremely low as compared to the other AGN, which is easily explain by a rapid formation of their stars and SMBHs. When a galaxy has lost almost all its gas, therefore, the grow in mass of its galaxy grow as slowly as the grow in mass of its SMBH. In other words, in the absence of a gas reservoir, both processes are near extinction. Here, the presence of OF in a galaxy increases both the sSFR and sBHAR, suggesting, since they formed in the same way, they could be at a less evolve phase of evolution. But can we impute this evolution to OF quenching the SFR?
 
\begin{figure*}[ht!]
\gridline{\fig{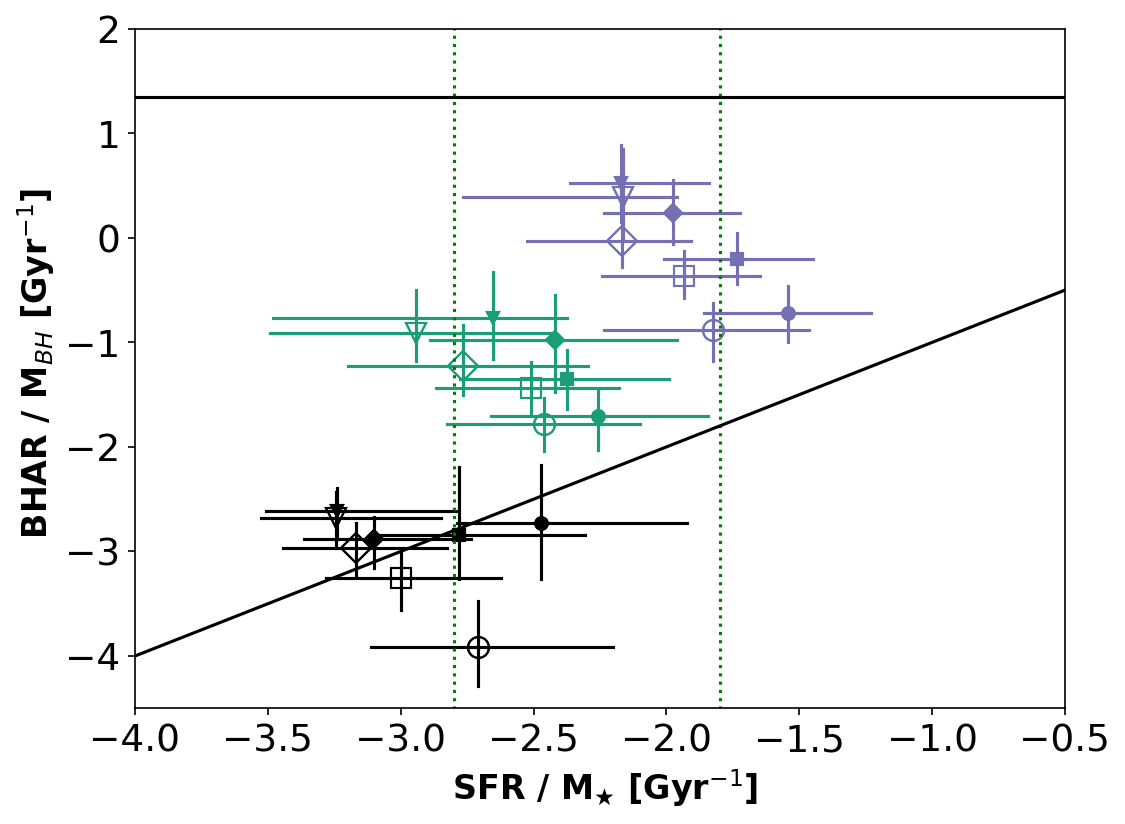}{0.4\textwidth}{(a)}
          \fig{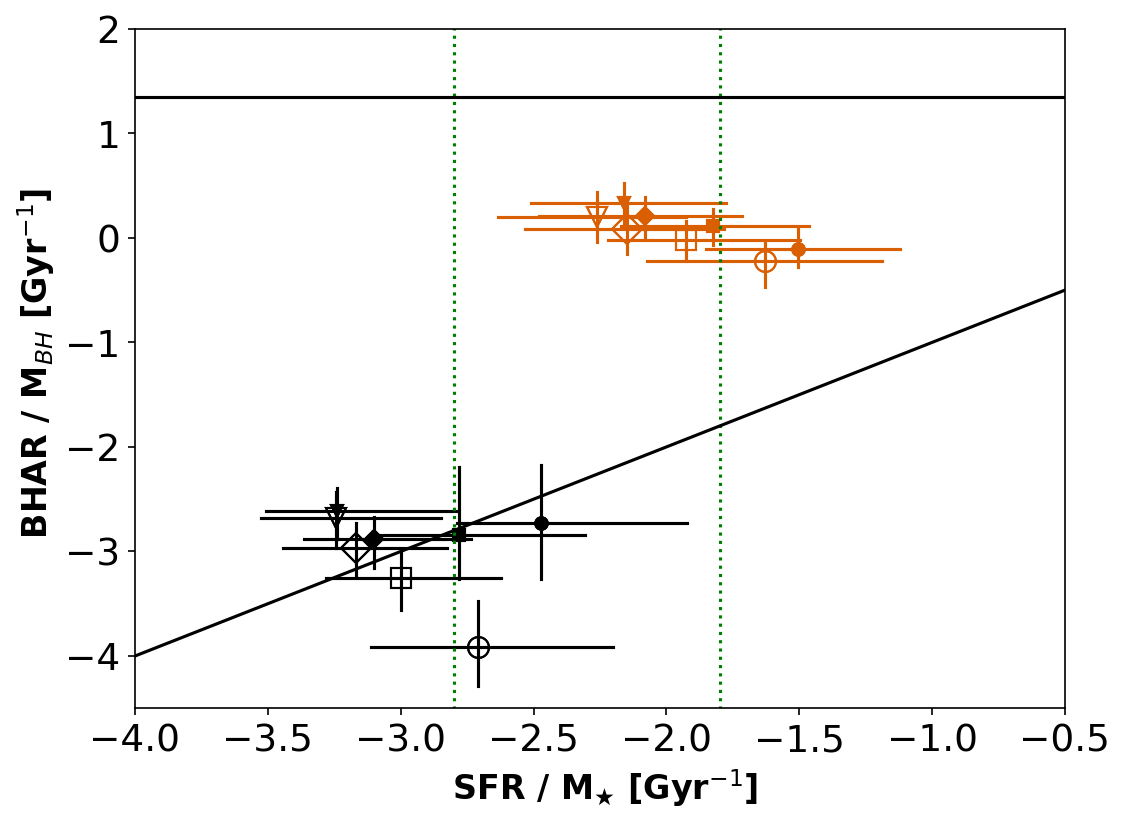}{0.4\textwidth}{(b)}
          }
\caption{Variation of sBHAR vs. sSFR in AGN with different classes (black, RG, green, LINER, blue Sy2, and red Sy1) at different redshifts (circle, redshift range (0,0.1], square (0.1,0.2], losange (0.2,0.3] and triangle, (0.3,0.4]; a filled symbol identify AGN with OF.   \label{fig24:OF_AGNvs4binz}
}
\end{figure*}
\begin{figure*}[ht!]
\gridline{\fig{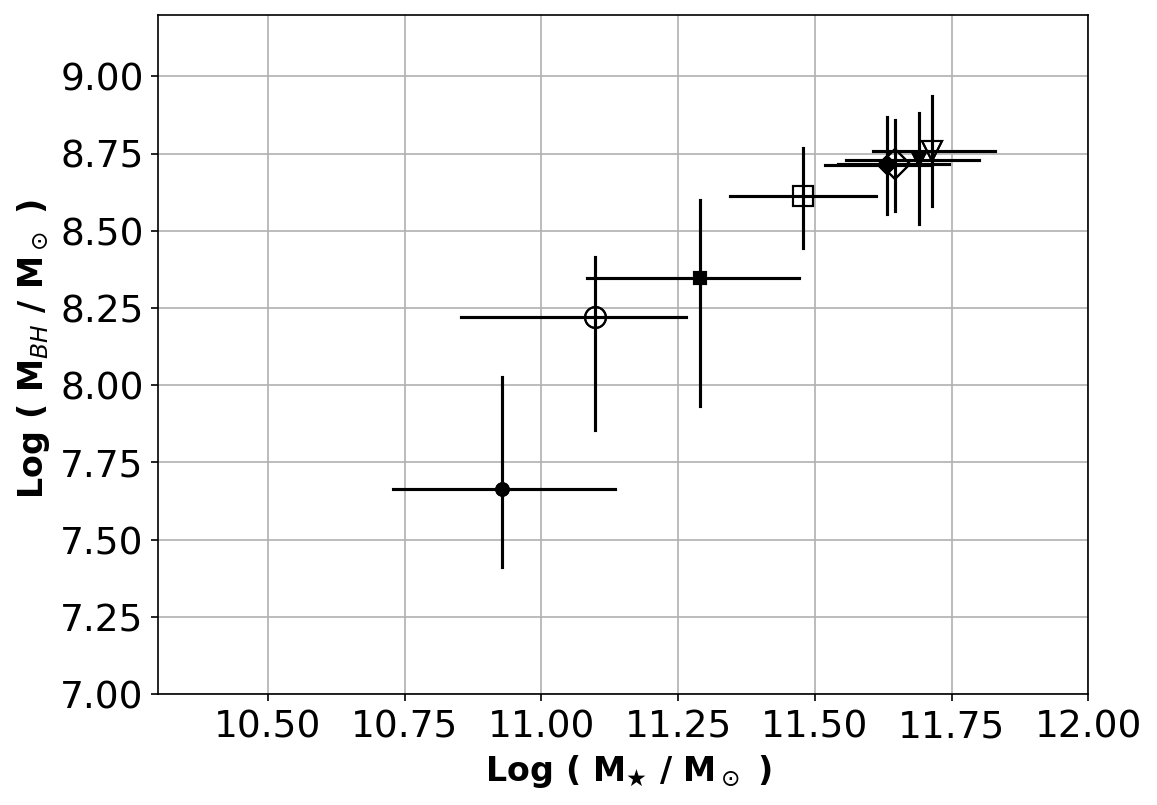}{0.4\textwidth}{(a)}
          \fig{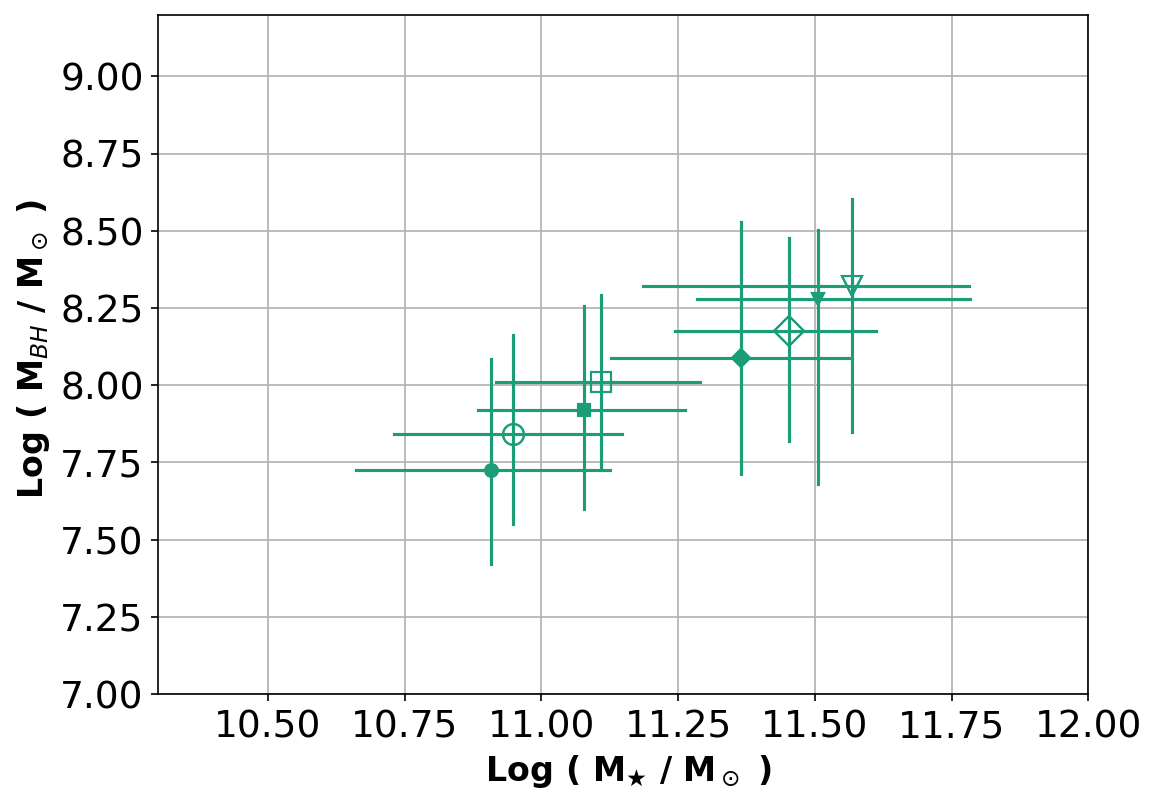}{0.4\textwidth}{(b)}
          }
 \gridline{\fig{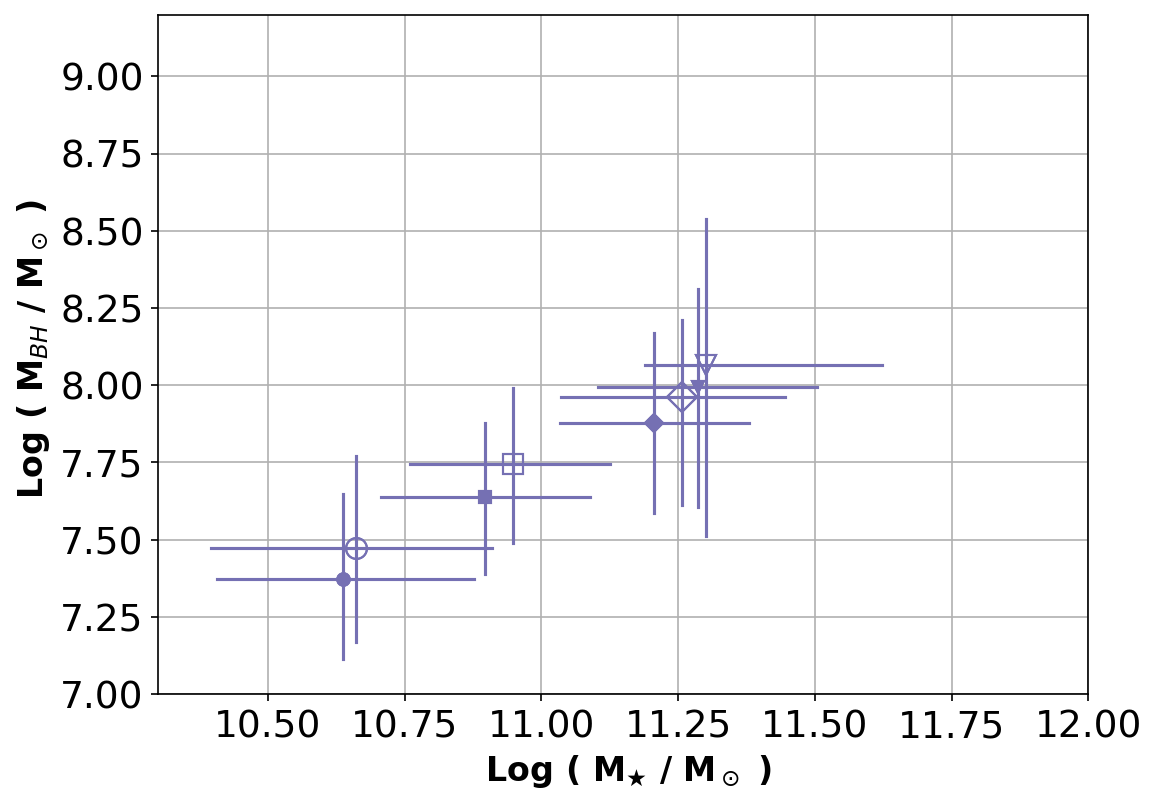}{0.4\textwidth}{(c)}
          \fig{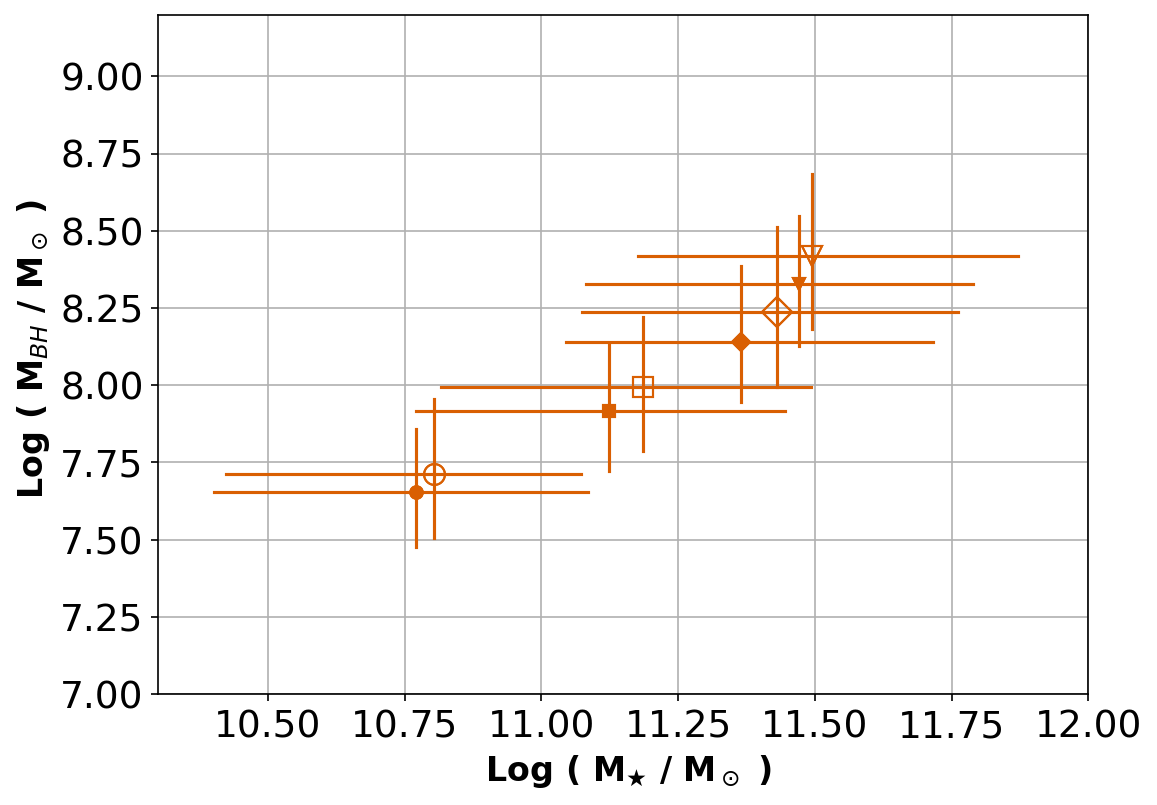}{0.4\textwidth}{(d)}
          }
\caption{Increases in mass of BHs and host stellar populations with the redshift. The colors and symbols have same meanings than in Figure~\ref{fig24:OF_AGNvs4binz}: circle, (0,0.1], square (0.1,0.2], losange, (0.2,0.3] and triangle, (0.3,0.4] \label{fig25:MMvs4binz}}
\end{figure*} 
If there is evolution, this should appear in relation to the redshift, which is the only parameter related to time. In Figure~\ref{fig24:OF_AGNvs4binz} we trace once again the specific diagnostic diagrams, separating each AGN class in four redshift bins. Since there is very few Sy1, Sy2 and LINER detected in radio, we only distinguish between AGN with OFs and without OF. As the redshift decreases, sBHAR decreases but sSFR increases. So, what we observe is a decrease with time in accretion rates (or increase in BH masses) but an increase of SFR (or decrease in stellar masses). Moreover, these changes happen on parallel paths for the Sy1, Sy2 and LINER. This is consistent with our observation in Table~\ref{tab:OF} that at each redshift the fraction of galaxies with OF is almost constant for the Sy1, Sy2 and LINER. For RG, one can observe that the gap between OF and non OF galaxies increases at lower redshifts, which is consistent with a cosmological change, RGs with OFs appearing in late-type spirals mostly at low redshifts.

In Figure~\ref{fig25:MMvs4binz} we show how M$_{BH}$ and M$_{*}$ vary at different redshifts. The variation is the same in all the AGN spectral classes: the mass of the BH decreases at the same time as the stellar mass. Moreover, in each bin of redshift, galaxies with OF have lower masses (BH and stellar) than their counterpart without OF. This suggests that in each AGN class, galaxies with OFs form the lower tails in mass of the galaxy mass distribution. 


Going back to Figure~\ref{fig24:OF_AGNvs4binz}, what we see therefore is that 
independent of their AGN class and whether or not an OF is present, galaxies at higher redshifts have higher accretion rates but lower SFRs. The evolution pattern is inverted, which is as we expect from downsizing: more massive galaxies form first at high redshifts, building more massive bulges and massive SMBHs, exhausting their gas reservoir more rapidly than galaxies forming at lower redshifts. Since this happens also for galaxies with OFs, this suggests their presence is not the cause of any change in the different processes of galaxy formation, but rather the products of these different processes. Consequently, the differences in OF detection in AGN can be explained by assuming the formation process of galaxies is skewed towards massive early-type galaxies in group-E and lower mass late-type galaxies in group-L, the excess of late-type spirals rich in gas in group-L favoring the formation of OFs. 
\begin{deluxetable*}{lrcccccccccc}
\tablecaption{Variation of ages and masses at different redshifts \label{tab:ages}}
\tablewidth{0pt}
\tablehead{
\colhead{ Sample} & \colhead{bin in z} & \multicolumn{7}{c}{Log ( $t_{old}$ / $yr$)} & $\langle Ltt \rangle$ & $\langle \Delta t_{old} \rangle$ & $\langle {\rm M}_{*} \rangle$\\
\cline{3-9}
\colhead{} & \colhead{} & \multicolumn{3}{c}{+OF} & & \multicolumn{3}{c}{-OF} & \colhead{} & \colhead{} & \colhead{}\\
\cline{3-5} \cline{7-9} 
\colhead{ } & \colhead{ } & \colhead{1Q} & \colhead{2Q} & \colhead{3Q} & & \colhead{1Q} & \colhead{2Q} & \colhead{3Q} & \colhead{ (Gyr)}& \colhead{ (Gyr)} &\colhead{ (M$_\odot$)} 
}
\startdata
RG        & (0.0-0.1] & 8.8 & 9.3 &  9.0 & & 9.0 & 9.5 &  9.8 & 0.7 & 0.3 & 11.08 \\
          & (0.1-0.2] & 9.2 & 9.5 &  9.7 & & 9.7 & 9.8 &  9.9 & 1.9 & 0.3 & 11.43 \\
          & (0.2-0.3] & 9.5 & 9.6 &  9.8 & & 9.7 & 9.9 &  9.9 & 3.0 & 0.2 & 11.61 \\
          & (0.3-0.4] & 9.6 & 9.8 &  9.9 & & 9.7 & 9.8 &  9.9 & 3.9 & 0.0 & 11.69 \\
          \hline
LINER     & [0.0-0.1] & 8.2 & 8.6 &  9.6 & & 8.7 & 9.4 &  9.8 & 0.7 & 2.1 & 10.91 \\
          & (0.1-0.2] & 8.5 & 9.1 &  9.7 & & 9.1 & 9.6 &  9.9 & 1.9 & 2.7 & 11.08 \\
          & (0.2-0.3] & 9.3 & 9.6 &  9.9 & & 9.4 & 9.7 &  9.9 & 3.0 & 1.0 & 11.37 \\
          & (0.3-0.4] & 9.5 & 9.8 & 10.0 & & 9.7 & 9.9 & 10.0 & 3.9 & 1.6 & 11.48 \\
          \hline
Sy2       & (0.0-0.1] & 8.1 & 8.5 &  9.0 & & 8.6 & 8.8 &  9.4 & 0.7 & 0.3 & 10.62 \\
          & (0.1-0.2] & 8.4 & 8.8 &  9.1 & & 8.8 & 9.2 &  9.4 & 1.9 & 1.0 & 10.91 \\
          & (0.2-0.3] & 8.9 & 9.2 &  9.4 & & 9.1 & 9.3 &  9.6 & 3.0 & 0.4 & 11.20 \\
          & (0.3-0.4] & 9.1 & 9.4 &  9.7 & & 9.4 & 9.6 &  9.9 & 3.9 & 1.5 & 11.31 \\
\enddata
\tablecomments{ $\langle Ltt \rangle$, average light travel time, $\langle \Delta t_{old} \rangle$, average difference of ages between galaxies with and without OF, $\langle {\rm M}_{*} \rangle$, average mass in each redshift bin. All values are in dex.}
\end{deluxetable*}

In Table~\ref{tab:ages} the ages of the oldest stellar populations as determined by STARLIGHT in each redshift bin are reported. One can see that the ages of the stellar populations are systematically younger in galaxies with OFs than their counterpart without OF. Consistent with their late-type morphologies, this suggests that these galaxies form their stars more slowly than their counterpart without OFs, which, in the hierarchical model of galaxy formation imply they form in slightly less dense environments. In Table~\ref{tab:ages} the age difference, $\Delta t_{old}$, between galaxies with OF and galaxies without OF, decreases at high redshift, following the increase in mass at each redshift, which is fully consistent with downsizing and the differences in formation processes this induces.    



Based on the above analyses, we must conclude that the hypothesis with the highest probability is \#3: although AGN winds are ubiquitous and might have been more important during the formation of the galaxies \citep{Silk1998}, with time they would have much less impact on well formed galaxies at low redshifts (an OF affecting only the gas) and what we interpret today as quenching could simply be the results of different formation processes in the past of the galaxies. As an alternative, hypothesis \#2 could be in part valid, but that would make OFs one among multiple processes that explain why spiral galaxies can form stars over longer interval of time that what their reservoirs of gas would suggest \citep[e.g.,][]{1983Kennicutt}. As for the evolution of AGN in general, a better scenario would be that, like QSO, they eventually evolve into normal spirals galaxies, with a more or less dormant SMBH at their centers \citep{2019Bland-Hawthorn,2022Akiyama}.

\section{Summary and conclusions} \label{sec5:Con}

After the discovery of the M$_{BH}$-$\sigma$ relation in AGN \citep{Magorrian1998}, it was proposed that AGN outlfows (OFs) could play a key role in the evolution of galaxies, by quenching the star formation in the bulge of their host galaxies \citet{Silk1998}. This is the AGN feedback hypothesis. However, finding observational evidence in support of this hypothesis turned out to be more difficult than previously thought \citep{2023Harrison}. To better understand this problem, we have extended our search for OF previously done in Sy1  \citep[][]{2020Torres-Papaqui}, to a larger sample of type~2 AGN also at low redshifts, $z < 0.4$. Our new sample is composed of 18,585 Sy2, 25,656 LINER and 15,793 RG with spectra in SDSS DR8. Applying STARLIGHT, we have determined the physical characteristics of their host galaxies to compare them with the properties of their SMBHs looking for differences that could be interpreted as evidence of OF feedback effects. The most important results and conclusions of the present study are the following:

\begin{enumerate}

\item In our sample, Sy2 and LINER are systematically at lower redshifts than Sy1 and RG-AGN. For the emission line galaxies, this bias could be explain assuming the Sy1 class is typical of field galaxies while the Sy2 and LINER classes are typical of clusters and groups: as the volume surveyed increases with the redshift, the probability of finding a Sy1 surpasses the probability of finding Sy2/LINER, unless they are also detected in radio. The same difference in environment would also explain why no RG-AGN at high redshifts is classified as Sy1. 

\item In general, the galaxy host of AGN tend to be more massive at high redshift and to be slightly earlier in morphological type. This could be evidence of downsizing, that is, assuming more massive galaxies form first at higher redshifts.  

\item The probability of detecting a resolved OF increases with the AGN luminosity, decreasing along the sequence Sy1$\to$Sy2$\to$LINER/RG. Both LINER and RG-AGN have low OF frequency detection. At low redshifts LINER are more massive with more massive BH and have lower SFR and older stellar populations than Sy2. At high z, the same differences distinguish RG-AGN from Sy1, explaining the different frequencies of OF detection. 

\item Independently of the AGN spectral class or redshifts, galaxies with OFs have later morphological types, are less massive and have higher SFR than their counterpart without OF. 

From this observation we conclude that since late-type galaxies naturally have more gas in their disks than early-type galaxies, favoring star formation, the reason for observing resolved OFs in AGN at low redshifts ($z \le 0.4$) is because they have huge amount of gas extended over kpc (the NLR) in their host disks.

\item Based on W80, the intensity of OF increases with L$_{AGN}$, suggesting they are generally driven by radiation. However, in RG-AGN and other AGN spectral classes where radio emission is detected, W80 is systematically higher than in those without radio detection. Moreover, RG systematically show an excess of ionization for their continuum luminosity. This suggests that radio structures injected an excess of energy in the ISM of their host galaxies that could produce ionization or trigger OFs.  

Actually, in a few AGN, double OFs can be detected, their frequency significantly increasing when radio emission is present. From these observations we conclude that in any AGN, the two modes, radiative and jet can be active at the same time: the radiative mode, possibly primordial since only a few AGN emit in radio, dominating in emission-line galaxies, radio jets being too weak and compact to be detected, and the jet mode, when the conditions are favorable, becoming dominant in RG.

\item Comparing the characteristics of the galaxy hosts and their SMBHs in each AGN spectral classes, there are no detectable difference between galaxies with and without OFs. Similarly comparing the correlations between the hosts parameters with SMBH parameters, no difference is noted when OFs are present. This suggests that OFs had no effect in the formation process of their galaxy hosts.

\item In the specific diagnostic diagram, sBHAR vs. sSFR, separating our samples in four different redshift bins, we observe that at each redshift galaxies with OF, independently of the AGN class, have higher sSFR and sBHAR than their counterpart without OF, while galaxies detected in radio have systematically lower sBHAR than their counterpart without radio. The first observation supports the idea that galaxies with OFs represent different states of galaxy formation, where BHAR and SFR are high, while the second suggests the emission in radio follows an increase in mass of the SBHMs, producing lower sBHAR. 

\item Tracing the variation of M$_{BH}$ and M$_{*}$ in the different redshift bins in each AGN class, we found that in each bin of redshift, galaxies with OF have lower masses than their counterpart without OF. 

\end{enumerate}

In the case of RG, where the spectral type changes with the redshift, downsizing could also explain why the RG-AGN at high redshifts are more massive and evolved, with older stellar populations and lower SFR than the RG-LINER/Sy2 at intermediate redshifts and RG-SFG/TO at low redshifts. A difference in formation process would also explain the higher frequency of OF in RG-SFG/TO than in RG-AGN. 

The fact sBHAR increases at high redshift while sSFR decreases is also consistent with downsizing: more massive galaxies having form at earlier time than less massive galaxies, they have more massive bulges and SMBHs, and having exhaust their reservoir of gas more rapidly, having now lower sSFR and sBHAR.

In general, a difference in formation of the galaxy hosts would also explain why OFs in RG and LINER are not common: forming more massive bulge and exhausting their reservoir of gas more rapidly than the Seyfert galaxies, they lack the extended spiral disks necessary to form them. In fact, the most significant difference between LINER and RG seems to be their masses, RG being three times more massive, and this difference might also be the physical reason why RG are radio loud while LINER are radio quiet: a higher accretion of gas in their BH in the past triggered radio emission. 

Finally, because of the different formation processes of galaxies, OF feedback effects might also be expected to vary. Here is a possible scenario. After a rapid grow in mass parallel to the grow in mass of their hosts (at the origin of the M$_{BH}-\sigma_{*}$ relation), the SMBH continues to grow in mass by accretion but at slower pace and only as long as there is still gas funneling in its zone of influence. Since how much gas remains in a galaxy depends on how fast this galaxy form its stars, OFs might have played a more important role during the formation of galaxies (at high z), in particular in early-type galaxies, which today would not have enough gas to produce observable OFs. For spiral galaxies, which form their stars more slowly, the gas forming massive disks, the formation of OFs might be favored. However, since the formation of stars is spread over a long period of time, OF feedback might  not be expected to leave as significant marks on a disk where star formation is continuous or recurrent, explaining why it is so difficult to detect their effects. 


    

In \citet{2023Harrison} the authors summarize what is the state of our present knowledge about the role AGN outflows are expected to play in the formation/evolution of galaxies, insisting on three aspects: 1) according to multiple simulations of galaxy formation, AGN winds seem to be essential in order to explain the properties of galaxies, 2) as many observations suggest, AGN winds (radiative or radio jet) have sufficient energy to perturb the distribution of molecular gas in a galaxy, driving it away or exciting it, which could affect (quenching or triggering) their normal process of star formation, and 3) the problem with assuming a major quenching effect is that AGN with OFs at low redshifts are usually observed in galaxies which are rich in gas and where star formation is still very high. The third aspect fully agrees with our results. However, to this aspect we must now add one new element, which is that these AGN are preferentially late-type spirals. This supplementary facts points to a direct link with galaxy formation, where, contrary to the AGN feedback hypothesis, the quenching of SFR in AGN is the consequence of the variation in astration rates during the formation of their host galaxies and where the formation of OFs and their effects on galaxies depends on how much gas this astration process leaves in their disks. This implies that the problem of understanding what effect AGN OFs have on galaxies should first pass by defining better how these galaxies form. 
As the recent discoveries of massive galaxies forming and merging frenetically at very early epochs suggests \citep[e.g.,][]{2022Naidu,2023Labbe,2023Larson,2023Agazie},\citep[e.g.,][]{2022Naidu,2023Labbe,2023Larson,2023Agazie}, the process by which SMBHs become intertwined with their host galaxies could be more complex than previously thought and the AGN feedback hypothesis might not be the ultimate solution it was supposed to be \citep[e.g.][]{2024Silk}. 


\section{Acknowledgments}
\begin{acknowledgments}
The authors would like to thank Dr. Heinz Andernach for his diligent search in NVSS and FIRST for extra radio sources in our samples of LINER, Sy2 and Sy1, as reported in Table~\ref{RES_AVhost}. An anonymous reviewer is also acknowledge for the multiple comments and suggestions that helped us improve our study and the final version of our article. Finally, J.P.T.-P. and R.C. acknowledge DAIP-UGto (Mexico) for granted support (0077/2021). 
\end{acknowledgments}

\section{Software and third party data repository citations} \label{sec:cite}

This research has made use of the VizieR catalogue access tool, CDS, Strasbourg, France (DOI : 10.26093/cds/vizier), which original description was published in 2000, A\&AS 143, 23. This research have also made use of the cross-match service provided by CDS, Strasbourg. Funding for SDSS-III has been provided by the Alfred P. Sloan Foundation, the Participating Institutions, the National Science Foundation, and the U.S. Department of Energy Office of Science. The SDSS-III web site is http://www.sdss3.org/. SDSS-III is managed by the Astrophysical Research Consortium for the Participating Institutions of the SDSS-III Collaboration including the University of Arizona, the Brazilian Participation Group, Brookhaven National Laboratory, Carnegie Mellon University, University of Florida, the French Participation Group, the German Participation Group, Harvard University, the Instituto de Astrofisica de Canarias, the Michigan State/Notre Dame/JINA Participation Group, Johns Hopkins University, Lawrence Berkeley National Laboratory, Max Planck Institute for Astrophysics, Max Planck Institute for Extraterrestrial Physics, New Mexico State University, New York University, Ohio State University, Pennsylvania State University, University of Portsmouth, Princeton University, the Spanish Participation Group, University of Tokyo, University of Utah, Vanderbilt University, University of Virginia, University of Washington, and Yale University.

%






\appendix
\restartappendixnumbering
\section{Results of max-t tests}

\begin{figure*}[ht!]
 \gridline{\fig{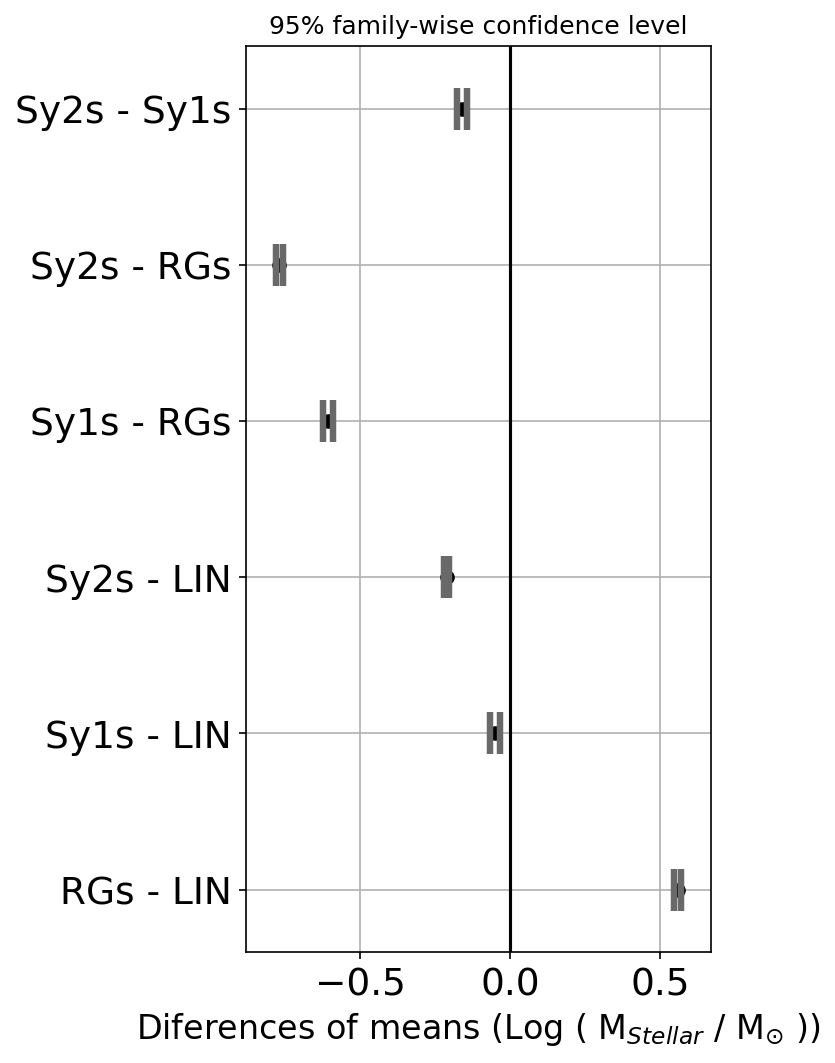}{0.4\textwidth}{(a)}
                \fig{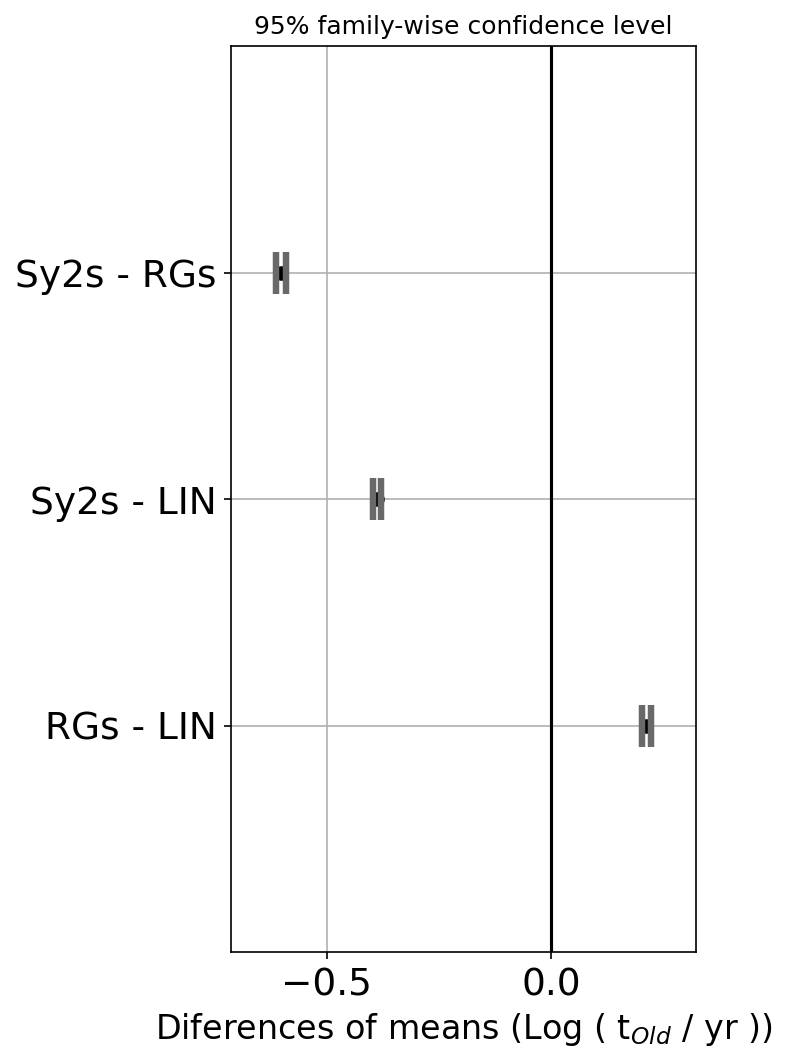}{0.36\textwidth}{(b)}
             }
\caption{Results of max-t tests comparing in 
(a) the Stellar mass of the host (M$_{stellar}$), and in (b) the ages of oldest stellar population ($t_{old}$).  
\label{figA1:maxt_host}
}
\end{figure*}

\begin{figure*}[ht!]
\gridline{\fig{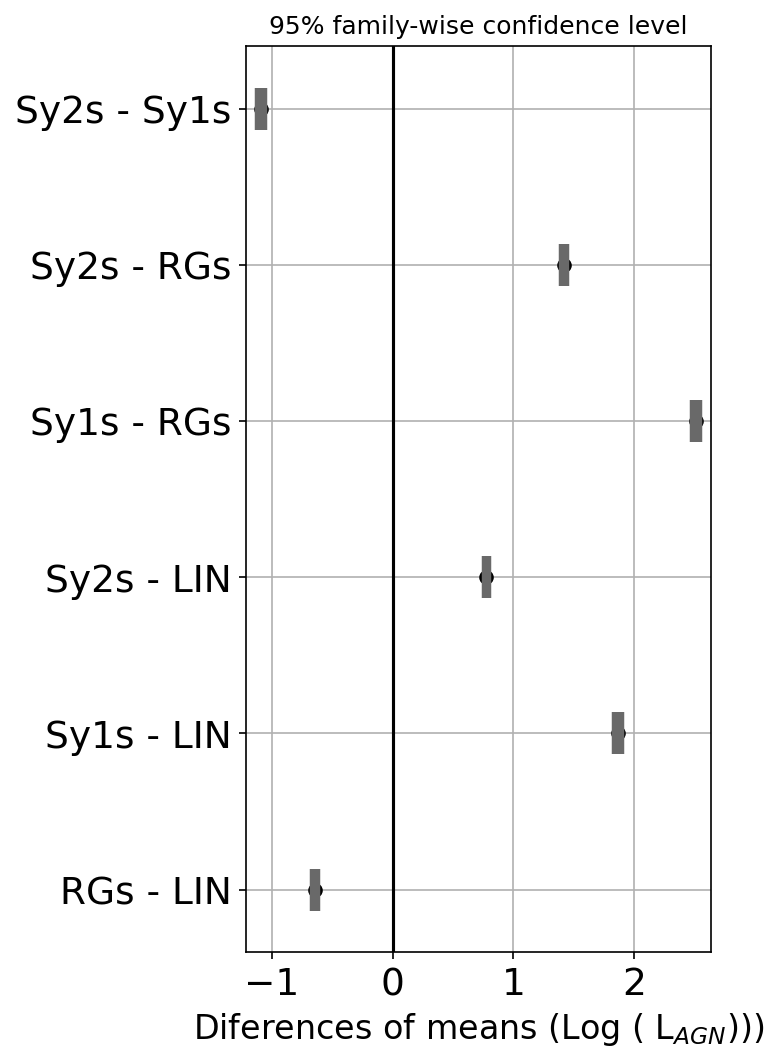}{0.35\textwidth}{(a)}
           \fig{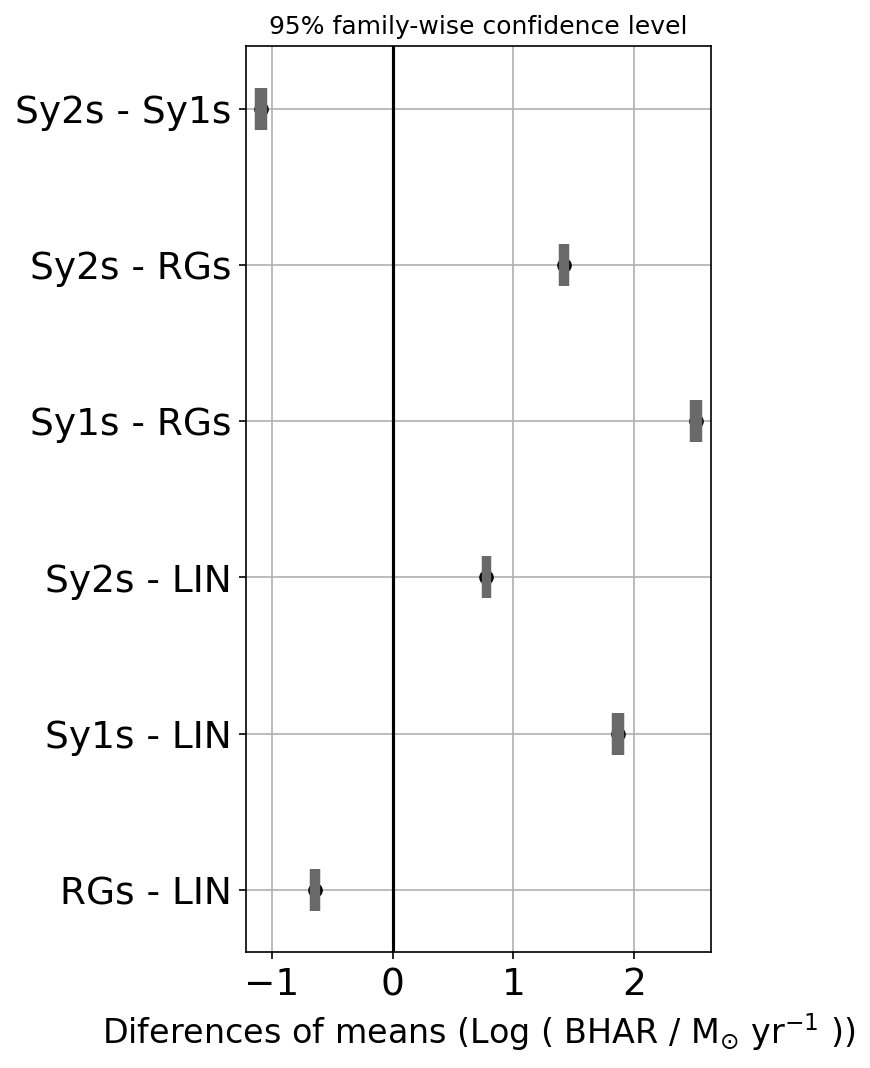}{0.39\textwidth}{(b)}
           }
\caption{Results of max-t tests comparing in (a) AGN luminosities, L$_{AGN} = \lambda {\rm L}(5100)$ and b) BH accretion rates (BHAR).
\label{figA2:maxt_BH}
}
\end{figure*}

\begin{figure*}[ht!]
\gridline{             
             \fig{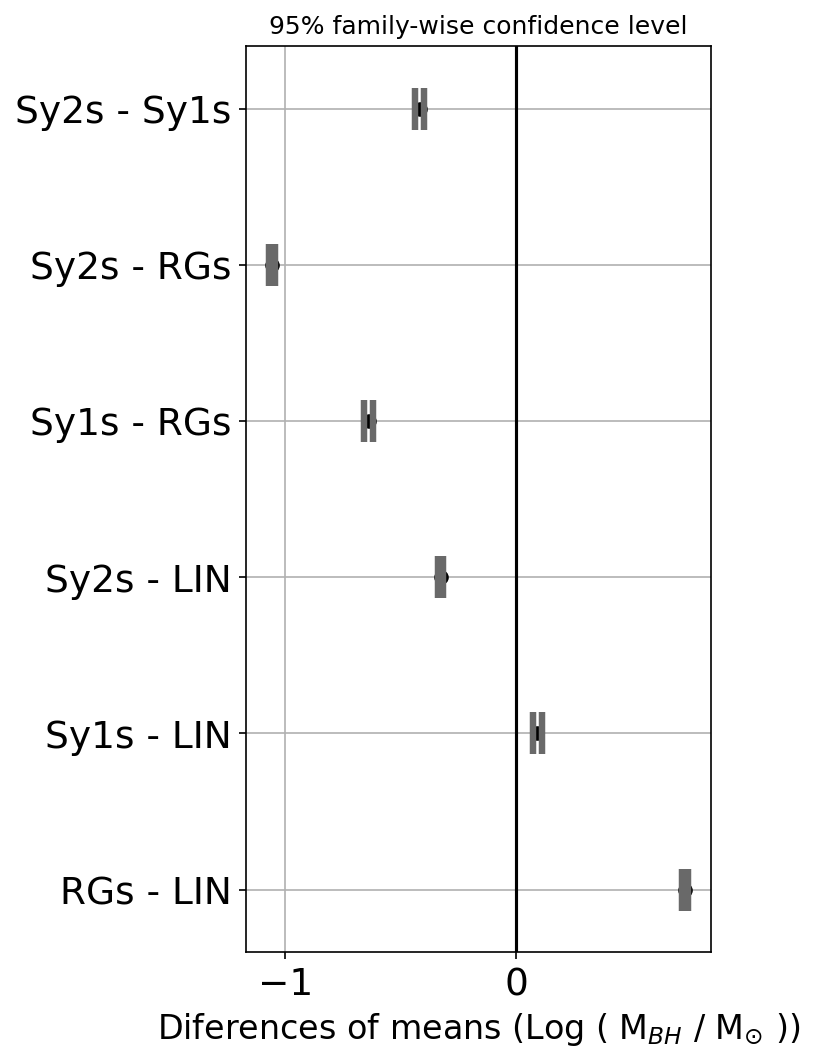}{0.35\textwidth}{(a)}
             \fig{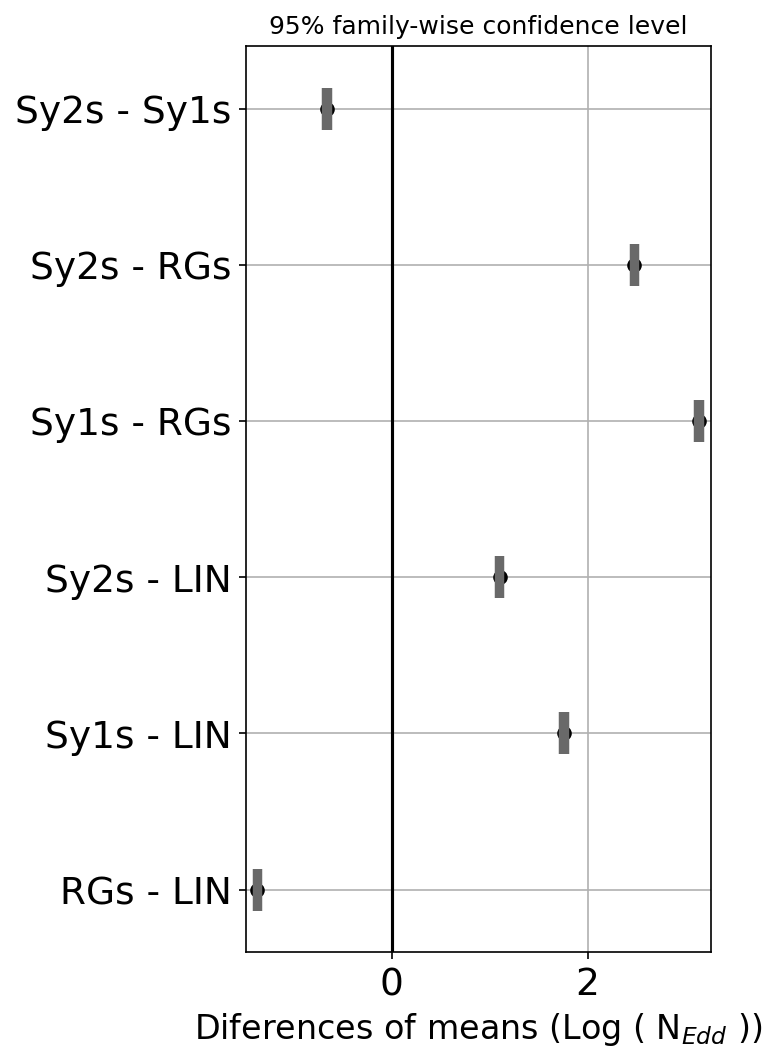}{0.35\textwidth}{(b)}
             }
\caption{Results of max-t tests comparing in (a) BH masses (M$_{BH}$) and (b) Eddington ratios (N$_{Edd}$)
\label{figA3:maxt_BH}
}
\end{figure*}

\begin{figure*}[ht!]
\gridline{\fig{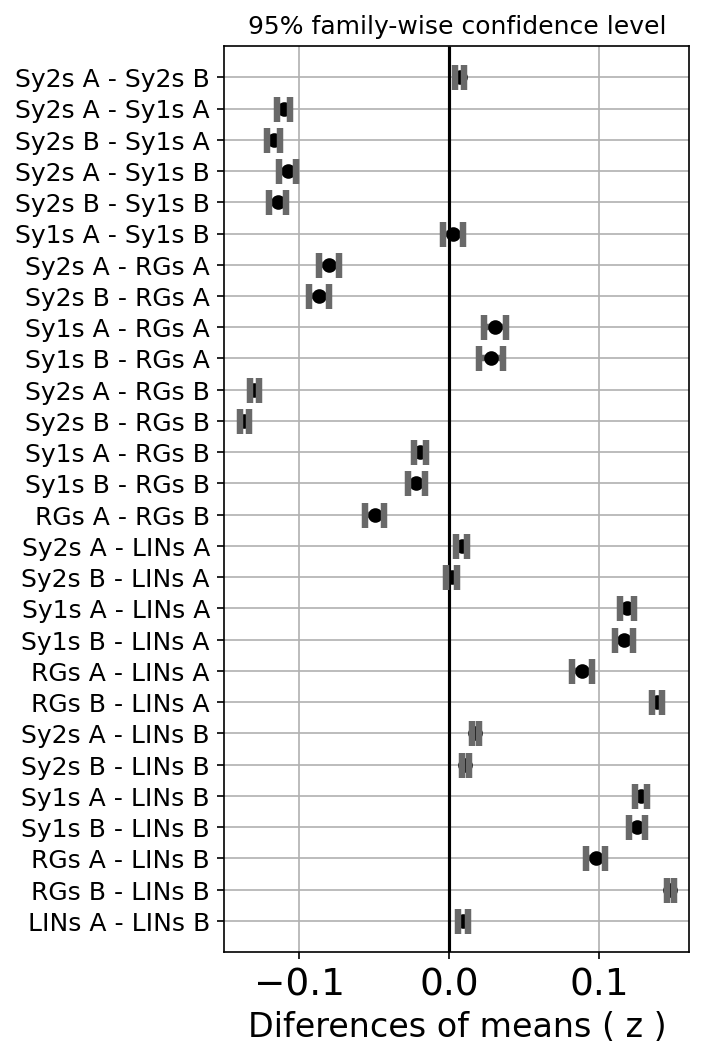}{0.3\textwidth}{(a)}
             \fig{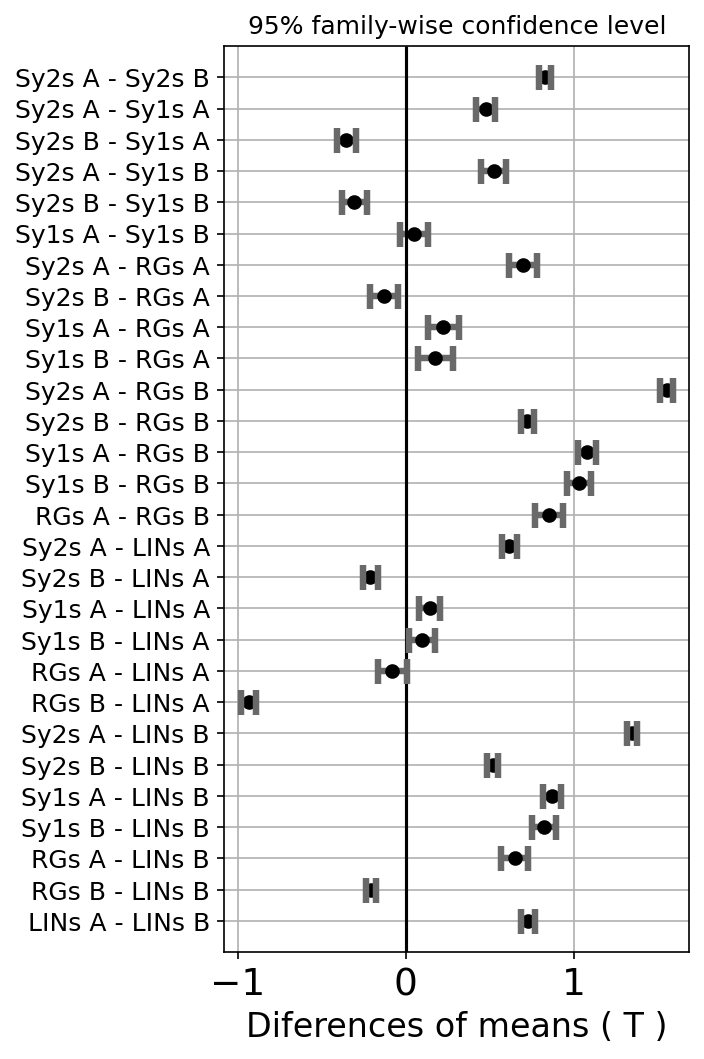}{0.3\textwidth}{(b)}
             \fig{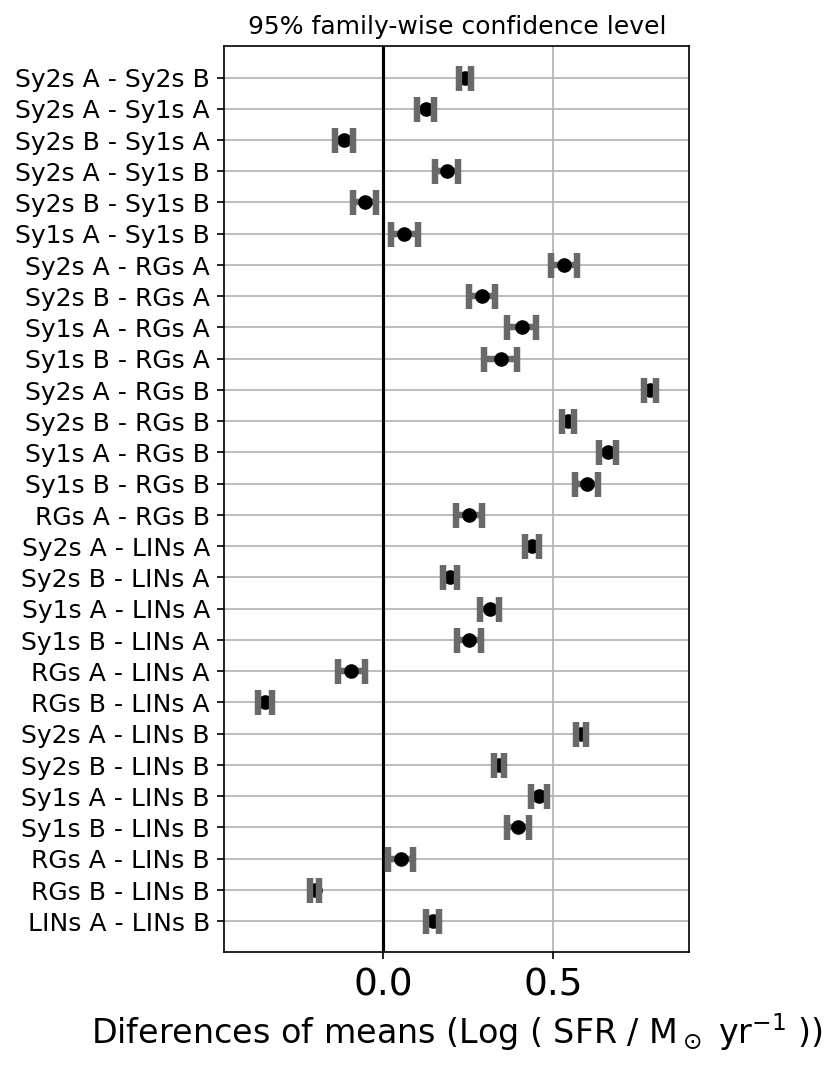}{0.35\textwidth}{(c)}
             }
\caption{Results of max-t tests comparing AGN with OF with those without OF: (a)  redshift, (b) morphology (b) SFR.
\label{figA4:maxt_OF1}
}
\end{figure*}

\begin{figure*}[ht!]
\gridline{\fig{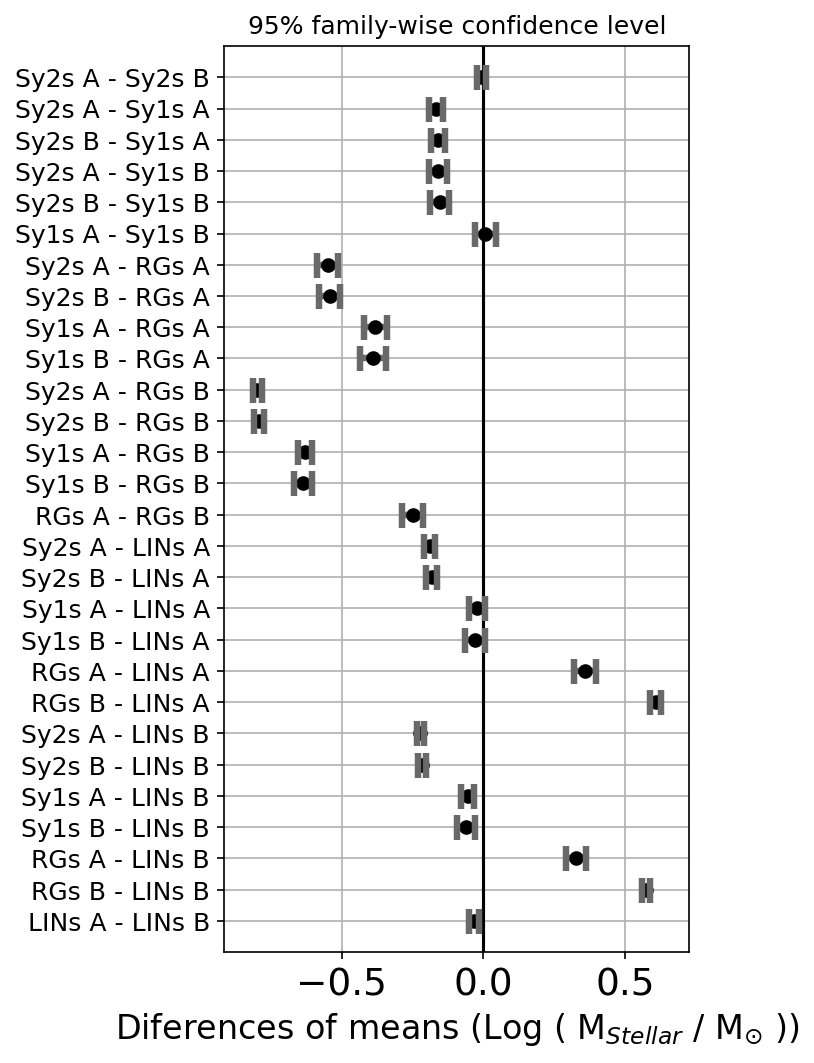}{0.35\textwidth}{(a)}
           \fig{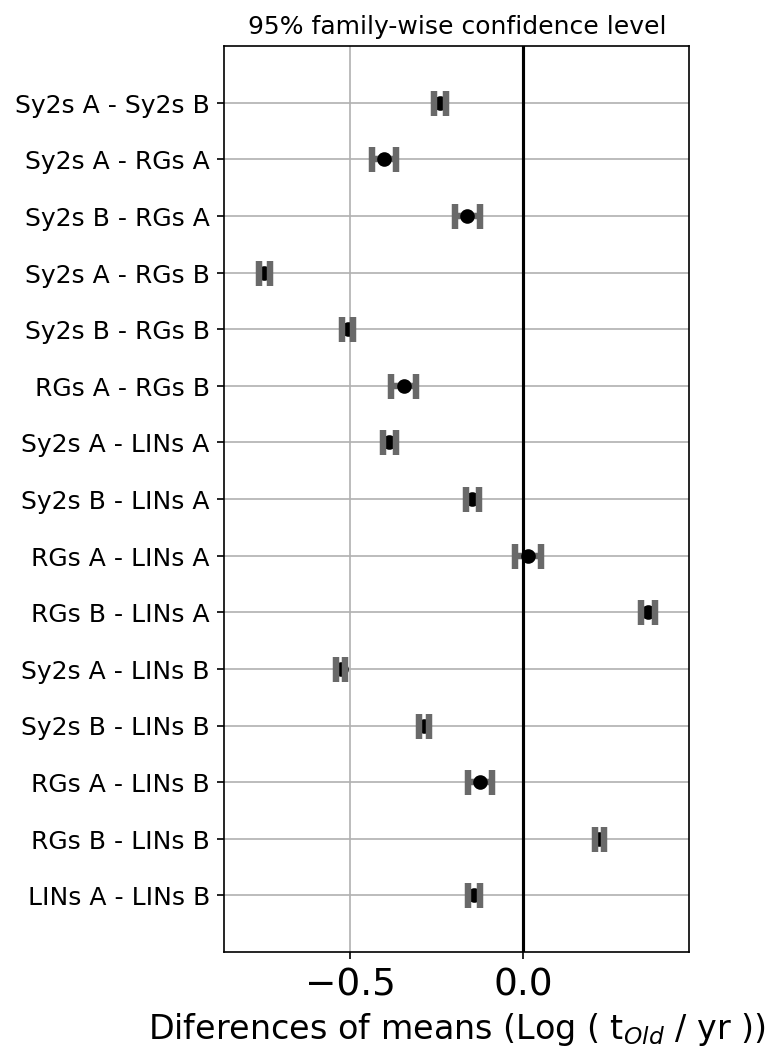}{0.34\textwidth}{(b)}  
             }
\caption{Results of max-t tests comparing AGN with OF with those without OF: (a) stellar mass of host (b) stellar ages.
\label{figA5:maxt_OF2}
}
\end{figure*}

\begin{figure*}[ht!]
\gridline{\fig{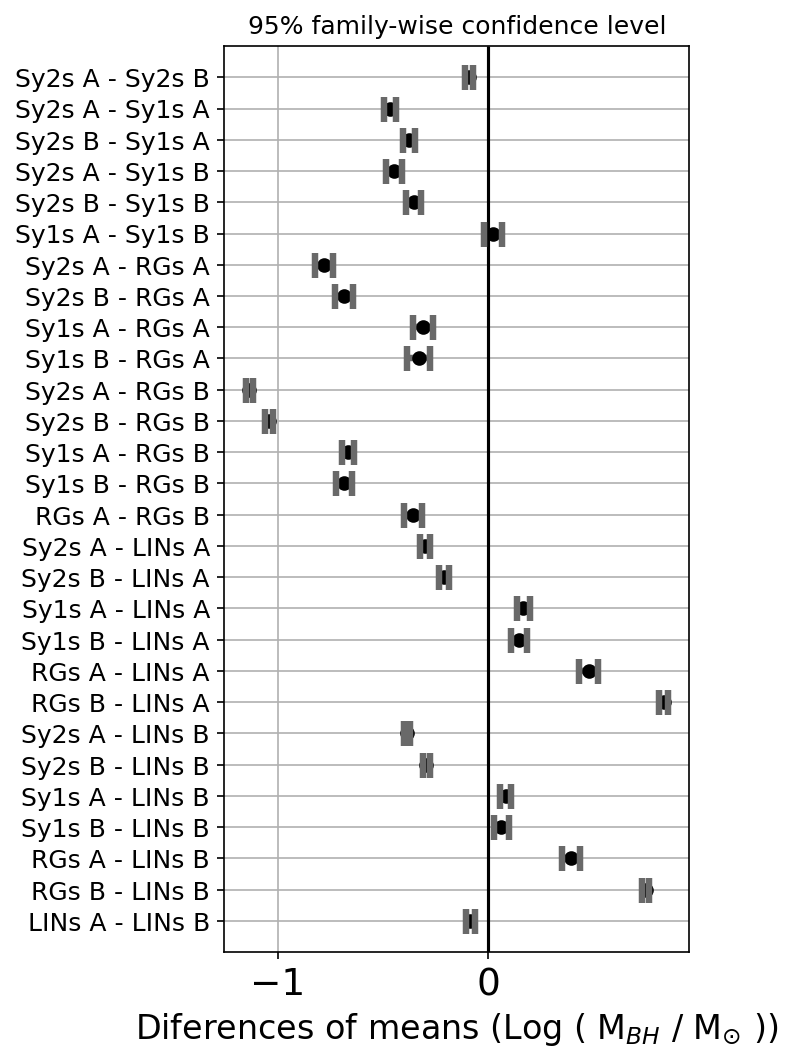}{0.32\textwidth}{(a)}
\fig{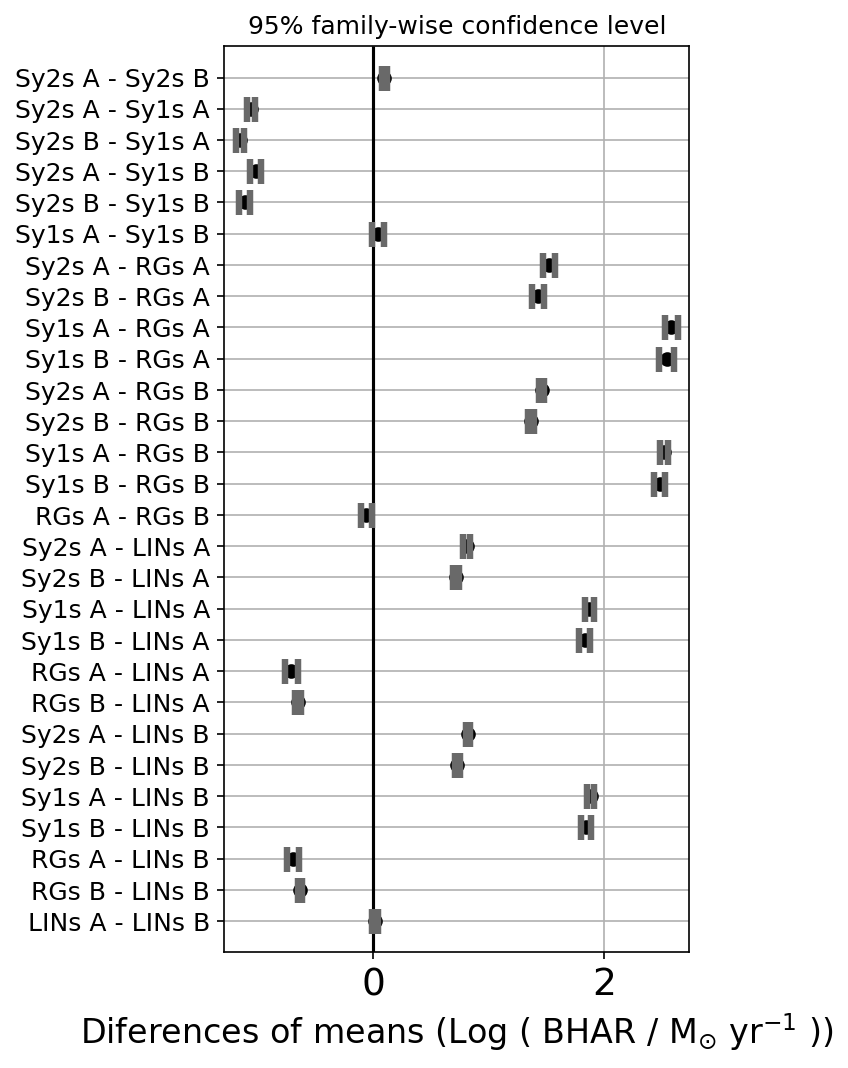}{0.34\textwidth}{(b)}
 \fig{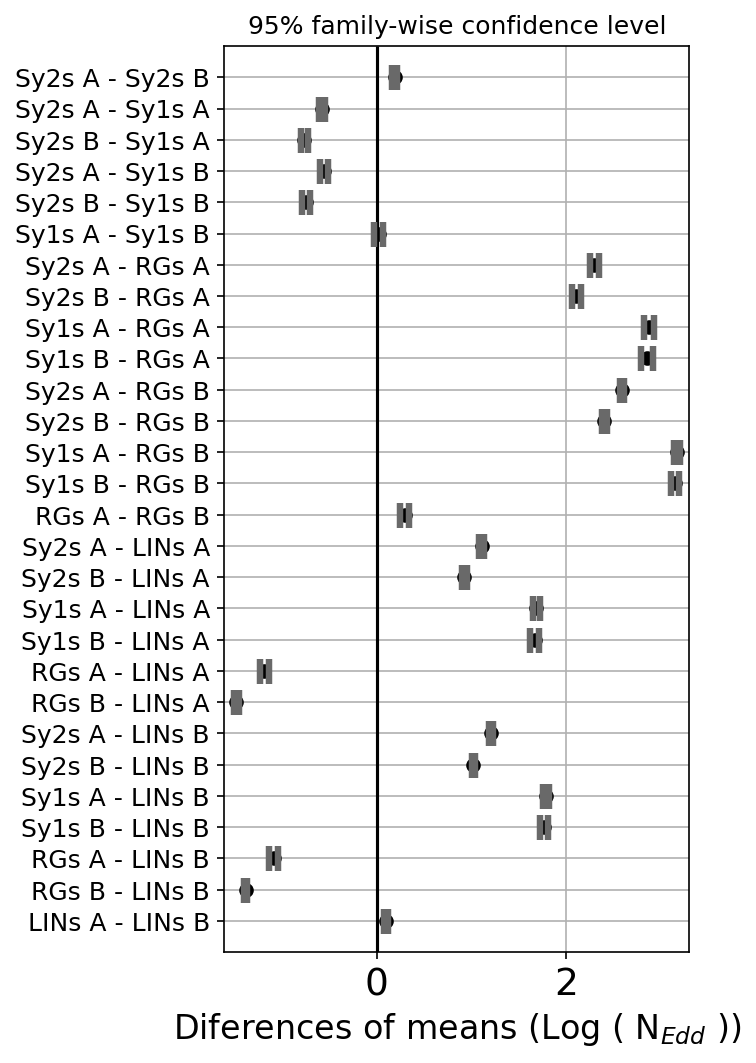}{0.31\textwidth}{(c)}                
             }
\caption{Results of max-t tests comparing AGN with OF with those without OF: (a) BH mass, (b) BHAR and (c) Eddington ratios.
\label{figA6:maxt_OF_BH}
}
\end{figure*}

\begin{figure*}[ht!]
\gridline{ \fig{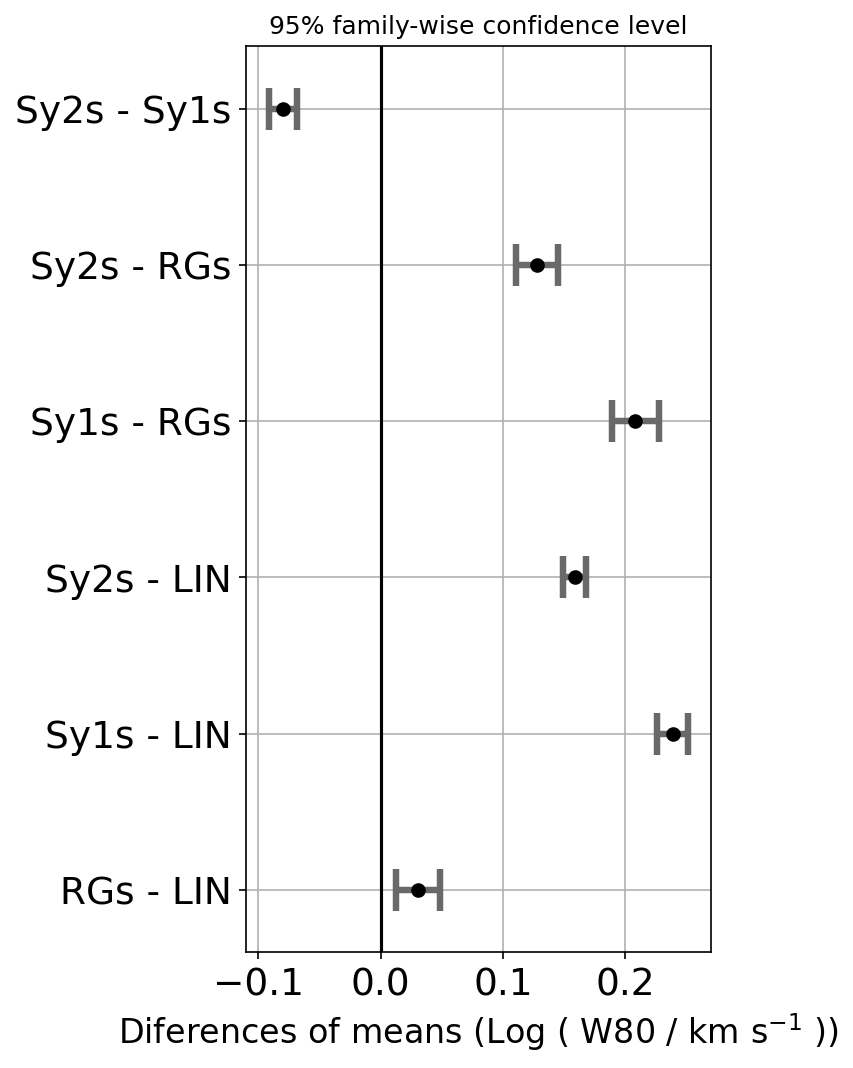}{0.4\textwidth}{(a)}
             \fig{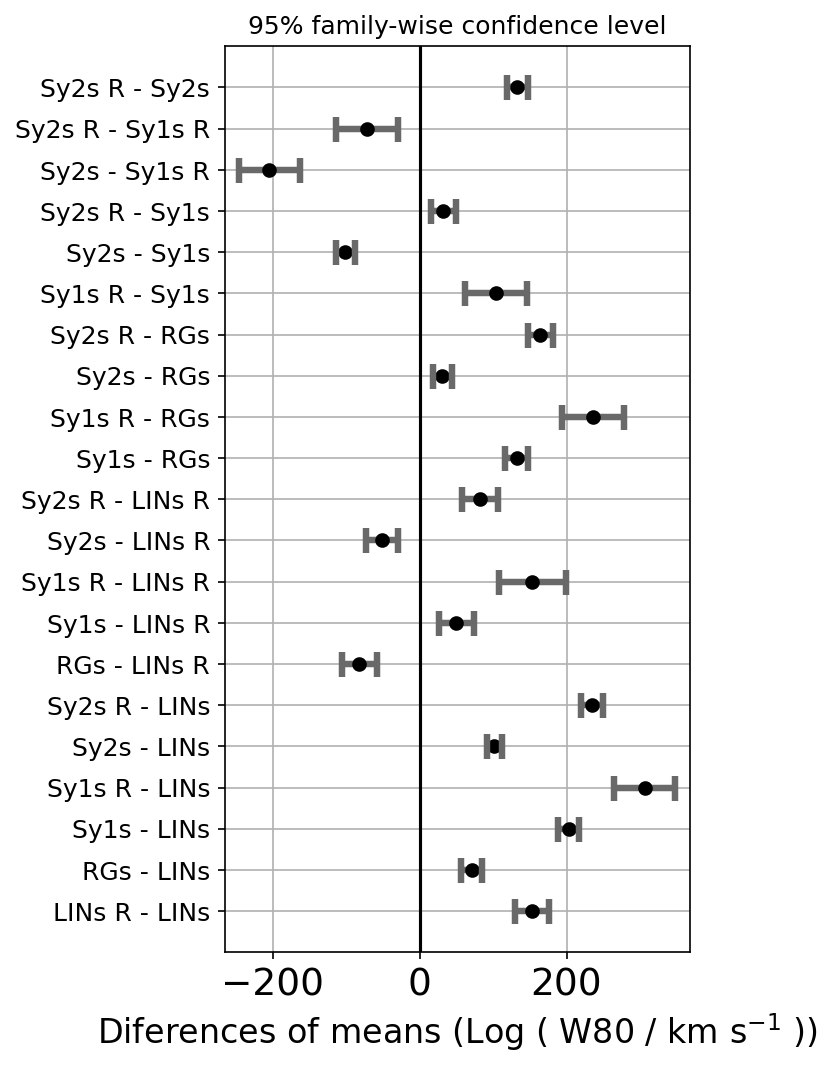}{0.39\textwidth}{(b)}
             }
\caption{Max-t test for the intensity of the wind W80 (a) in different AGN classes, (b) radio detected (R) and not detected.
\label{figA7:maxt_W80}
}
\end{figure*}

Using the max-t test \citep[explanations and references in][]{2012aJPTP}, the significance of the differences in characteristics of the galaxy host in each AGN spectral classes were determined in Figure~\ref{fig19:host1} (as example in the text) and Figure~\ref{figA1:maxt_host}, while the differences of the characteristics of their SMBHs are presented in Figure~\ref{figA2:maxt_BH} and Figure~\ref{figA3:maxt_BH} respectively. The results of the max-t tests comparing the characteristics of galaxy hosts with and without OF are presented in Figure~\ref{figA4:maxt_OF1} and Figure~\ref{figA5:maxt_OF2}, while the the results comparing the characteristics of BH are presented in Figure~\ref{figA6:maxt_OF_BH}. Finally, the results of max-t tests comparing the intensity of the in AGN are presented in Figure~\ref{figA7:maxt_W80}.

The results are presented under the form of simultaneous confidence intervals for all pairwise comparisons of the means in the four AGN types. Confidence intervals including zero indicate no statistically significant difference and the farther from zero the more significant the differences. A positive difference implies the first member of the pair has higher means than the second member (or vice versa when negative). Finally, the smaller the confidence interval and the more significant the difference.

\clearpage

\clearpage
\bibliography{JPTP_etal2022V0}{}
\bibliographystyle{aasjournal}



\end{document}